\documentclass[traditabstract]{aa}
\usepackage{graphicx}
\usepackage{txfonts}
\usepackage{color}
\usepackage{natbib}
\usepackage{epsfig}
\usepackage{hyperref}
\usepackage[switch]{lineno}
\usepackage{lineno}
	
\begin{document}
   \title{Radio to gamma-ray variability study of blazar S5 0716+714}

   \subtitle{ }

   \author{
          B. Rani \inst{1,}\thanks{Member of the International Max Planck Research School (IMPRS) for Astronomy and
Astrophysics at the Universities of Bonn and Cologne}
          \and T. P.\ Krichbaum \inst{1}
          \and L. Fuhrmann \inst{1}
          \and M. B{\"o}ttcher \inst{2,3}
          \and B. Lott \inst{4}
          \and H. D. Aller \inst{5}
          \and M. F. Aller \inst{5}
          \and E. Angelakis \inst{1} 
          \and U. Bach \inst{1}
          \and D. Bastieri \inst{6,7}
          \and A. D. Falcone \inst{8}
          \and Y. Fukazawa \inst{9}
          \and K. E. Gabanyi \inst{10,11}
          \and A. C.\ Gupta \inst{12}
          \and M. Gurwell \inst{13}
          \and R. Itoh \inst{9}
          \and K. S. Kawabata \inst{14}
          \and M. Krips \inst{15}
          \and A. A. L{\"a}hteenm{\"a}ki \inst{16}
          \and X. Liu \inst{17}
          \and N. Marchili \inst{1,7} 
          \and W. Max-Moerbeck \inst{18}
          \and I. Nestoras \inst{1}
          \and E. Nieppola \inst{16}
          \and G. Quintana-Lacaci \inst{19,20}
          \and A. C. S. Readhead \inst{18}
          \and J. L. Richards \inst{21}
          \and M. Sasada \inst{14,22} 
          \and A. Sievers \inst{19} 
          \and K. Sokolovsky \inst{1,23}
          \and M. Stroh \inst{8}
          \and J. Tammi \inst{16}
          \and M. Tornikoski \inst{16}
          \and M. Uemura \inst{14}
          \and H. Ungerechts \inst{19} 
          \and T. Urano \inst{9}
          \and J.A. Zensus \inst{1} \\
          }
   \institute{ 
              Max-Planck-Institut f$\ddot{u}$r Radioastronomie (MPIfR), Auf dem H{\"u}gel 69, D-53121 Bonn, Germany 
          \and
             Astrophysical Institute, Department of Physics and Astronomy, Ohio University Athens, OH 45701, USA 
          \and
             Centre for Space Research, North-West University, Potchefstroom Campus, Potchefstroom, 2531, South Africa
          \and 
            Universit{\'e} Bordeaux 1, CNRS/IN2p3, Centre d'Etudes Nucl{\'a}ires de Bordeaux Gradignan, 
            33175 Gradignan, France
          \and 
             Astronomy Department, University of Michigan, Ann Arbor, MI 48109-1042, USA
          \and 
             Istituto Nazionale di Fisica Nucleare, Sezione di Padova, I-35131 Padova, Italy 
          \and 
              Dipartimento di Fisica e Astronomia, Universit\`a di Padova, I-35131 Padova, Italy
          \and 
             Dept. of Astronomy and Astrophysics, Penn State University, University Park, PA 16802, USA 
          \and 
             Department of Physical Sciences, Hiroshima University, 1-3-1 Kagamiyama, Higashi-Hiroshima, Hiroshima 739-8526
          \and 
             F{\"O}MI Satellite Geodetic Observatory, PO Box 585, 1592 Budapest, Hungary  
          \and 
             Konkoly Observatory, Research Centre for Astronomy and Earth Sciences, Hungarian Academy of Sciences, H-1121 Budapest, Hungary
          \and     
              Aryabhatta Research Institute of Observational Sciences (ARIES), Manora Peak, Nainital, 263 129, India
          \and 
             Harvard-Smithsonian Center for Astrophysics, Cambridge, MA 02138 USA
          \and 
             Hiroshima Astrophysical Science Center, Hiroshima University, 1-3-1 Kagamiyama, Higashi-Hiroshima, Hiroshima 739-8526
          \and 
             IRAM, 300 rue de la piscine, F-38406 Saint-Martin d'H{\'e}res, France
          \and 
             Aalto University Mets{\"a}hovi Radio Observatory, Kylm{\"a}l{\"a}, Finland
          \and 
             Xinjiang Astronomical Observatory, Chinese Academy of Sciences, 150 Science 1-Street, Urumqi
             830011, P.R. China
          \and 
             Cahill Center for Astronomy and Astrophysics, California Institute of Technology, Pasadena, CA 91125, USA
          \and
              Instituto de Radioastronom{\'i}a Milim{\'e}trica (IRAM), Avenida Divina Pastora 7, Local 20, 18012 Granada, Spain
          \and 
              CAB, INTA-CSIC, Ctra. de Torrej\'on a Ajalvir, km 4, E-28850, Torrej\'on de Ardoz, Madrid, Spain
          \and 
              Purdue University, Department of Physics, 525 Northwestern Ave, West Lafayette, IN 47907, USA
          \and 
             Department of Astronomy, Kyoto University, Kitashirakawa-Oiwake-cho, Sakyo-ku, Kyoto 606-8502
          \and 
             Astro Space Center of Lebedev Physical Institute, Profsoyuznaya 84/32, 117997 Moscow, Russia
           }

   \date{Received ---------; accepted ----------}

 
  \abstract
{ We present the results of a series of radio, optical, X-ray and $\gamma$-ray observations of the BL Lac object 
S50716+714 carried out between April 2007 and January 2011. The multi-frequency observations 
were obtained using several ground and space based facilities. 
The intense optical monitoring of the source reveals faster repetitive variations superimposed on a long-term 
variability trend at a time scale of $\sim 350$~days. Episodes of fast variability recur on time scales of 
$\sim 60$ -- 70 days. The intense and simultaneous activity at optical and $\gamma$-ray frequencies favors 
the SSC mechanism for the production of the high-energy emission. Two major low-peaking radio flares were 
observed during this high optical/$\gamma$-ray activity period. The radio flares are characterized by a rising and a 
decaying stage and are in agreement with the formation of a shock and its evolution. We found that the evolution 
of the radio flares requires a geometrical variation in addition to intrinsic variations of the source.    
Different estimates yield a robust and self-consistent lower limits of 
$\delta \geq 20$ and equipartition magnetic field $B_{eq} \geq$ 0.36 G. Causality arguments constrain 
the size of emission region $\theta \leq 0.004$ mas. 
We found a significant correlation between flux variations at radio frequencies with those at optical 
and $\gamma$-rays. The optical/GeV flux variations lead the radio variability by 
$\sim 65$~days. The longer time delays between low-peaking 
radio outbursts and optical flares imply that optical flares are the precursors of radio ones.  
An orphan X-ray flare challenges the simple, one-zone emission models, rendering them too simple.
Here we also describe the spectral energy distribution modeling of the source from  simultaneous data taken 
through different activity periods.  
}



\maketitle

\section{Introduction}
The BL Lac object S5 0716+714 is among the most  extremely variable blazars. 
The optical continuum of the source is so featureless that it is hard to estimate its redshift.
\citet{nilsson2008} claimed a value of $z = 0.31\pm0.08$ based on the photometric detection of 
the host galaxy. Very recently, the detection of intervening Ly$\alpha$ systems in the ultra-violet 
spectrum of the source confirms the earlier estimates with a redshift value 0.2315 $< z < $ 0.3407 
\citep{danforth2012}. 
This source has been classified as an intermediate-peaked blazar (IBL) by \citet{giommi1999}, as the 
frequency of the first spectral energy distribution (SED) peak varies between 10$^{14}$ and 10$^{15}$ Hz, and thus does not 
fall into the wavebands specified by the usual definitions of low and high energy peaked blazars 
(i.e. LBLs and HBLs). 

S5~0716+714 is one of the brightest BL Lacs in the optical bands and has an optical intraday variability (IDV) duty
cycle of nearly one \citep{wagner1995}.  Unsurprisingly, this source has been the subject of several 
optical monitoring campaigns on intraday (IDV) timescales \citep[e.g.][and references therein]{wagner1996, 
rani2011, montagni2006, gupta2009, gupta2012}. The source has shown five major optical 
outbursts  separated roughly by $\sim$3.0$\pm$0.3 years \citep{raiteri2003}. High optical polarization  
$\sim$ 20$\%$ -- 29$\%$ has also been observed in the source \citep{takalo1994, fan1997}. 
\citet{gupta2009} analyzed the excellent intraday optical light curves of the source observed by 
\citet{montagni2006} and reported good evidence of nearly periodic oscillations ranging 
between 25 and 73 minutes on several different nights. Good evidence for the presence of  
$\sim 15$-min periodic oscillations at optical frequencies has been reported by \citet{rani2010b}. 
A detailed multiband short-term optical flux and color variability study of the source is reported 
in \citet{rani2010a}. There we found that the optical spectra of the source tend to be bluer 
with increasing brightness.

There is a series of papers covering flux variability studies 
\citep[][and references therein]{quirrenbach1991, wagner1996} and morphological/kinematic studies 
at radio frequencies \citep[][and references therein]{witzel1988, jorstad2001, antonucci1986, 
bach2005, rastorgueva2011}. Intraday variability at radio wavelengths is likely to be 
a mixture of intrinsic and extrinsic (due to interstellar scintillation) mechanisms. A significant 
correlation between optical -- radio flux variations at day-to-day timescales has been reported by 
\citet{wagner1996}. The frequency dependence of the variability index at radio bands is not found 
to be consistent with interstellar scintillation \citep{fuhrmann2008}, which implies the presence 
of very small emitting regions within the source. However, the IDV time scale does show evidence 
in favor of an annual modulation, suggesting that the IDV of S5~0716+714 could be dominated by 
interstellar scintillation \citep{liu2012}.

EGRET on board the {\it Compton Gamma-ray Observatory (CGRO)} detected  high-energy 
$\gamma$-ray ($>$100 MeV) emission from S5~0716+714 several times from 1991 to 1996 \citep{hartman1999, 
lin1995}. Two strong $\gamma$-ray flares on September and October 2007 were detected in the 
source  with AGILE \citep{chen2008}. These authors have also carried out SED modeling of the source 
with two synchrotron self-Compton (SSC) emitting components, representative of a slowly
and a rapidly variable component, respectively.  Observations by BeppoSAX 
\citep{tagliaferri2003} and XMM-Newton \citep{foschini2006} provide evidence for a concave 
X-ray spectrum in the 0.1 -- 10~keV band, a signature of the presence of both the steep tail 
of the synchrotron emission and the flat part of the Inverse Compton (IC) spectrum.      
Recently, the MAGIC collaboration reported the first detection of VHE $\gamma$-rays from the 
source at the $5.8 \sigma$ significance level \citep{anderhub2009}. The discovery of S5~0716+714 as a VHE 
$\gamma$-ray BL Lac object was triggered by its very high optical state, suggesting a possible correlation 
between VHE $\gamma$-ray and the optical emissions. This source is also among the bright blazars in the 
{\it Fermi}/LAT (Large Area Telescope) Bright AGN Sample (LBAS) \citep{abdo2010LBAS}, whose GeV spectra are governed by a broken 
power law. The combined GeV -- TeV spectrum of the source displays absorption-like features in 10 -- 100~GeV 
energy range \citep{senturk2011}.

The broadband flaring behavior of the source is even more complex.  
A literature study reveals that the broadband flaring activity of the source is not 
simultaneous at all frequencies \citep{chen2008, villata2008, vittorini2009, ostorero2006}.
Also, it is hard to explain both the slow modes of variability at radio and hard X-ray bands and 
the rapid variability observed in the optical, soft X-ray, and $\gamma$-ray bands using 
a single component SSC model \citep[see][]{villata2008,  giommi2008, chen2008, vittorini2009}. 
The X-ray spectrum of the source contains contributions from both synchrotron and IC 
emission \citep{foschini2006, ferrero2006} and the simultaneous optical-GeV variations favor an
SSC emission mechanism \citep{chen2008, vittorini2009}.

Despite of several efforts to understand the broadband flaring activity of the source, we still 
do not have a clear knowledge of the emission mechanisms responsible for its origin. 
In particular, the location and the mechanism responsible for the high-energy emission and the 
relation between the variability at different wavelengths are not yet well understood.  
Therefore, it is important to investigate whether a correlation exists between optical and 
radio emissions, which are both ascribed to synchrotron radiation from relativistic electrons in 
a plasma jet. If the same photons are up-scattered to high energies, simultaneous variability 
features could be expected at optical -- GeV frequencies. But the observed variability often challenges 
such scenarios. To explore the broadband variability features and to understand the underlying emission 
mechanism, an investigation of long-term variability over several decades of frequencies is 
crucial. The aim of the broadband variability study reported in this paper is to provide a 
general physical scenario which allows us to put each observed variation of the source across 
several decades of frequencies in a coherent context.

Here, we report on a broadband variability study of the source spanning 
a time period of April 2007 to January 2011. The multi-frequency observations comprise 
GeV monitoring by {\it Fermi}/LAT, X-ray observations by Swift-XRT, as well as optical and radio 
monitoring by several ground 
based telescopes.  More explicitly, we  investigate the correlation of $\gamma$-ray activity 
with the emission at lower frequencies, focusing on the individual flares observed between 
August 2008 and January 2011. The evolution of radio (cm and mm) spectra is tested in the context of a standard 
shock-in-jet model. The broadband SED of the source is investigated using a one-zone  synchrotron 
self-Compton (SSC) model and also with a hybrid SSC plus external radiation Compton model. 
In short, this study allows us to shed light on the broadband radio-to-$\gamma$-ray flux and 
spectral variability during a flaring activity phase of the source over  this period. 

This paper is structured as follows. Section 2 provides a brief description of the multi-frequency 
data we employed. In Section 3, we report the statistical analysis and its results. Our discussion is 
given in Section 4 and we present our conclusions in Section 5.


\section{Multi-frequency Data }

From April 2007 to February 2011, S5~0716+714 was observed using both ground and space 
based observing facilities. These multi-frequency observations of the source extend over 
a frequency range between radio and $\gamma$-rays including optical and X-rays. 
In the following subsections, we summarize the observations and data 
reduction.

  \begin{figure*}
   \centering
\includegraphics[scale=0.95]{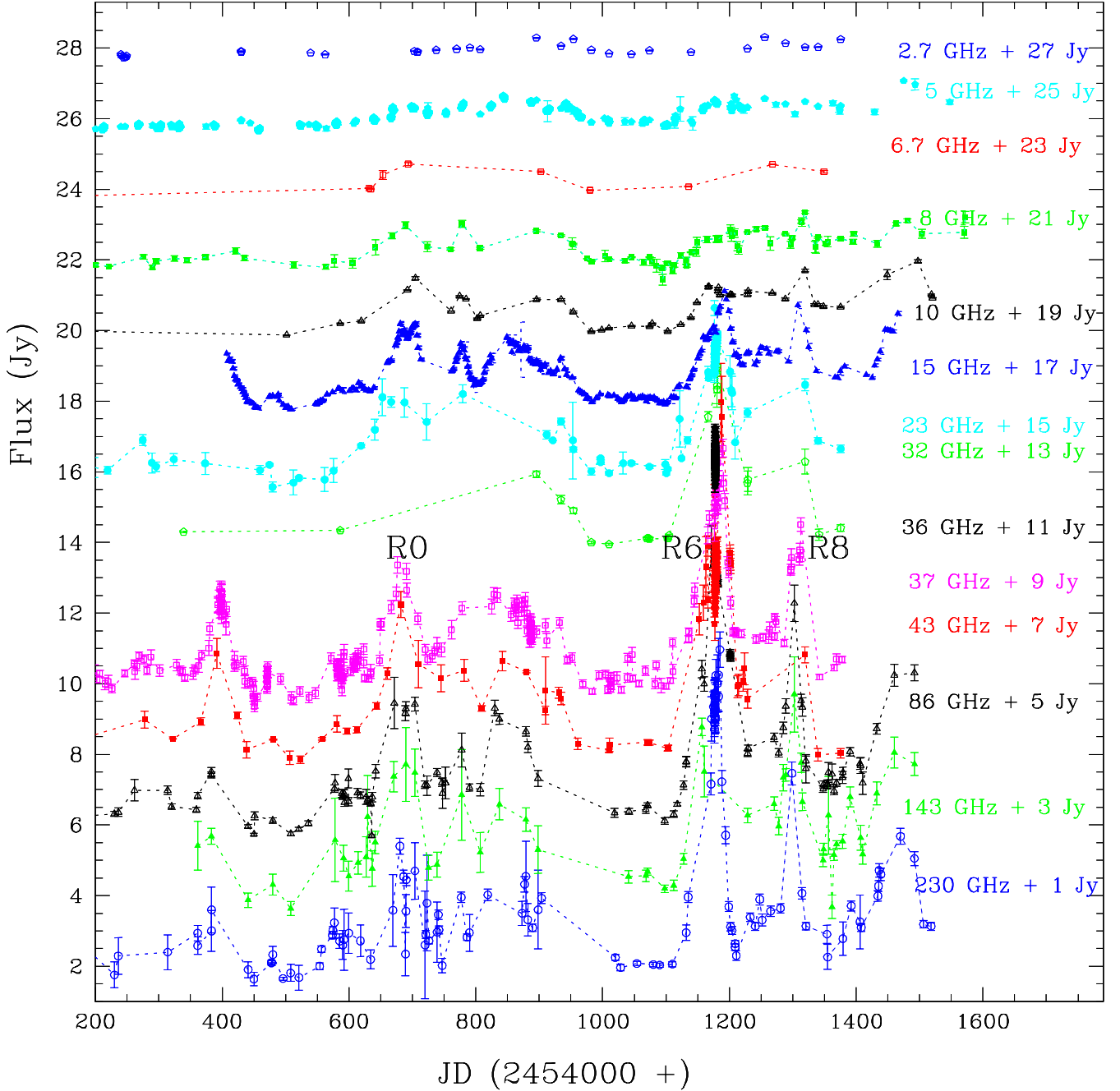} 
  \caption{Radio to mm wavelength light curves of S5 0716+714 observed over the past $\sim$3 years. 
For clarity, the light curves at different frequencies are shown with arbitrary offsets 
(indicated by a ``Frequency + x Jy" label).  The major radio flares are labeled as ``R0", ``R6" 
and ``R8" (see Fig. \ref{plot_flx_total_LC} for the details of labeling). }
\label{plot_flx_rad}
\end{figure*}

\begin{table}
\caption{ Ground based radio observatories }
\begin{tabular}{l c c} \hline  
Observatory          &  Tel. dia.      &  Frequency  (GHz)  \\\hline 
Effelsberg, Germany  & 100 m                    & 2.7, 4.8, 6.7 \\
                     &                          & 8.3, 10.7, 15, 23  \\
                     &                          & 32, 43    \\                   
UMRAO, USA           & 26 m                     & 4.8, 8, 14.5  \\
NOTO, Italy          & 32 m                     & 5, 8, 22, 43  \\
Urumqi, China        & 25 m                     & 4.8    \\
OVRO, USA            & 40 m                     & 15      \\
Metsahovi, Finland   & 14 m                     & 37        \\
PdBI, France         & 6$\times$15 m            & 86, 143, 230 \\
Pico Veleta, Spain   & 30 m                     & 86, 143, 230  \\
SMA, USA             & 8$\times$6 m             & 230, 345    \\\hline
\end{tabular} \\
\end{table}

\subsection{Radio and mm data}
We collected 2.7 to 230 GHz radio wavelength data of the source over a time 
period of April 2007 to January 2011 (JD'\footnote{JD' = JD$-$2454000} = 200 to 1600) using the 9 radio 
telescopes listed in Table 1. The cm/mm radio light curves of the source were observed as a part 
of observations within the framework of F-GAMMA program \citep[{\it Fermi}-GST related 
monitoring program of $\gamma$-ray blazars, ][]{fuhrmann2007, angelakis2008}.
The overall frequency range spans from 2.7 GHz to 230 GHz using the Effelsberg
100 m telescope (2.7 to 43 GHz) and the IRAM 30 m Telescope at the Pico Veleta 
(PV) Observatory (86 to 230 GHz).  
These flux measurements were performed quasi-simultaneously using the  
cross-scan method slewing over the source position, in azimuth and elevation
direction with adaptive number of sub-scans for reaching the desired sensitivity. 
Subsequently, atmospheric opacity correction, pointing off-set correction, gain 
correction and sensitivity correction have been applied to the data.

This source is also a part of an ongoing blazar monitoring program at 15 GHz at the 
Owens Valley Radio Observatory (OVRO) 40-m radio telescope which provides the radio 
data sampled at 15 GHz. 
We have also used the combined data from the University of Michigan Radio Astronomy 
Observatory \citep[UMRAO; 4.8, 8 and 14.5 GHz,][]{aller1985} and the Mets{\"a}hovi 
Radio Observatory \citep[MRO; 22 and 37 GHz;][]{terasranta1998, terasranta2004},
which provide us with radio light curves at 5, 8, 15 and 37 GHz. Additional flux 
monitoring  at 5, 8, 22 and 43 GHz radio bands is obtained using 32 m telescope at 
NOTO radio observatory. The Urumqi 25 m radio telescope monitors the source at 5 GHz. 
The 230 and 345 GHz data
are provided by the Submillimeter Array (SMA) Observer 
Center\footnote{http://sma1.sma.hawaii.edu/callist/callist.html} data base \citep{gurwell2007},
complemented by some measurements from PV and the Plateau de Bure Interferometer (PdBI). 
The radio light curves of the source are shown in Fig \ref{plot_flx_rad}. 
The mm observations are closely coordinated with the more general flux density monitoring 
conducted by IRAM, and data from both programs are also included.

 \begin{figure*}
\includegraphics[scale=0.95]{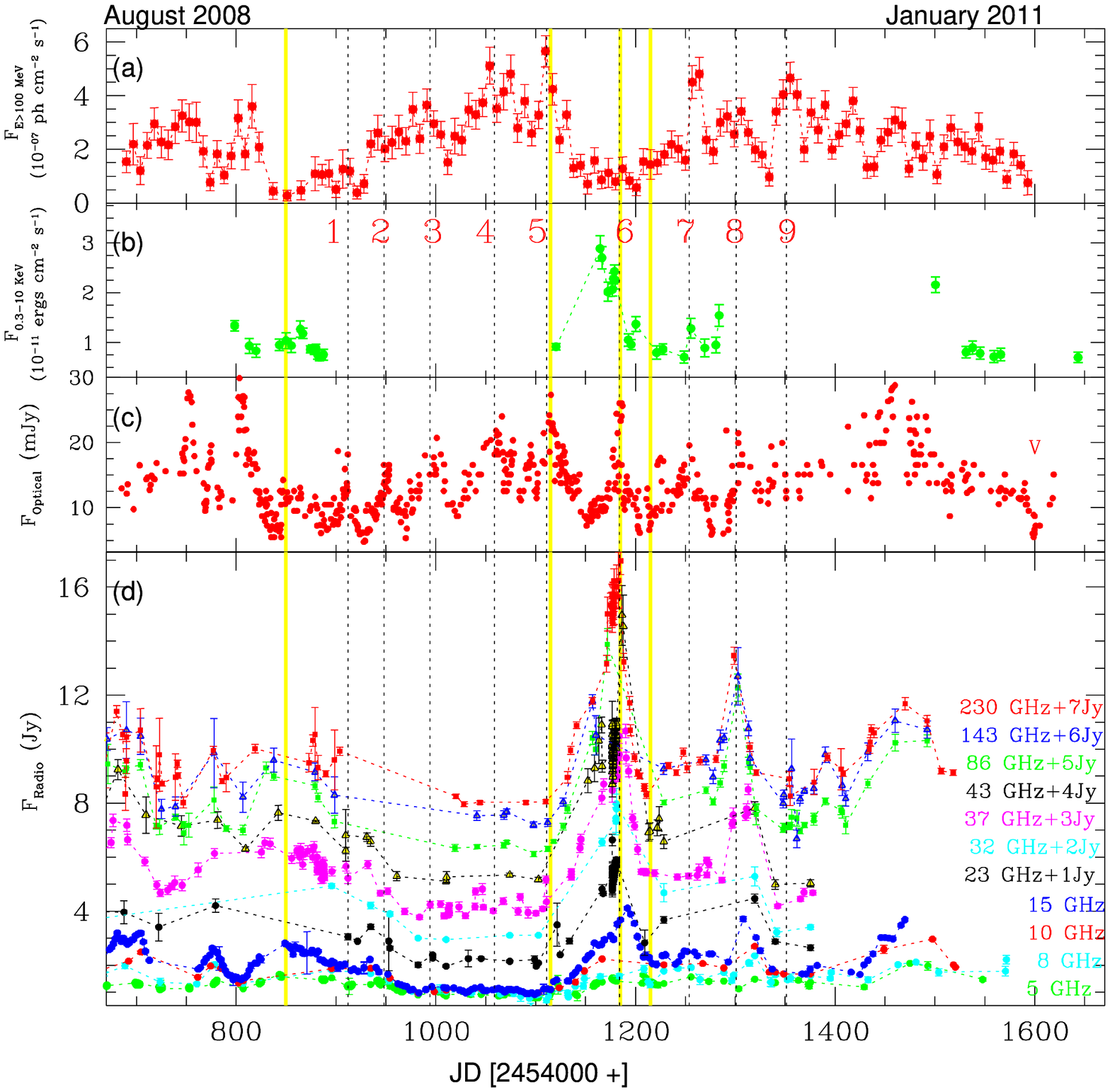}
   \caption{Light curves of 0716+714 from $\gamma$-ray to radio wavelengths (a): GeV light curve at E$>$100 MeV, 
(b): X-ray light curve at 0.3$-$10 KeV, (c): optical V passband light curve and (d) : 5 to 230 GHz radio 
light curves. Vertical dotted lines are marked w.r.t. different optical flares labeling the broadband flares 
as ``G" for $\gamma$-rays, ``X" for X-rays, ``O" for optical and ``R" for Radio followed by the number close to flare. 
The yellow area represents the period for which we construct the broadband SEDs of the source (see Section \ref{spectra_all} 
for details).       }
\label{plot_flx_total_LC}
    \end{figure*}

\subsection{Optical data}
\label{data_opt}
Optical V passband data of the source were obtained from the observations at 
the 1.5-m Kanata Telescope located on Higashi-Hiroshima Observatory over a time period of February 14, 2009 
to June 01, 2010 (JD' = 877 to 1349). 
The Triple Range Imager and SPECtrograph \citep{watanabe2005} was used for the observations 
from JD' = 612 to 1228. Then, the HOWPol \citep[Hiroshima One-shot Wide-field 
Polarimeter,][]{kawabata2008} 
was used from JD'~1434 to 2233 with the V-band filter. Exposure times for an image
ranged from 10 to 80~s, depending on the magnitude of the
object and the condition of sky. The photometry on the CCD images
was performed in a standard procedure: after bias subtraction and flat-field division, 
the magnitudes were calculated using the aperture photometry technique.

Additional optical V-passband data were obtained from the 2.3~m Bok Telescope of Steward 
Observatory from April 28, 2009 through June 2, 2010 (JD' = 950--1350).  These data are from the public 
data archive that provides the results of polarization and flux monitoring of  
bright $\gamma$-ray blazars selected from the {\it Fermi}/LAT-monitored blazar 
list\footnote{http://james.as.arizona.edu/$\sim$psmith/Fermi}. We have also 
included optical V passband archival data extracted from the {\it American Association 
of Variable Star Observers} (AAVSO; see http://www.aavso.org/ for more information)
over a period September 2008 to January 2011 (JD' = 710--1600). The combined optical V passband flux light 
curve of S5~0716+714 is shown in Figure \ref{plot_flx_total_LC} (c).

\subsection{X-ray data}
The X-ray (0.3 -- 10~keV) data were obtained by {\it Swift}-XRT over a time period of September 
2008 to January 2011 (JD' = 710--1600). The {\it Swift}-XRT data were processed using the most recent versions of the 
standard {\it Swift} tools: {\it Swift} Software version 3.8, FTOOLS version 6.11 and XSPEC 
version 12.7.0 \citep[see][and references therein]{kalberla2005}.

All of the observations were obtained in photon counting (PC) mode. Circular and annular regions 
are used to describe the source and background areas, respectively, and the radii of both regions 
depend on the measured count rate using the FTOOLS script {\it xrtgrblc}. Spectral fitting was done 
with an absorbed power-law with $N_H = 0.31 \times 10^{21} cm^{-2}$ set to the Galactic value found 
by \citet{kalberla2005}. For 
three of the observations, there were more than 100 counts in the source region, and a chi-squared 
statistic is used with a minimum bin size of 20 cts/bin. For the bin centered on JD'=1214, 
only 62 counts were found in the source region, and the unbinned data are fitted using C-statistics 
as described by \citet{cash1979}. One sigma errors in the de-absorbed flux were calculated assuming 
that they share the same percentage errors as the absorbed flux for the same time and energy range.
The X-ray light curve of the source is shown in Fig. \ref{plot_flx_total_LC}(b).

\subsection{$\gamma$-ray data}
We employ $\gamma$-ray (100 MeV -- 300 GeV) data collected between
a time period August 08, 2008 to January 30, 2011 (JD'= 686--1592) in 
survey mode by {\it Fermi}/LAT. The LAT data are analyzed using the standard 
ScienceTools (software version v9.23.1) and the instrument response function
 P7V6\footnote{http://fermi.gsfc.nasa.gov/ssc/data/analysis/scitools/overview.html}. 
Photons in the Source event class are selected for this analysis. 
We select $\gamma$-rays with zenith angles less than 100 deg to avoid contamination from $\gamma$-rays 
produced by cosmic ray interactions in the upper atmosphere.
The diffuse emission from our Galaxy 
is modeled using a spatial model\footnote{http://fermi.gsfc.nasa.gov/ssc/data/access/lat/BackgroundModels.html} 
(gal$\_$2yearp7v6$\_$v0.fits) which was derived with {\it Fermi}/LAT 
data taken during the first two years of operation. The extragalactic diffuse and residual 
instrumental backgrounds are modeled as an isotropic component (isotropic$\_$p7v6source.txt)  with 
both flux and spectral photon index left free in the model fit. 
The data analysis is done   with an 
unbinned maximum likelihood technique \citep{mattox1996}  using  the publicly available 
tools\footnote{http://fermi.gsfc.nasa.gov/ssc/data/analysis/scitools/likelihood$\_$tutorial.html}.  
We analyzed a Region of Interest (RoI) of 10$^\circ$ in radius, centered 
on the position of the $\gamma$-ray source associated with S5~0716+714, using the 
maximum-likelihood algorithm implemented in {\it gtlike}. In the RoI model, we include all 
the 24 sources within 10$^\circ$ with their model parameters fixed to their 2FGL catalog values, as 
none of these sources are reported to be variable \citep[see][for details]{2fgl_cat}. 
The contribution of these 24 sources within the RoI model to the observed variability of the 
source is negligible as they are very faint compared to S5~0716+714. 
The LAT instrument-induced variability is tested with bright pulsars and is $\sim$2$\%$ and is much 
smaller than the statistical errors reported for the source \citep{ackermann2012}.

Source variability is investigated by producing the light curves (E $>$ 100 MeV) by 
likelihood analysis with time bins of 1, 7 and 30 days. 
The bin-to-bin exposure time variation is less than 7$\%$. 
  The light curves are produced by 
modeling the spectra over each bin by a simple power law which provide a good fit over 
these small bins of time. Here, we set both the photon index and integral flux as free parameters. 
We will use the weekly averaged light curves for the multi-frequency analysis, as the daily 
averaged flux points have a low TS (Test Statistics) value ($<$9).  In a similar 
way, we construct the GeV spectrum of the source over different time periods (see Section \ref{spectra_all} 
for details). We split the 0.1 to 100~GeV spectra into 5 different energy bins : 
0.1 -- 0.3, 0.3 -- 1.0, 1.0 -- 3.0, 3.0 -- 10.0 and 10.0 -- 100.0~GeV. A simple power 
law with constant photon index $\Gamma$ (the best fit value obtained for the entire energy range)  
provides a good estimate of integral flux over 
each energy bin. The GeV spectrum of the source is investigated over four different time periods, which represent 
different brightness states of the source and will 
be used in broadband spectral modeling.  The details of the broadband spectral analysis are given in 
Section \ref{spectra_all}.

\section{Analysis and Results}

\subsection{Light Curve Analysis}

\subsubsection{Radio frequencies }    
Fig. \ref{plot_flx_rad} displays the 2.7 -- 230~GHz radio band light curves of the source observed 
over the past $\sim 3$~years. The source exhibits significant variability, being more rapid and pronounced 
towards higher frequencies over this period with two major outbursts. 
Towards lower frequencies ($<$10 GHz) the individual flares appear less pronounced and broaden in time. 

\begin{figure}
   \centering
\includegraphics[scale=0.45]{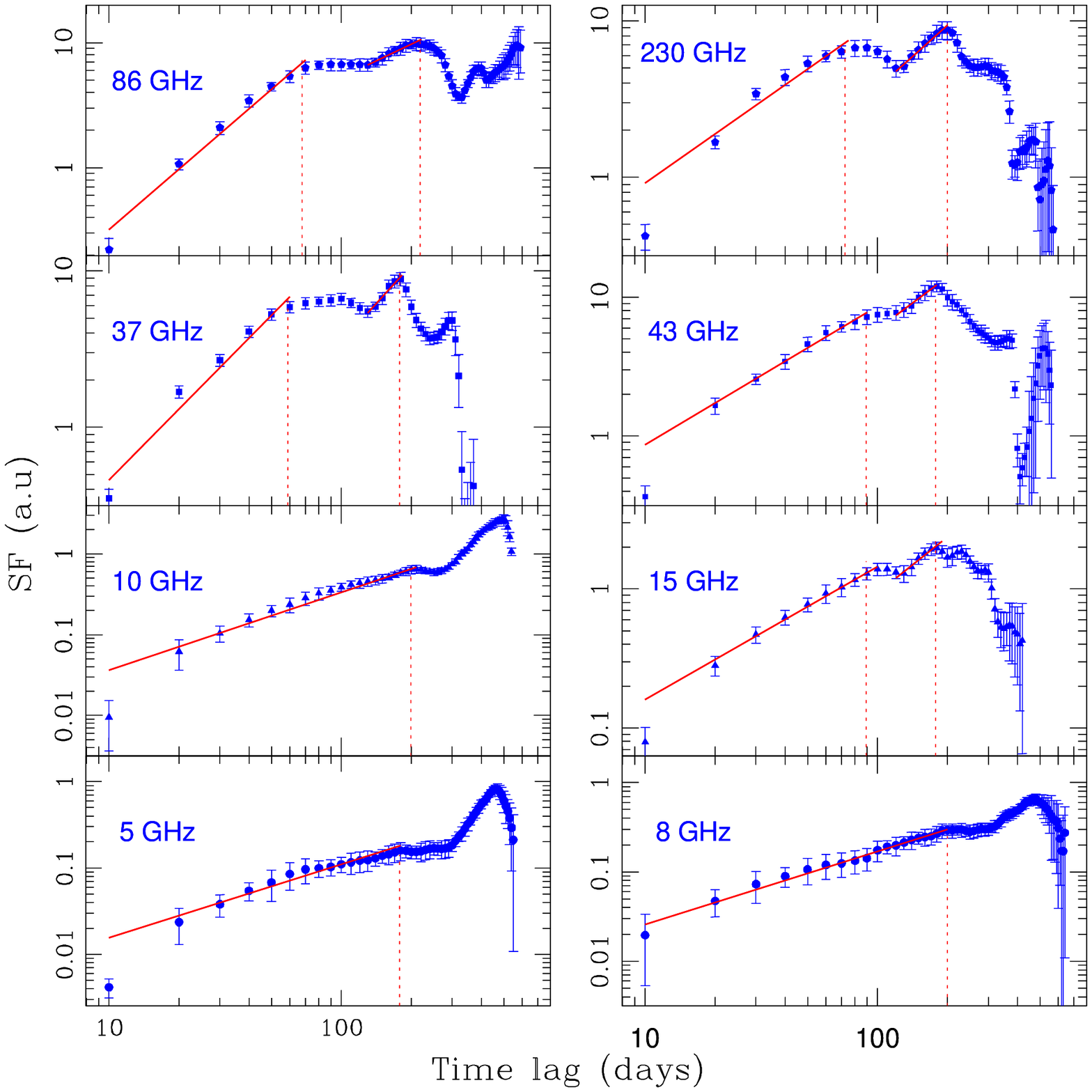}
\includegraphics[scale=0.31,angle=-90]{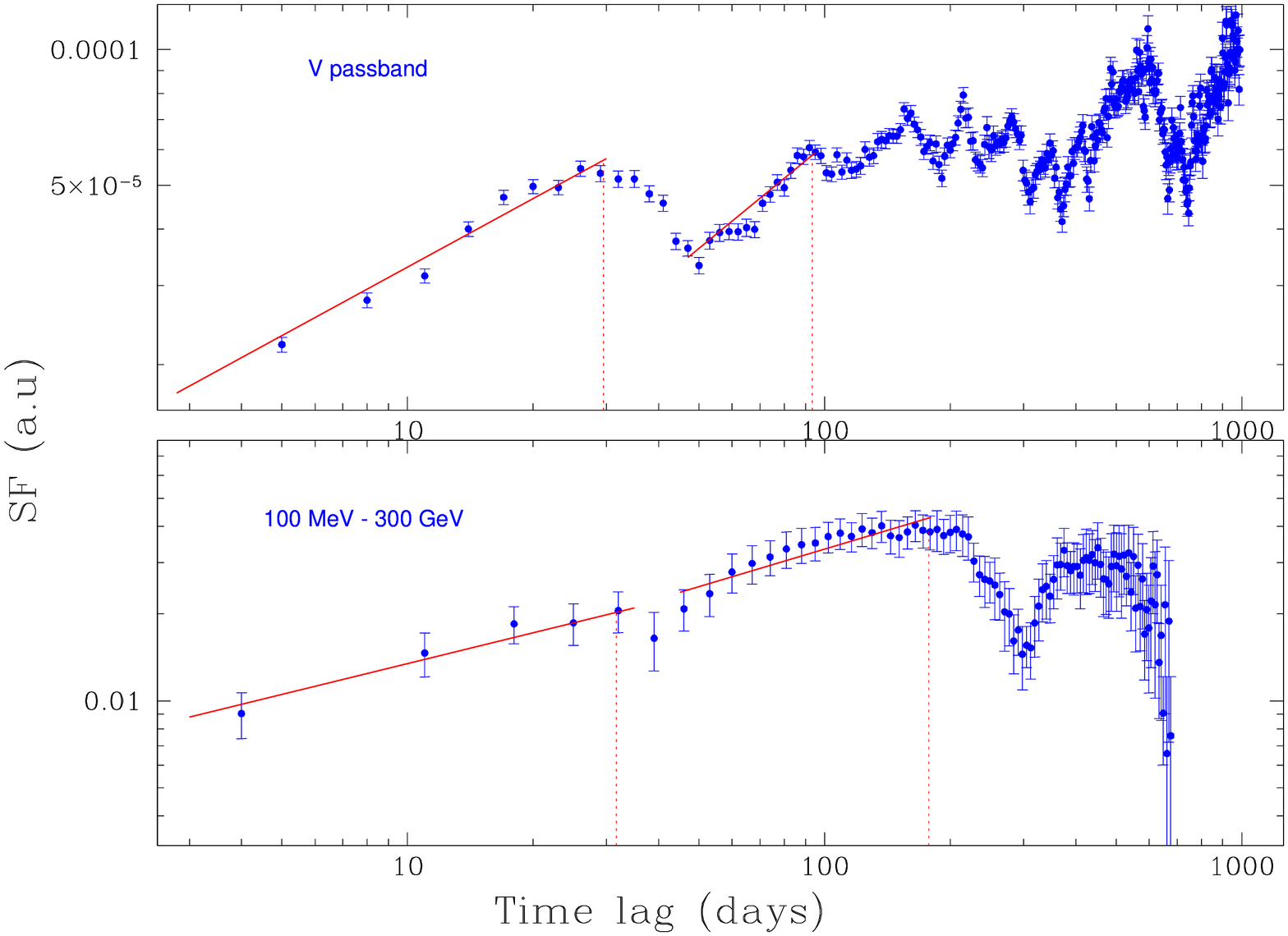}
   \caption{Top: Structure functions at radio frequencies. The solid curves represent the best fitted power laws. 
The dotted lines in each plot indicate the timescale of variability ($t_{var}$).  Bottom: Structure functions at optical 
and $\gamma$-ray frequencies. }
\label{sf_all}
    \end{figure}

To quantify the strength of variability at different radio frequencies, we extract 
the time scale of variability ($t_{var}$) from the observed light curves. 
We employed the structure function (SF) \citep{simonetti1985} analysis method 
following  \citet{rani2009}. The radio SF curves are shown in Fig. \ref{sf_all}.

\begin{table}
\caption{Variability time scales at radio wavelengths}
\begin{tabular}{l c c c c} 
Frequency & $\beta1^*$ &t$_{var1}$ (days) & $\beta2$ & t$_{var2}$ (days)      \\\hline
15 GHz    &0.95$\pm$0.03 & 100$\pm$5   &1.32$\pm$0.04  & 195$\pm$5      \\
37 GHz    &1.50$\pm$0.13 & 100$\pm$5   &1.64$\pm$0.10  & 200$\pm$5      \\
43 GHz    &0.99$\pm$0.02 &  90$\pm$5   &1.23$\pm$0.04  & 180$\pm$5       \\
86 GHz    &1.60$\pm$0.10 &  90$\pm$5   &0.89$\pm$0.02  & 180$\pm$5       \\
230 GHz   &1.04$\pm$0.03 &  90$\pm$5   &1.31$\pm$0.03  & 180$\pm$5       \\\hline
\end{tabular} \\
$^*$ : P(f) $\sim$ $f^{\beta}$, $\beta$ is the slope of the power law fit.  
\label{tab_sf}
\end{table}

 \begin{figure}
\includegraphics[scale=0.3,angle=-90]{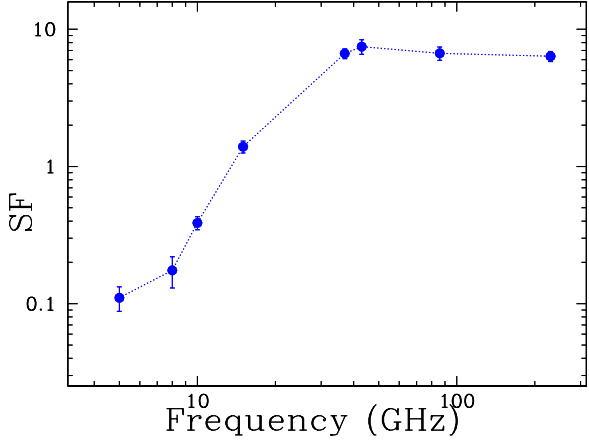}
   \caption{SF vs frequency at GHz frequencies for a structure function time = 100 days. }
\label{sf_radio}
    \end{figure}

At 15 GHz and higher radio frequencies, the SF curve follow a continuous rising trend showing a peak at $t_{var1}$, 
following a plateau and again reaching a maximum at $t_{var2}$. However,  the SF curve at 10 GHz and lower frequencies 
do not reveal a sharp break in the slope as the variability features seem to be milled out at these frequencies. 
We do not consider variability 
features at time lags longer than half of the length of the observations due to the increasing
statistical uncertainty of the SF values in this region.
To extract the variability timescales, 
we fitted a power law ($P(f) \sim f^{\beta}$) to the two rising parts of the SF curves. The variability 
timescale is given by a break in the slope of the power-law fits.  In Fig. \ref{sf_all} (top), the red 
curves represent the fitted power-law to rising trend of SF curves and the vertical dotted lines 
stand for $t_{var1}$ and $t_{var2}$. The best fitted values of the power-law slopes and the estimated 
timescale of variability are given in Table \ref{tab_sf}.  
Thus, the SF curve reveals 
two different variability time scales, one which reflects the short-term variability ($t_{var1}$) 
while the other one refers to the long-term variability ($t_{var2}$).

The SF value is proportional to the square of the amplitude of variability, so to compare the strength of 
the variability at different frequencies, we produce SF vs frequency plots at different time lags. In 
fig. \ref{sf_radio}, we show the SF vs frequency plot at a time lag of 100 days. 
The source displays more pronounced and faster variability at higher 
radio frequencies compared to lower ones, peaking at a frequency of 43 GHz,
and have similar amplitude at higher frequencies.  It seems that the radio variability saturates above this 
frequency. We notice a similar trend at 50 and 200 day time lags.  Thus, for different time lags the variability 
strength exhibits a similar frequency dependence.  \\

 \begin{figure}
\includegraphics[scale=0.4,angle=0]{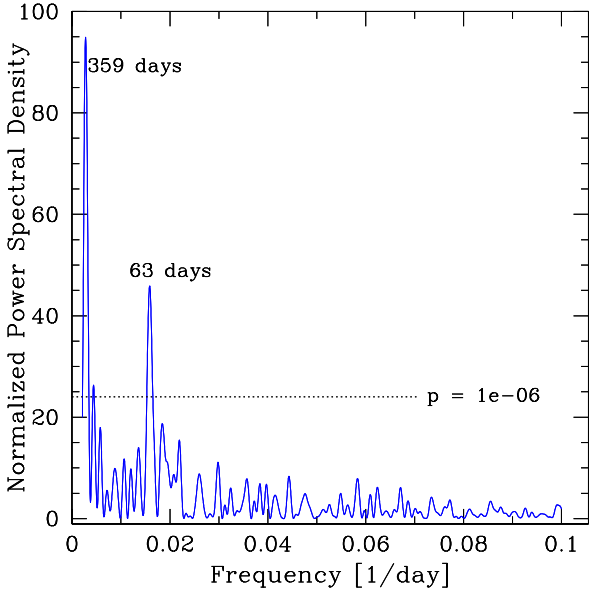}
   \caption{LSP analysis curve showing a peak at a period of 63 and 359 days. }
\label{lsp_opt}
    \end{figure}

 \begin{figure}
   \centering
\includegraphics[scale=0.32,angle=-90]{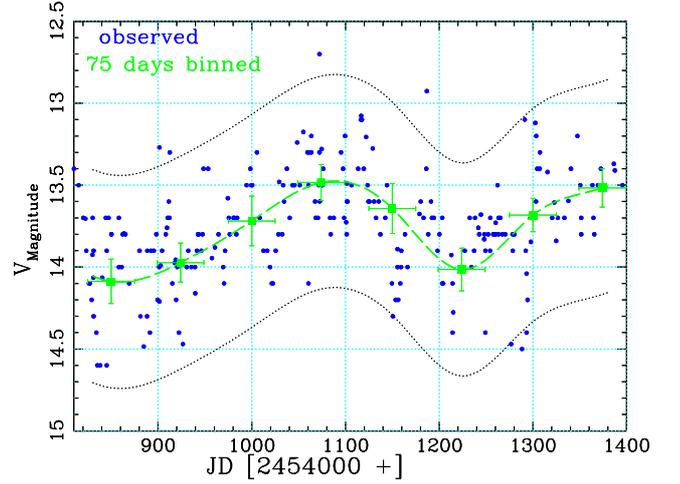}
   \caption{Optical V passband light curve of S5~0716+714 with different time binning. The light curve appears as a 
superposition of fast flares on a modulated base level varying on a (350$\pm$9) day timescale. These slower  
variations can be clearly seen in 75 days binned light curve (error bar represents variations in flux over the binned 
period). The dashed line represents a spline interpolation through the 75 day binned light curve. Dotted lines are 
obtained by shifting the spline by $\pm$0.65 mag. }
\label{plot_flx_optR}
    \end{figure}

\subsubsection{Optical frequencies }  
The source exhibits multi-flaring behavior at optical frequencies 
with each flare roughly separated by 60 -- 70~days (see Fig. \ref{plot_flx_total_LC}(c)). 
The optical V band SF curve in Fig. \ref{sf_all} (bottom) shows rapid variability with 
multiple cycles of rises and declines. The first peak in SF curve appears at a timescale 
$t_{var} \sim 30$~days which is followed by a dip at $\sim 60$~days. This peak corresponds 
to the fastest variability timescale. The peaks in the optical SF curve 
at $t_{var} = 90$, 180, 240 and 300~days represent the long-term variability timescales.  
This indicates a possible superposition of a short 30-40 day time scale variability with 
the long-term variability trend.

Multiple cycles in the optical SF curve represent the nearly periodic variations at $\sim$60 days timescale.
We used the Lomb-Scargle Periodogram (LSP) \citep{lomb1976, rani2009} method to test the presence of this 
harmonic. The LSP analysis of the whole data set is displayed in Fig. \ref{lsp_opt}. 
The LSP analysis reveals two significant ($>$99.9999$\%$ confidence level) peaks at 359 and 63 days. 
The peak at 359 days is close to half of the duration of observations, so it is hard to claim this 
frequency due to limited number of cycles. The nearly annual periodicity could also be effected by 
systematic effects due to annual observing cycle.  
A visual inspection of the light curve indicates a total of 
7 rapid flares separated by 60 -- 70~days. Also, the LSP is only sensitive for a dominant white-noise 
process ($P_{N}(f) \propto f^{0}$). 
It is for this reason that we further inspect the significance of 
this frequency with the Power Spectral Density (PSD) analysis method \citep{vaughan2005} which is a 
powerful tool to search for periodic signals in time series, including those contaminated by white noise 
and/or red noise. To achieve a uniform sampling in the optical data, we adopt a time-average binning of 3 days.
We found that the significance of the period at $\sim 60$~days is only 2 $\sigma$. We therefore 
can not claim a significant detection of a quasi-periodic variability feature at this frequency. 


We also notice that during the course of our optical monitoring, the peak-to-peak amplitude of the short-term variations 
remains almost constant, $\sim 1.3$~magnitudes. Hence, the variability trend traced by the minima is very similar to 
that by the maxima of the light curve during the course of $\sim 2$ years of our monitoring. The constant variability 
trend is displayed in Fig. \ref{plot_flx_optR}. 
In this figure, the dashed line denotes a spline through the 75-days binned light curve and the dotted lines 
are obtained by shifting the spline by $\pm 0.65$~mag. So, the observed variations fall within a constant 
variation area. A constant variability amplitude in magnitudes implies that the flux variation amplitude is 
proportional to the flux level itself (following $m_1 -m2 \propto log_{10}(f1/f2)$).  This can be easily 
interpreted in terms of variations of the Doppler boosting factor, $\delta~=~[\Gamma(1-\beta~cos~\theta)]^{-1}$ 
\citep{raiteri2003}. In such a scenario, the observed flux is relativistically boosted by a factor 
of $\delta^3$ and requires a variation in $\delta$ by a factor of $\sim 1.2$. Such a change in $\delta$ can be due to 
either a viewing angle ($\theta$) variation or a change of the bulk Lorentz factor ($\Gamma$) or may 
be a combination of both. A change in $\delta$ by a factor of 1.2 can be easily interpreted as a few 
degree variation in $\theta$ while it requires a noticeable change of the bulk Lorentz factor. Therefore, it 
is more likely that the long-term flux base-level modulations are dominated by a geometrical effect 
than by an energetic one.

Hence, we consider that the optical variability amplitude remains almost constant during 
our observations. A similar variability trend was also observed in this source by \citet{raiteri2003}. 
They also noticed a possibility of nearly periodic oscillations at a timescale of $3.0 \pm 0.3$~years, but
due to the limited time coverage this remains uncertain. The optical light curve of the source also displays fast 
flares with a rising timescales of  $\sim 30$~day. However, the rising timescale of the radio flares is of the order 
of $\sim 60$~days (see Table \ref{tab_R6model}). Thus, the optical variability features are very rapid compared 
to those at radio wavelengths.    \\

\subsubsection{X-ray frequencies }  
Fig. \ref{plot_flx_total_LC}(b) displays the 0.3 --- 10~keV light curve of S5~0716+714.  
Although the X-ray light curve is not as well sampled as those 
at other frequencies, the data indicate a flare at keV energies between JD' = 1122 to 1165. 
Due to large gap in the data train, it is not possible to locate 
the exact peak time of this flare. Still, it is interesting to note that this orphan X-ray 
flare is not simultaneous with any of the GeV flares nor with the optical flares.  
Its occurrence coincides with the optical -- GeV minimum after the major flares at these frequencies  
(O5/G5, see Fig. \ref{plot_flx_total_LC}).  \\

\subsubsection{Gamma-ray frequencies } 
\begin{figure}
\includegraphics[scale=0.34,angle=-90]{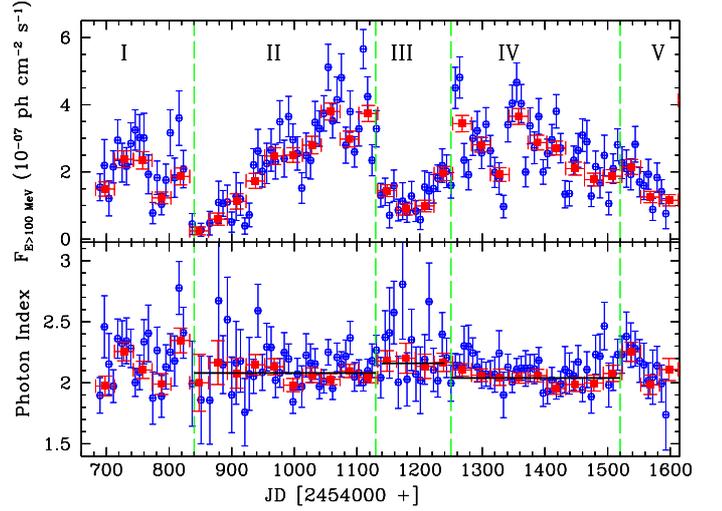}
   \caption{Gamma-ray flux and photon index light curve of S5~0716+714 measured with the {\it Fermi}/LAT for a 
time period between August 04, 2008 to February 07, 2011. The blue symbols show the weekly averaged flux while monthly averaged values are 
plotted in red. Different activity states of the source are separated by vertical green lines.  A marginal softening 
of spectrum over the quiescent state can be seen here. }
\label{flx_gamma}
    \end{figure}

The GeV light curve observed by {\it Fermi}/LAT is extracted over the time period of August 2008 
to February 2011. Fig. \ref{flx_gamma} shows the weekly and monthly averaged $\gamma$-ray light curves 
integrated over the energy range 100 MeV to 300 GeV. There is a significant enhancement in the weekly 
averaged $\gamma$-ray flux over the time period of JD' = 900 to 1110, peaking at JD' $\sim$ 1110, 
with a peak flux value of $(0.57 \pm 0.05) \times10^{-6}$~ph~cm$^{-2}$s$^{-1}$, which is  $\sim 6$
times brighter than the minimum flux and $\sim 3$ times brighter than the average flux value. 
Later it decays, reaching a minimum at JD' = 1150, and then it remains in a quiescent state until 
JD' = 1200. The quiescent state is followed by another sequence of flares. 
The source displays similar variability features in the constant uncertainty light curve obtained using the adaptive 
binning analysis method following \citet{lott2012}. A more detailed discussion of GeV variability is given 
in Rani et al. (2012).  

The source exhibits a marginal spectral softening during the quiescent state in the monthly averaged light 
curve.  The $\gamma$-ray photon index ($\Gamma$) changes from $\sim 2.08 \pm 0.03$ (II) to $2.16 \pm 0.02$ (III), 
then again to $2.04 \pm 0.04$ (IV). Different activity states of the source are separated 
by vertical lines in Fig. \ref{flx_gamma}. We notice no clear spectral variation in the weekly 
averaged light curve due to large uncertainty and scatter of individual data points. \\

The SF curve of the source at GeV frequencies is shown in Fig. \ref{sf_all} (bottom). 
The variability timescales are extracted using the power-law fitting method as described above, which 
gives a break at $\sim 30$ and 180~days.   
{\it We notice that the variability features at $t_{var} \sim 180$~days are also observed at radio and optical 
frequencies. However, the faster variability ($t_{var} \sim 30$~days) observed at optical and $\gamma$-ray frequencies
does not extend to radio wavelengths. A similar long-term variability timescale at $\gamma$-ray, optical and radio 
frequencies provides a possible hint of a co-spatial origin.} In the following sections, we 
will search for such possible correlations.

 \begin{figure*}
   \centering
\includegraphics[scale=0.75]{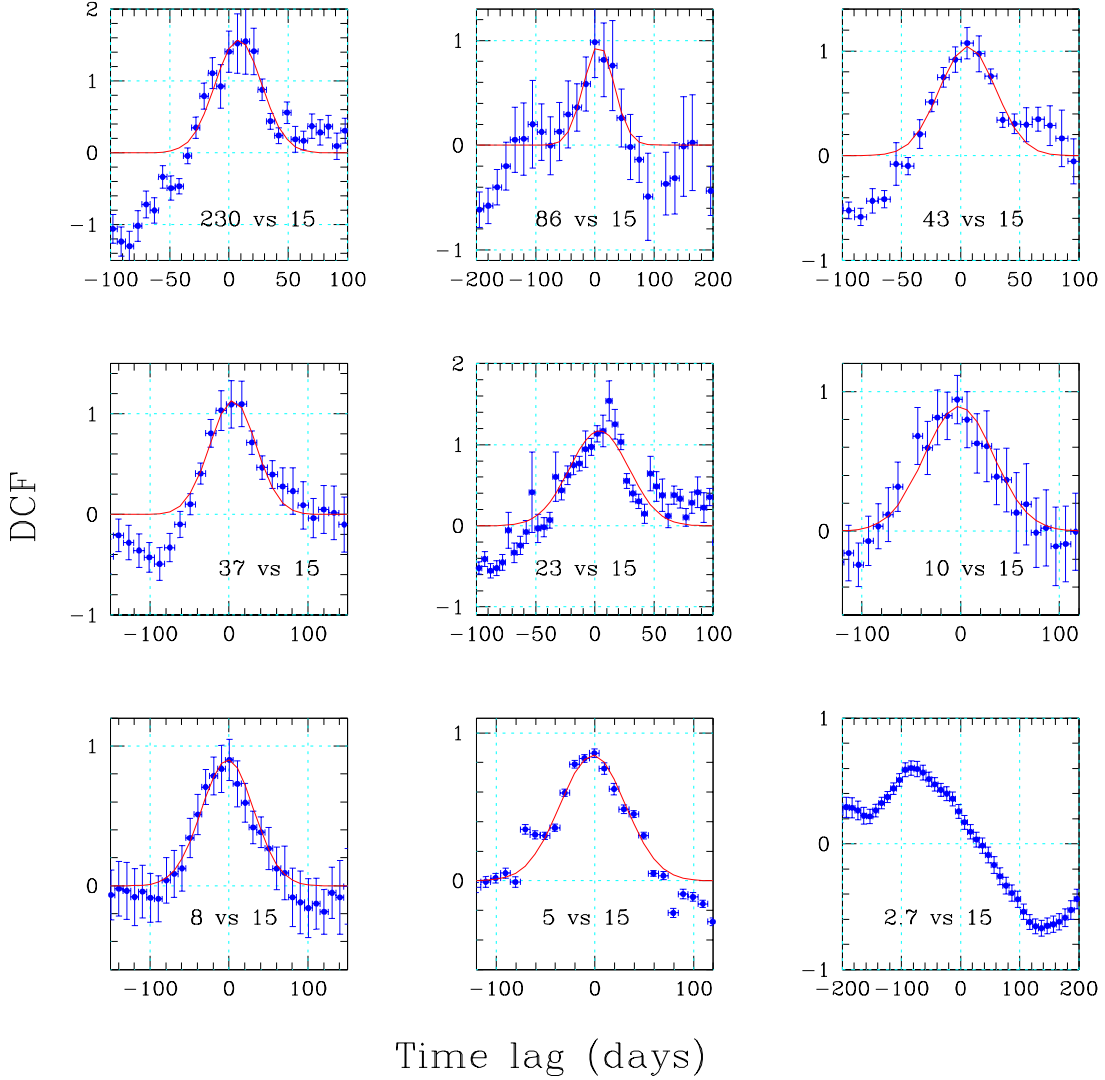}
  \caption{The DCF curves among the different radio frequency light curves. 
The solid curves represent the best-fit Gaussian function.}
\label{radio_dcf}
    \end{figure*}

\subsection{Correlation Analysis}
\label{sec_corr}
The multi-frequency light curves of S5~0716+714 are presented in Figure \ref{plot_flx_total_LC}. 
The analysis presented in this section is focused on a time period between 
JD' = 800 to 1400, which covers the two major radio flares and the respective optical-to--$\gamma$-ray 
flaring activity. 
The source displayed multiple flares across the whole electromagnetic spectrum over this period, which we label as 
follows. We mark the vertical dotted lines w.r.t. different optical flares, labeling the broadband 
flares as ``G" for $\gamma$-rays, ``X" for X-rays, ``O" for optical and ``R" for radio followed by 
the number adjacent to them.  For example, the optical flares should be read as ``O1" to ``O9".

To search for possible time lags and to quantify the correlation among the multi-frequency light 
curves of the source,  we computed discrete cross-correlation 
functions (DCFs) following the method described by \citep{edelson1988}. 
The details of this method are also discussed by \citet{rani2009}. 
In the following sections, we 
will discuss in detail the correlation between the observed light curves.

\subsubsection{Radio correlation} 
\label{sec_radio_dcf}
At radio wavelengths, the source exhibits significant flux variability, being more rapid and pronounced 
towards higher frequencies. We label the two major outbursts as ``R6" and ``R8" (see Fig. \ref{plot_flx_total_LC}). 
Apparently, the mm flares are observed a few days earlier than the cm flares.  
Our dense frequency coverage facilitates a cross-correlation analysis between the different observing bands. 
Owing to its better time sampling, we have chosen the 15~GHz light curve as a reference. Fig. \ref{radio_dcf} 
shows the DCF curves adopting a binning of 10 days.

\begin{table}
\caption{ Correlation analysis results among radio frequencies}
\begin{tabular}{l c c c } \hline
Frq. (GHz) &$a$             &$b$ (days)         &$c$ (days)         \\\hline
230 vs 15 & 1.56$\pm$0.15  &    7.96$\pm$2.23   &    19.81$\pm$2.24  \\
86 vs 15  & 0.94$\pm$0.13  &    6.65$\pm$3.28   &    25.80$\pm$4.28   \\
43 vs 15  & 1.04$\pm$0.11  &    5.95$\pm$2.08   &    23.86$\pm$3.09   \\
37 vs 15  & 1.13$\pm$0.09  &    4.95$\pm$2.21   &    29.39$\pm$2.81    \\
23 vs 15  & 1.17$\pm$0.10  &    3.74$\pm$1.50   &    25.00$\pm$2.50    \\
10 vs 15  & 0.89$\pm$0.09  &    -1.01$\pm$1.09  &    35.07$\pm$4.49     \\
8 vs 15   & 0.84$\pm$0.08  &    -1.09$\pm$1.01  &    35.96$\pm$4.10     \\
5 vs 15   & 0.84$\pm$0.10  &    -1.23$\pm$1.25  &    33.13$\pm$4.15      \\
2.72 vs 15& 0.59$\pm$0.12  &    -78.75$\pm$12.39&    53.54$\pm$13.86      \\\hline
\end{tabular} \\
$a$ :  peak value of the DCF, \\
$b$ :  time lag at which the DCF peaks, and \\
$c$ :  width of the Gaussian function (see text for details) \\
\label{tab_dcf}
\end{table}

We used a Gaussian profile fitting technique to estimate the time lag and respective cross-correlation 
coefficient value for the DCF curves. Around the peak, the DCF curve as a function of time lag $t$
can be reasonably well described by a Gaussian of the form: 
$DCF(t) = a \times ~exp [\frac{-(t-b)^{2}}{2c^{2}}]$, where $a$ is the peak value of 
the DCF, $b$ is the time lag at which the DCF peaks and $c$ characterizes the width of the Gaussian 
function. The calculated parameter ($a$, $b$ and $c$) values for each frequency are listed in 
Table \ref{tab_dcf}.  The solid curve represents the fitted Gaussian function in 
Fig. \ref{radio_dcf}. 

Our analysis confirms the existence of a significant correlation across all observed 
radio-band light curves till 15 GHz with formal delays listed in Table \ref{tab_dcf} (parameter $b$). 
However, no pronounced delayed flux variations are observed at 10 GHz and lower frequencies.  
Also, the flux variations at 2.7 GHz seem to be 
less correlated with those at 15 GHz, which is obvious as the flaring behavior is not 
clearly visible below 15 GHz.

The long term radio light curve shows three major radio flares, labeled as R0, R6 and R8 in 
Fig. \ref{plot_flx_rad}. In the correlation analysis of the entire light curves, these 
flares are blended and folded into a single DCF. In order to separate the flares from each other,  
we performed a correlation analysis over three different time bins which cover the time ranges of the individual radio 
flares: JD' = 500 to 750 (R0 flare), JD' = 1000 to 1210 (R6 flare) and JD' = 1210 
to 1400 (R8 flare). 
The time lags of these flares relative to the 15 GHz data are estimated as above.  
However due to sparse data sampling, it was not possible to estimate the time lags for R8  directly using the DCF method. 
So, for this flare, we first interpolate the data through a spline function and then perform the DCF analysis, except at 
23 GHz and 33 GHz (due to long data gaps). 
The calculated time lags of each flare are given in Table \ref{tab_dcf_flare}.

\begin{table}
\caption{ Radio correlation analysis results for individual flares }
\begin{tabular}{c c c c} \hline 
Frequency       &                 & Time lag$^{*}$ (days)     &                 \\
(GHz)        & R0 Flare     &  R6 Flare            & R8 Flare           \\\hline
23           & 2.9$\pm$1.4  & 4.06$\pm$1.6      & ---                \\
33           & 2.1$\pm$1.0  & 5.88$\pm$2.1      & ---                 \\
37           & 4.2$\pm$2.6  & 3.15$\pm$2.1      & 4.0$\pm$1.0         \\
86           & 4.2$\pm$1.0  & 6.15$\pm$2.6      & 5.0$\pm$1.8          \\
143          & 5.8$\pm$1.8  & 6.82$\pm$2.5      & 7.0$\pm$1.5          \\
230          & 6.0$\pm$2.4  & 8.54$\pm$1.9      & 9.0$\pm$2.0           \\\hline
\end{tabular} \\
$*$ : relative to the respective flares at 15 GHz \\
\label{tab_dcf_flare}
\end{table}

\begin{figure}
   \centering
\includegraphics[scale=0.4]{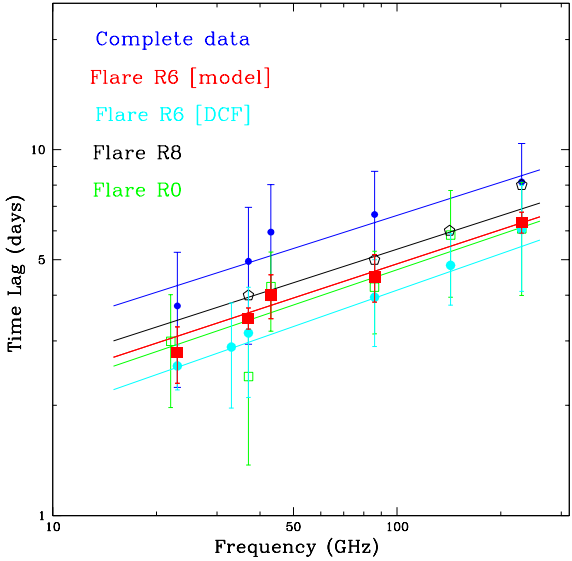}
  \caption{ The plot of time lag vs frequency, using 15 GHz as the 
reference frequency. Time lag vs frequency curves for individual flares are 
shown with different colors. The solid 
lines represent the best fitted power law in each case.
              }
\label{radio_corr_A}
    \end{figure}

In Fig. \ref{radio_corr_A}, we report the calculated time lags as a function of frequency with 
15 GHz as the reference frequency. The estimated time lag using the entire light curves are 
shown with blue circles. As we see in Fig. \ref{radio_corr_A}, the time lag increases with an increase 
in frequency and seems to follow a power law. Consequently, we fitted a power law, $P(f) = N f^{\alpha}$ 
to the time lag vs frequency curve. The best fitted parameters are $N =1.71 \pm 0.43$, 
$\alpha = 0.30 \pm 0.08$. For the individual flares, we find:
$N = 1.07 \pm 0.06$,  $\alpha = 0.32 \pm 0.01$ for the R0, $N = 1.45 \pm 0.61$, 
$\alpha = 0.32 \pm 0.08$ for R6, and $N = 1.33\pm0.01$, $\alpha = 0.29\pm0.03$ for R8. 
We conclude that a common trend in the time lag (with 15 GHz as the reference frequency)  
vs. frequency curve is seen for all the three radio flares (R0, R6 and R8) with an average 
slope of 0.30.   \\

\begin{figure}
   \centering
\includegraphics[scale=0.35,angle=-90]{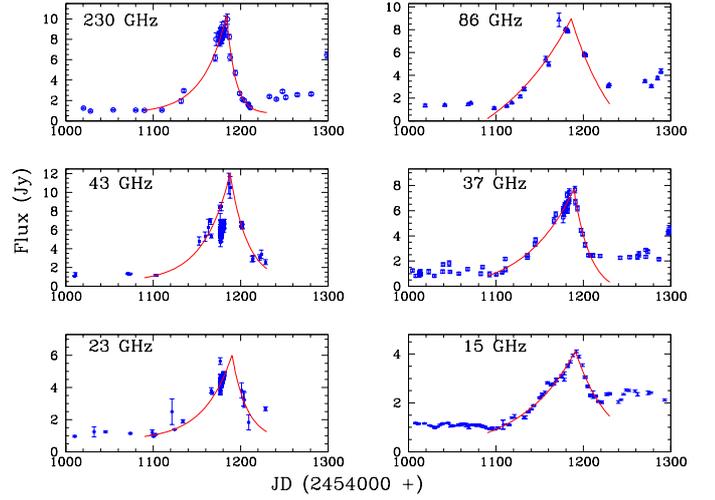}
  \caption{The modeled radio flare, R6. The blue points are the observed data while 
the red curve represent the fit.               }
\label{radio_corr_B}
    \end{figure}

We also followed an alternative approach to estimate the time shift of the radio flares at each 
frequency w.r.t. 15~GHz. The flares at each frequency can be modeled with an 
exponential rise and decay of the form:
\begin{equation}
f(t)=f_0 + f_{max}~exp[(t-t_0)/t_r], ~~~~for ~t<t_0, and
\end{equation}
\begin{displaymath}
   ~~~~~~=f_0 + f_{max}~exp[-(t-t_0)/t_d] ~~~~ for ~t>t_0
\end{displaymath}
where f$_0$ is the background flux level that stays constant over the corresponding interval, 
${\rm f_{max}}$ is the amplitude of the flare, ${\rm t_0}$ is the epoch of the peak, and ${\rm t_r}$ 
and ${\rm t_d}$ are the rise and decay time scales, respectively. 
Since R6 is the most pronounced and best sampled flare, we model this flare in order to 
cross-check the frequency vs time lags results obtained by the DCF method. 
As the flaring behavior is not clearly visible below 15 GHz, so we restrict this analysis
to frequencies above 15~GHz.  
As there is no observation available during the flaring epoch at 23 and 86 GHz,  
we fix t$_{r}$ and t$_{d}$ using the fit parameters from the adjacent frequencies.  
The best fit of the function f(t) for the R6 flare is shown 
in Fig. \ref{radio_corr_B} and the parameters are given in Table \ref{tab_R6model}. 
The estimated time shift around the R6 flare at each frequency w.r.t. 15~GHz are shown in Fig. \ref{radio_corr_A} 
(red symbols) and the fitted power law parameters are $N = 1.17 \pm 0.13$, $\alpha = -0.31 \pm 0.03$. 
Thus, this alternative estimate of the power law slope using the model fitting technique confirms the 
results obtained by the DCF analysis.

\begin{table}
\scriptsize
\caption{ Fitted model parameters for R6 flare}
\begin{tabular}{l c c c c c } \hline
Frq. &  f$_0^{*}$&  f$_{max}$   &     t$_r$       &        t$_d$     &   t$_0$  \\
 GHz&            &               &                 &                  & JD' [JD--2454000] \\\hline
15  & 0.02$\pm$0.07   &4.15$\pm$0.14 & 61.4$\pm$6.2  &  37.9$\pm$3.5  &1191.4 $\pm$0.9  \\
23  & 0.71$\pm$0.23   &5.30$\pm$0.15 & 32.3$\pm$4.8  &  17.3$\pm$2.1  &1190.2 $\pm$0.1  \\
37  & 0.58$\pm$0.11   &8.20$\pm$0.59 & 55.5$\pm$11.0 &  18.6$\pm$3.2  &1189.1 $\pm$0.7  \\
43  & 0.45$\pm$0.35   &9.50$\pm$0.62 & 60.5$\pm$9.4  &  20.1$\pm$2.9  &1188.0 $\pm$0.8  \\
86  & 0.64$\pm$0.18   &10.6$\pm$2.48 & 60.9$\pm$28.8 &  25.1$\pm$25.6 &1186.0 $\pm$0.5  \\
230 & 0.72$\pm$0.11   &12.64$\pm$0.29 & 50.3$\pm$2.4  &  9.9 $\pm$0.6  &1184.2 $\pm$0.4   \\\hline
\end{tabular} \\
$^{*}$ : see text for the extension of labels. \\
\label{tab_R6model}
\end{table}

\noindent
\subsubsection{Radio vs Optical correlation} 
\label{dcf_opt_rad}
\begin{figure}
   \centering
\includegraphics[scale=0.29,angle=-90]{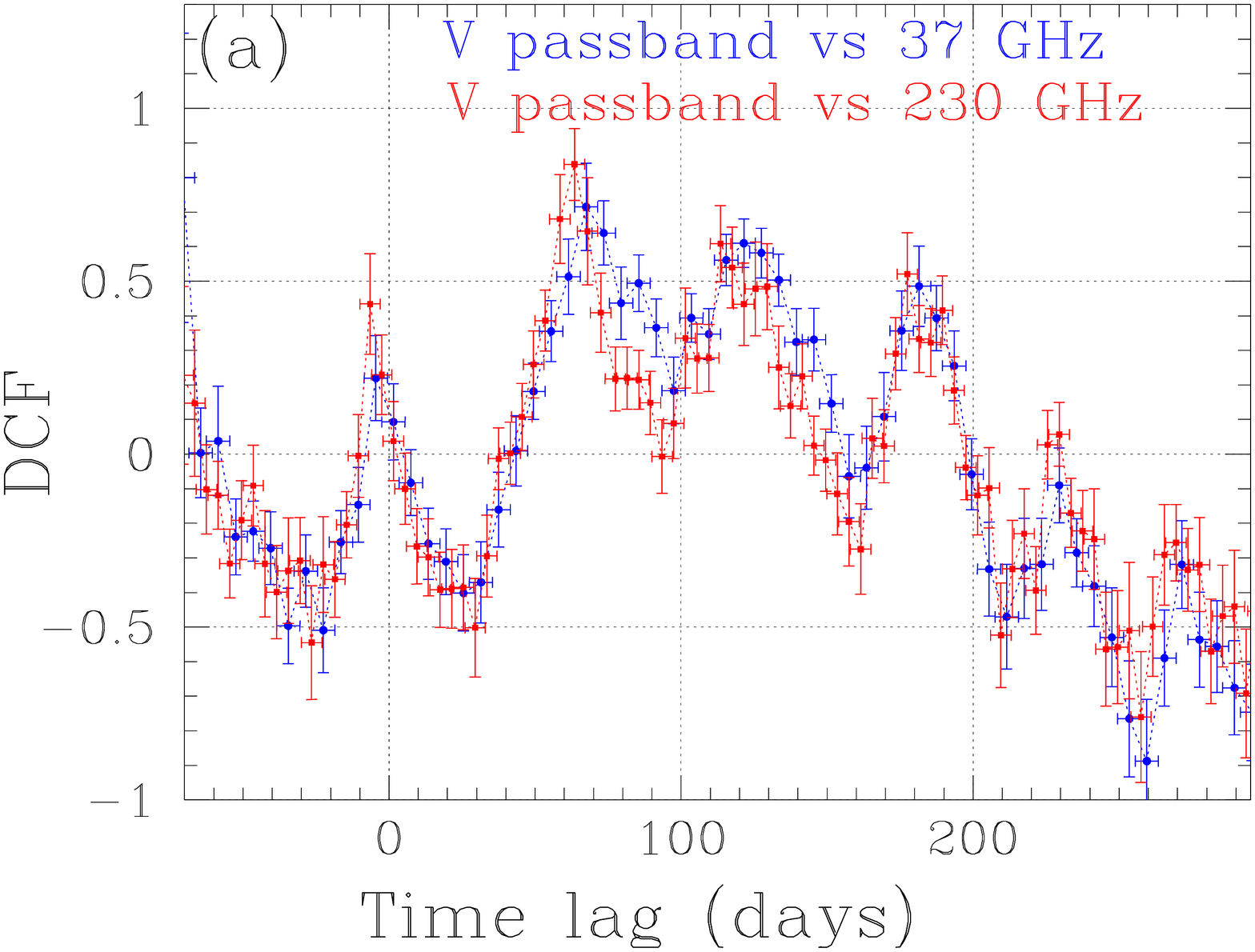} 
\vspace{0.2in}
\includegraphics[scale=0.29,angle=-90]{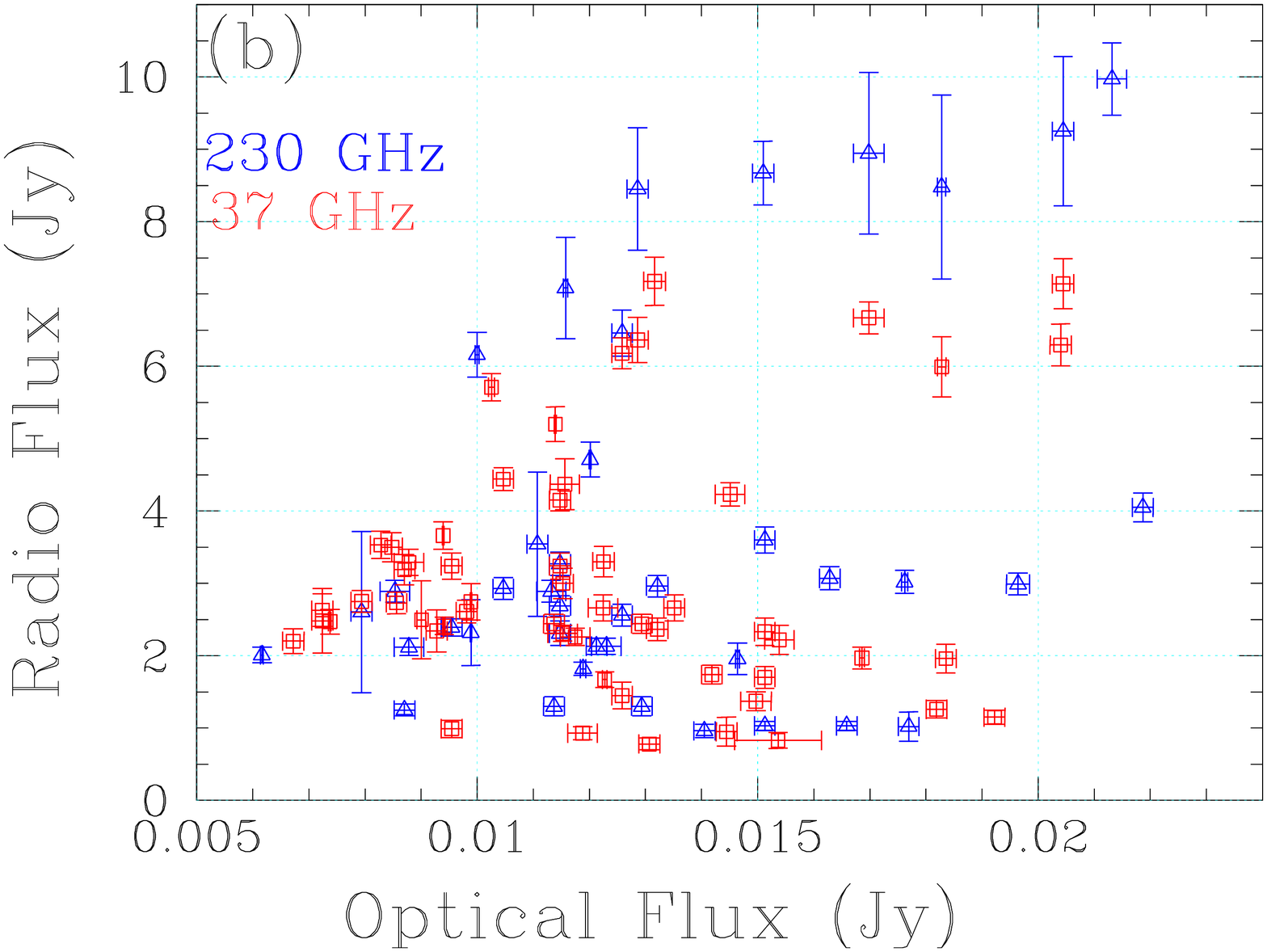}

\includegraphics[scale=0.29,angle=-90]{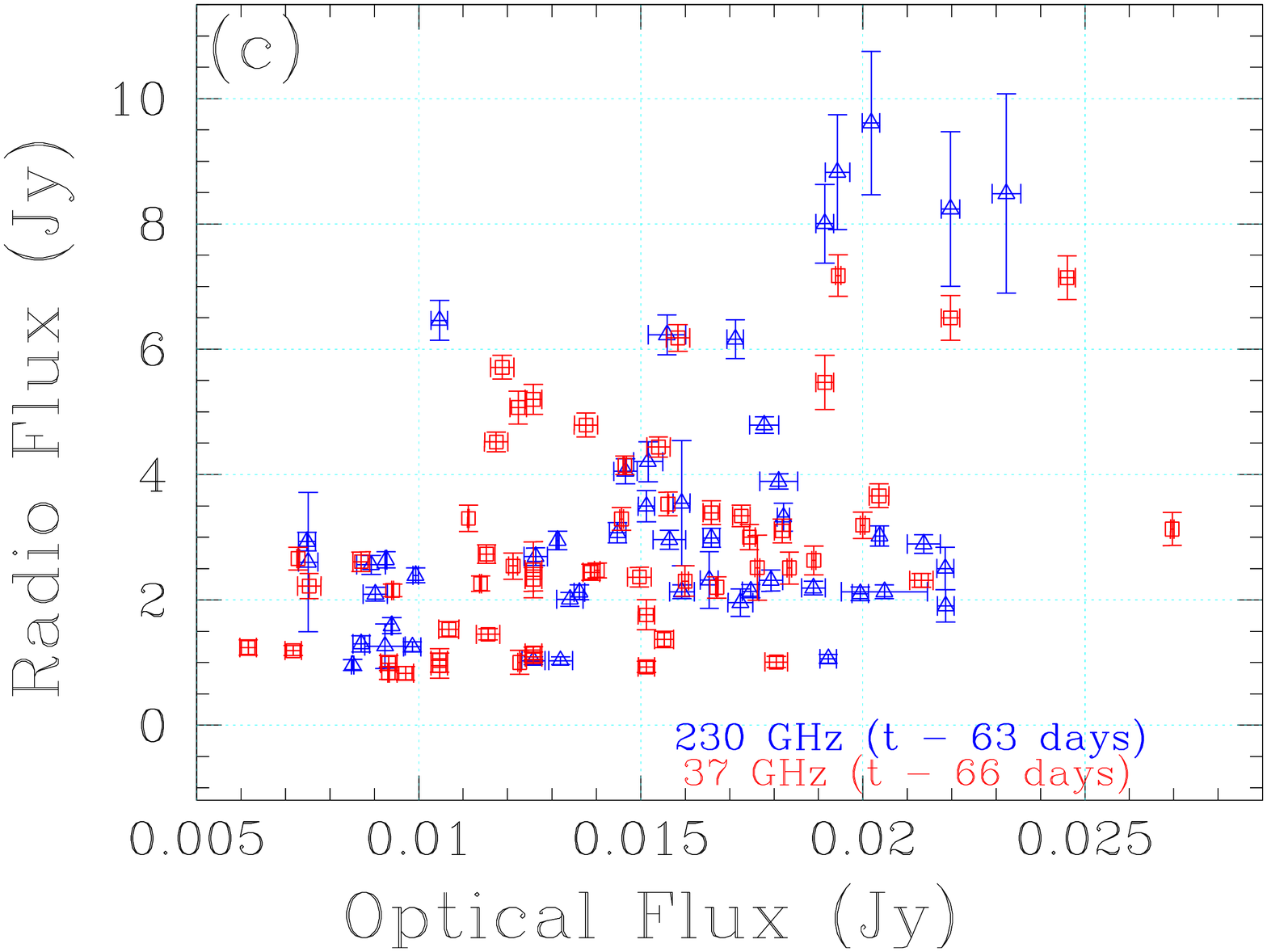}
   \caption{(a) The DCF curve for optical V passband vs 37~GHz (in blue) and 230~GHz (in red) flux with a bin 
size of 5 days.  (b) Radio flux vs optical V-band flux. (c) Time shifted radio flux vs optical 
V-band flux. The blue symbols show the time shifted 230~GHz (t-65 days) data while 37~GHz (t-68 days) data are 
shown in red. }
\label{dcf_Rvs37}
    \end{figure}

S5~0716+714 exhibits multiple flares at optical frequencies. The flares 
are roughly separated by 60 -- 70~days.  We labeled the different 
optical flares as O1 -- O9 as shown in 
Fig. \ref{plot_flx_total_LC}.  During this multi-flaring activity period two 
major flares are observed at radio wavelengths. The radio flare R6 apparently coincides in time with O6 and R8 with O8.
To investigate the possible correlation among the flux variations at optical and radio frequencies, 
we perform a DCF analysis using the 2-year-long simultaneous optical and radio data trains between 
JD' = 680 to 1600 (see Fig. \ref{plot_flx_total_LC}). 
Note that the strength of flux variability increases towards higher frequencies, peaking at 43~GHz 
(see Fig. \ref{sf_all}). Therefore, we choose two radio frequencies, 37~GHz and 230~GHz, in order 
to compare the strength of radio -- optical correlation above and below the saturation frequency (43~GHz).    
The optical vs radio DCF analysis curves are shown in Fig. \ref{dcf_Rvs37} (a).  Multiple peaks in the DCF  
may reflect a QPO behavior at optical frequencies. As the formal errors, we use half of the binning time. 
We summarize the optical vs 230~GHz and 37~GHz DCF analysis results in Table \ref{tab_opt_rad}. 

\begin{table}
\scriptsize
\caption{ Fitted model parameters for R6 flare}
\begin{tabular}{l c c c c c } \hline
\multicolumn {2} {c} {V vs 230 GHz} & \multicolumn {2} {c} {V vs 37 GHz} \\\hline 
 lag (days)   &   DCF Peak value  &  lag (days) &     DCF Peak value    \\\hline  
 -4$\pm$2.5   &  0.43$\pm$0.10    &  -2$\pm$2.5 &    0.28$\pm$0.11      \\
 63$\pm$2.5   &  0.83$\pm$0.11    &  66$\pm$2.5 &    0.76$\pm$0.08      \\
 120$\pm$2.5  &  0.60$\pm$0.08    &  124$\pm$2.5&    0.60$\pm$0.09      \\
 181$\pm$2.5  &  0.51$\pm$0.07    &  183$\pm$2.5&    0.49$\pm$0.08      \\\hline
\end{tabular} \\
\label{tab_opt_rad}
\end{table}

The maximum correlation of the optical V passband with the 230~GHz light curve occurs at a 65 day time lag.
However a second peak with lower peak coefficient also occurs close to zero time lag (see Fig. \ref{dcf_Rvs37}). 
The analysis shows that the cross-correlation coefficient of the simultaneous radio -- optical flare peaks O6-R6 
and O8-R8 is lower than the cross-correlation coefficient of the O5-R6 and O7-R8 flare peaks. In both cases the 
optical flares O5 and O7 are observed $\sim 65$~days earlier than the radio flares R6 and R8, respectively. The 
observed time lag between the optical and radio flares is consistent with the extrapolated frequency dependent 
time shift (as shown in Fig. \ref{radio_corr_A}) to optical wavelengths. In Section \ref{broad_corr}, we will discuss 
this in detail.   

In order to quantify the correlation among optical and radio data, we generate flux -- flux 
plots, which are shown in Fig. \ref{dcf_Rvs37} (b) -- (c). For the following analysis we used a 
1 day binning. Fig. \ref{dcf_Rvs37} (c) shows the time shifted 230~GHz (t-63) and 37~GHz (t-66)
flux plotted vs the optical V-band flux. The time-shifted radio and optical V-band fluxes fall
on a straight line, indicating a correlation. A Pearson correlation analysis reveals a significant 
correlation between the two data trains. We obtain the following values: $r_P = 0.59 $ and 99.93~\% 
confidence level for 230~GHz (t-65) vs V-band and $r_P =0.43 $ and 99.3~\% confidence level for 37~GHz 
(t-65) vs V-band, where $r_P$ is the linear Pearson correlation coefficient.  Thus, we found  a 
significant correlation among the time shifted radio vs optical V-band flux at a confidence 
level $> 99$~\%. 

In contrast to this, the radio (with no time shift) vs optical V-band correlations are found 
to be not significant. 
Fig. \ref{dcf_Rvs37} (c) shows 230~GHz and 37~GHz vs V-band flux-flux plots, and the correlation 
statistics are: 230~GHz vs V-band: $r_P =0.40 $, 91~\% confidence level and  
37~GHz vs V-band: $r_P =0.15 $, 74~\% confidence level. Thus, the confidence level of the correlations 
is lower than 95~\% in these cases. 
We also check the significance of the correlation statistics with a time shift of 
$\sim 120$ and 180~days and do not find a correlation to have a significance greater than 2$\sigma$
in any case.  

Hence, using DCF and linear Pearson correlation statistics, we have found a significant correlation among the flux 
variations at optical and radio frequencies with the optical V-band leading the radio fluxes at 230 and 37~GHz 
by $\sim 63$ and $\sim 66$~days, respectively.  We therefore conclude that flux variations at optical and radio 
frequencies are correlated such that the optical variability is leading the radio with a time lag of about two 
months.

\noindent
\subsubsection{Radio vs Gamma-ray correlation} 
\begin{figure}
   \centering
\includegraphics[scale=0.3, angle=-90]{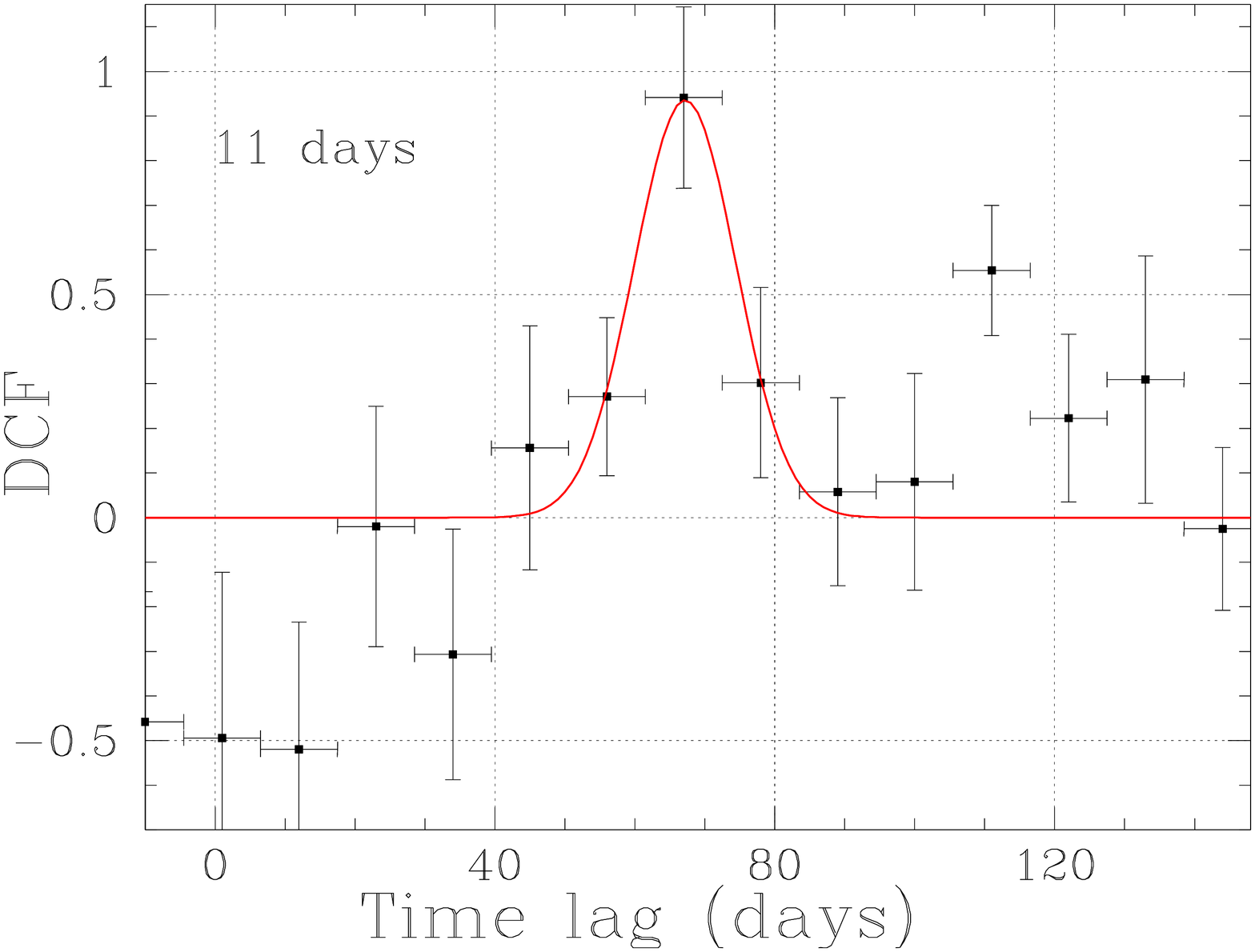}
\includegraphics[scale=0.4]{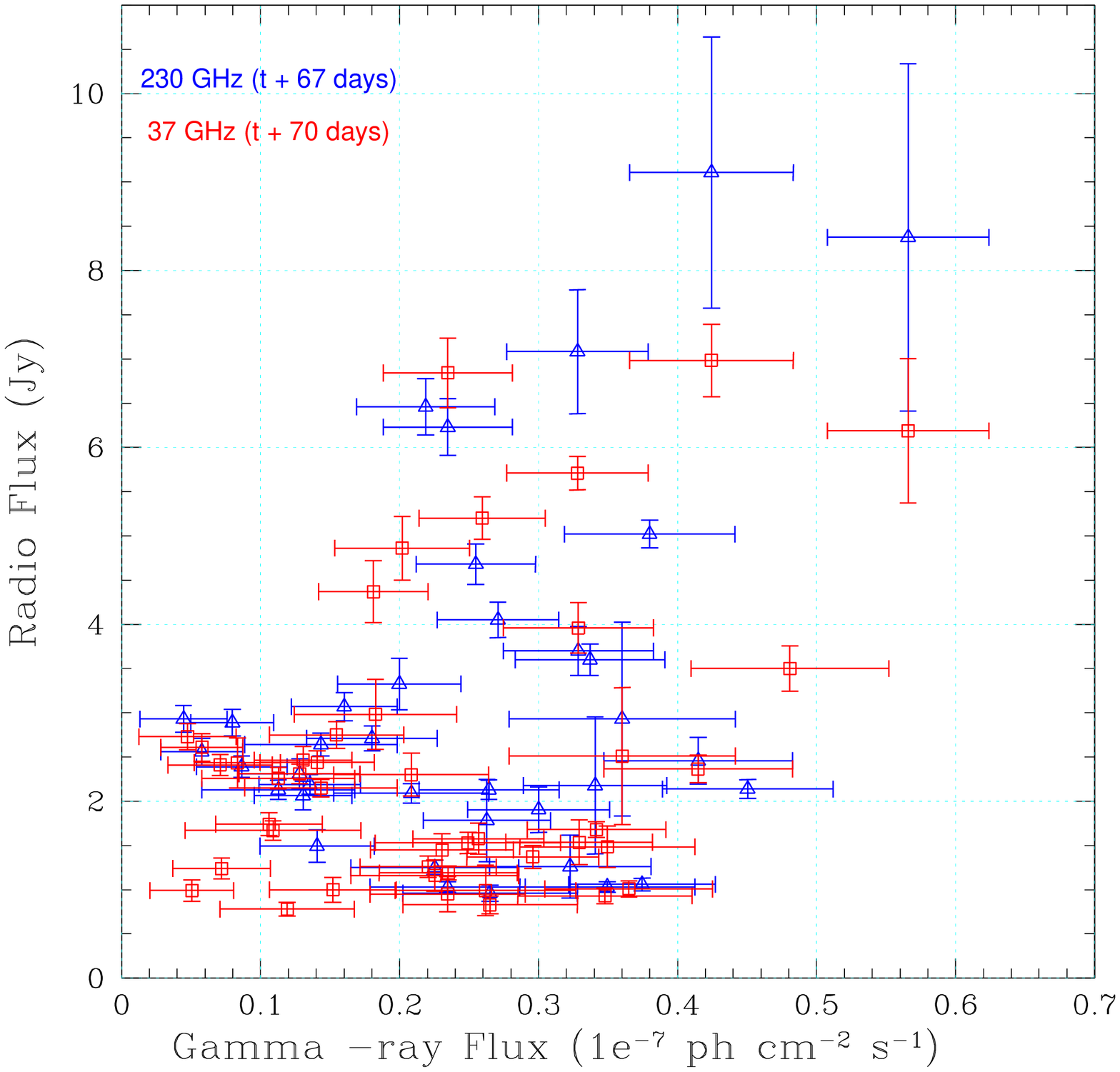}
   \caption{Top: DCF curve of the $\gamma$-ray light curve w.r.t. the 230~GHz radio light 
curve.  The solid curve is the best 
fitted Gaussian function to the 11 day binned DCF curve. 
Bottom: Flux-flux plot of the shifted radio vs $\gamma$-ray data. The blue symbols show the time 
shifted 230~GHz data while 37~GHz data are shown in red.    
              }
\label{dcf_gamma_230}
    \end{figure}

We apply the DCF analysis method to investigate a possible correlation among flux variations at radio and 
$\gamma$-ray frequencies. In Fig. \ref{dcf_gamma_230}, we report the DCF analysis results of the weekly 
averaged $\gamma$-ray light curve with the 230~GHz radio data with a time bin size of  11 days. 
As the flux variations at 37~GHz are delayed by $\sim 3$ days w.r.t. those at 230~GHz, we only 
show the DCF analysis curve w.r.t 230~GHz. To estimate the possible peak DCF value and respective time lag, 
we fit a Gaussian function to the DCF curve with a bin size of 11 days. The best-fit function is shown in 
Fig. \ref{dcf_gamma_230} and the fit parameters are $a = 0.94 \pm 0.30$, $b = (67 \pm 3)$~days and 
$c = (7 \pm 2)$~days. This indicates a clear correlation between the $\gamma$-ray and 230~GHz radio 
light curves of the source with the GeV flare leading the radio flare by $(67 \pm 3)$~days. 

To check the significance of the $\gamma$-ray vs radio correlation, we produce flux-flux plots 
of the time shifted radio vs $\gamma$-ray flux. Since the $\gamma$-ray flux is weekly averaged, we 
use a time binning equal to seven days. The weekly averaged flux-flux plots of the time shifted 230~GHz 
(t + 67 days) and 37~GHz (t + 70 days) vs $\gamma$-ray are shown in Fig. \ref{dcf_gamma_230} (bottom) and the 
correlation statistics are: 230~GHz (t + 67 days) vs $\gamma$-ray: $r_P$ = 0.37, 97.7~\% confidence level
and 37~GHz (t + 70 days) vs $\gamma$-ray: $r_P$ = 0.33, 97.3~\% confidence level. Thus, in each case the 
confidence level of the correlation is higher than 95~\%. This supports a possible correlation among the 
flux variations at $\gamma$-ray and radio frequencies with $\gamma$-rays leading the radio emission by 
$\sim 67$~days. We also note that the time shifts are very similar to the time shifts observed between 
radio and optical bands (see  Section \ref{dcf_opt_rad}). We therefore expect a very short or no time 
delay between the flux variations at optical and $\gamma$-ray frequencies. This will be investigated 
in the next section.

\begin{figure*}
\centering
\includegraphics[scale=0.7, angle=-90, trim = 200 0 0 0, clip]{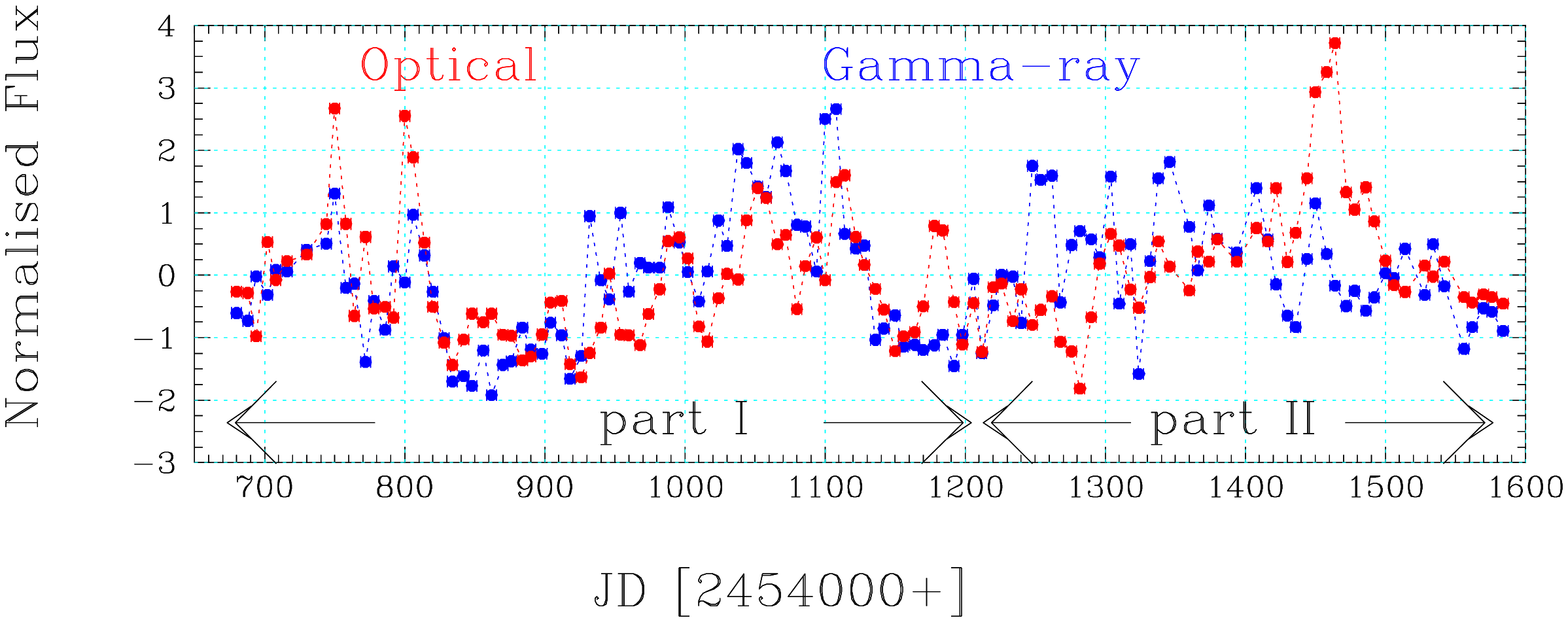}
\includegraphics[scale=0.34,angle=-90]{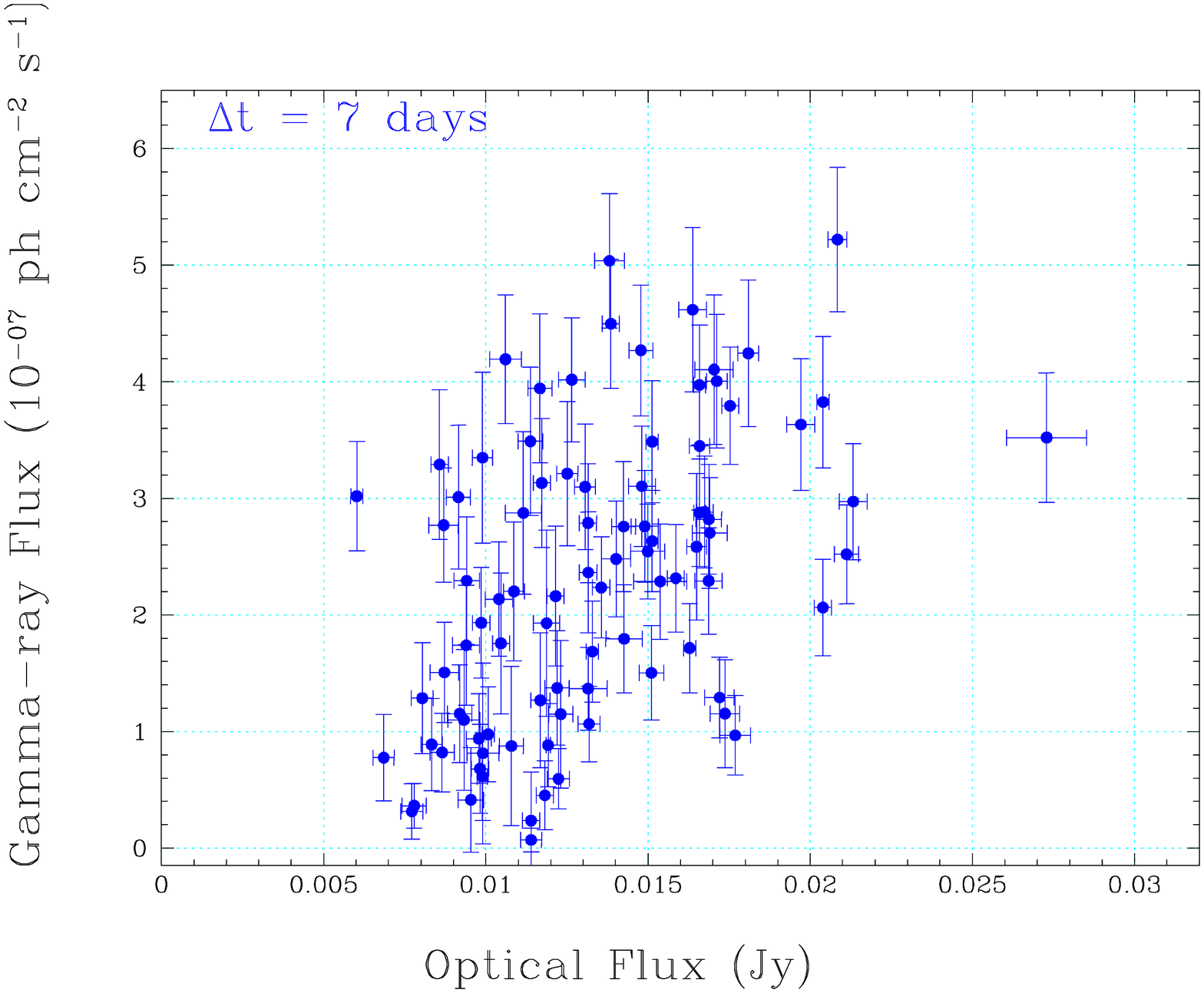}
\vspace{0.1in}
\includegraphics[scale=0.3, angle=-90]{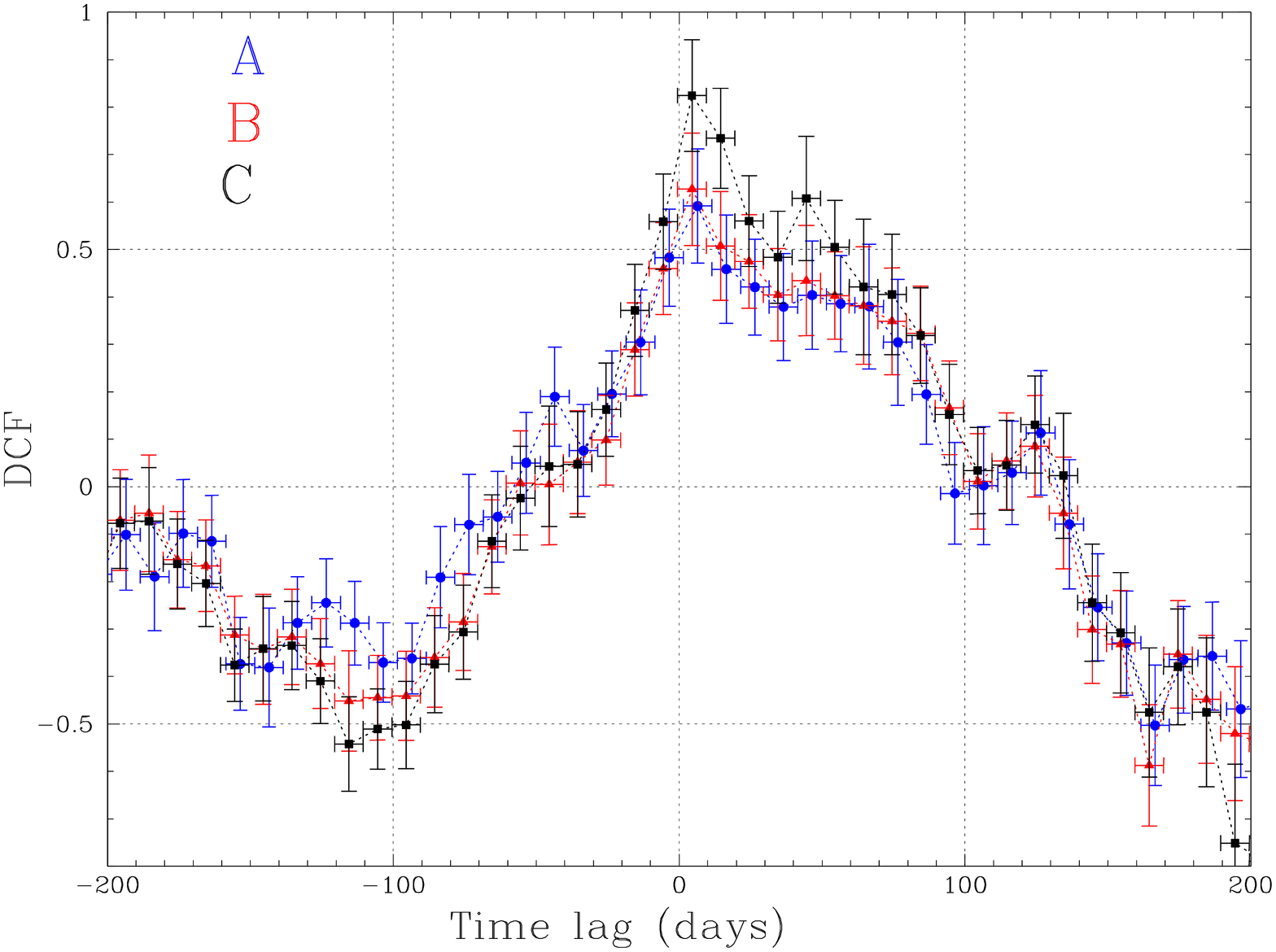}
   \caption{Top: The weekly averaged normalized flux at $\gamma$-ray and optical V 
band frequencies plotted vs. time. The flux variations at these two frequencies seem to have 
a one-to-one correlation with each other. Bottom left : $\gamma$-ray vs optical flux. 
Bottom right: The DCF curve of $\gamma$-ray vs optical V passband flux using 
a bin size of 10 days in each case. A: using the complete data as shown in Fig. 
\ref{plot_flx_total_LC}; B: after removing the data covering the duration of the 
optical flare O6; C: using the data before flare O6. }
\label{norm_optvsgamma}
    \end{figure*}

\noindent
\subsubsection{Optical vs Gamma-ray correlation} 
Visual inspection of variability curves in Fig. \ref{plot_flx_total_LC} shows an apparent correlation 
between the various flux density peaks of the GeV light curve and the optical peaks (O1 to O9, except O6). 
The flaring pattern at $\gamma$-rays is similar to the QPO-like behavior observed at optical 
frequencies. In addition, the long-term variability features are also simultaneous at the 
two frequencies. To compare the flaring behavior of the source at optical and $\gamma$-ray frequencies, 
we plot the normalized weekly averaged optical and $\gamma$-ray light curves  
on top of each other (see Fig. \ref{norm_optvsgamma} top). A  
consistent and simultaneous flaring behavior can be seen between JD' = 680 to 1200, however 
the $\gamma$-ray variability is less correlated later. 
In Fig. \ref{norm_optvsgamma} (middle), we show a flux-flux plot of the weekly averaged $\gamma$-ray vs 
optical V-band data. A clear correlation among the two can be seen,  which is confirmed by a linear 
correlation analysis, yielding $r_{P} = 0.36$ and 99.996~\% confidence level. The 
correlation is even stronger in part I, for which we find $r_{P} = 0.66$ and 99.9999~\% confidence level. 
Here, we have used the weekly averaged optical flux for the analysis, and the uncertainty represents 
variation of the flux over this period. 

In Fig. \ref{norm_optvsgamma} (bottom), we show the cross-correlation analysis results of the  
$\gamma$-ray  and optical data trains. We consider three different cases to investigate the possible 
correlation and summarize the results in Table \ref{tab_dcf_opt_gamma}. This analysis reveals that 
the two-year-long GeV and optical data trains are strongly correlated with each other with no time lag 
longer than one week. It is also important to note that the strength of correlation is higher before 
the end of the O5/G5 flares than after those flares.

\begin{table}
\scriptsize
\caption{ Optical vs. $\gamma$-ray cross-correlation analysis results}
\begin{tabular}{l c c c } \hline
Case     & Time duration     & Peak DCF value  & Time lag  \\
         & JD' [JD-2454000]     &                 & days       \\\hline         
A        & total (840 - 1350)              & 0.50$\pm$0.04   &0$\pm$5     \\
B        & removing O6 (1150 - 1220) flare & 0.61$\pm$0.04   &1$\pm$5      \\
C        & before O6 flare (840 - 1150)    & 0.80$\pm$0.08   &3$\pm$5     \\\hline              
\end{tabular} \\
\label{tab_dcf_opt_gamma}
\end{table}

\noindent
\subsubsection{The orphan X-ray flare } 
\begin{figure}
   \centering
\includegraphics[scale=0.3,angle=-90]{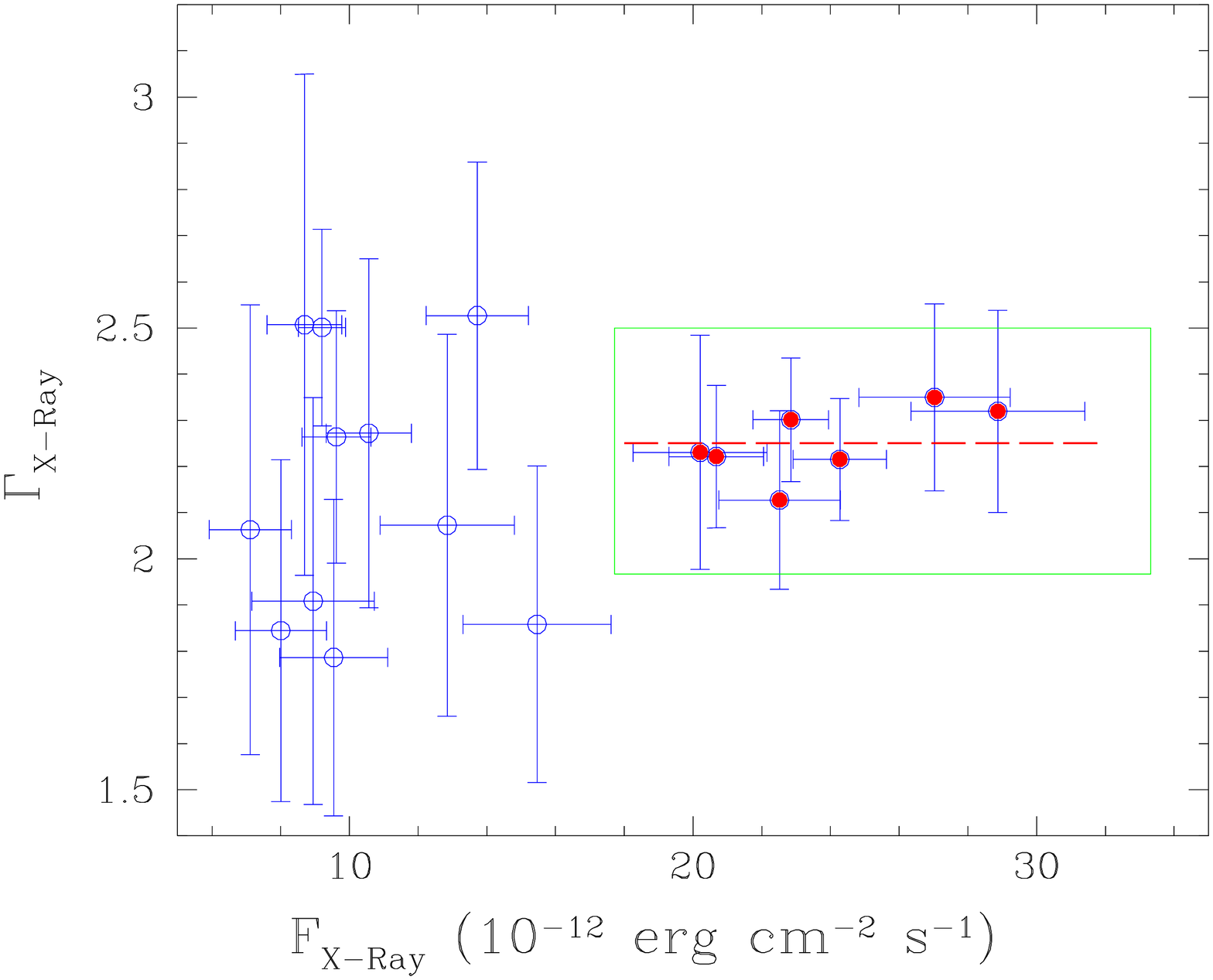}
\includegraphics[scale=0.29,angle=-90]{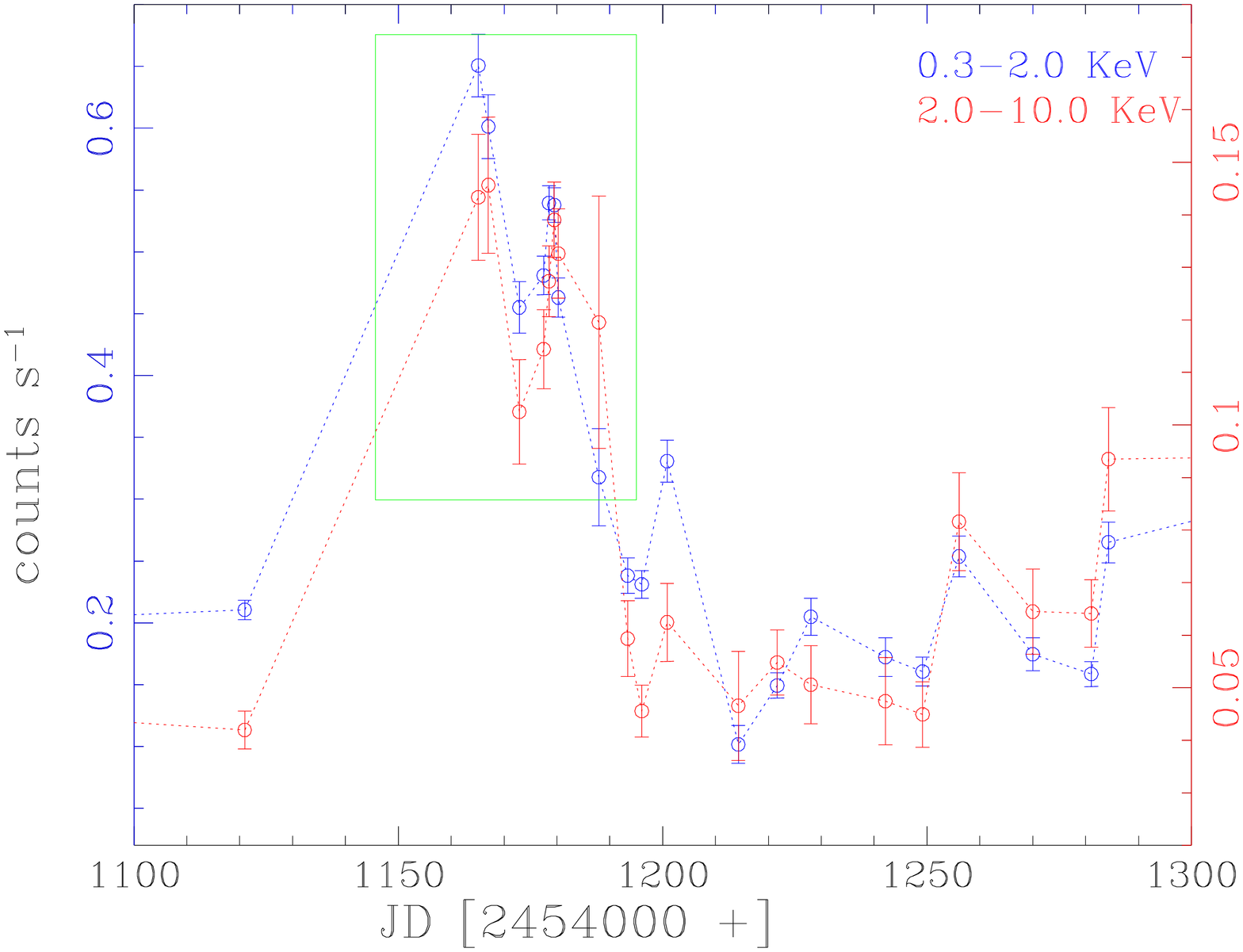}
   \caption{Top: X-ray photon index vs flux at 0.3 -- 10~keV.  The data points in the box belong 
to a phase of brightening shown in the bottom figure. The X-ray photon index of the source is almost 
constant at $2.25 \pm 0.25$ (shown by a dashed line) over the flaring period. 
 Bottom: Soft and hard X-ray light curves of the source over the period of high X-ray activity. The flaring activity is similar in the two X-ray bands. }
\label{orphan_xray}
    \end{figure}

In order to investigate the origin of the X-ray flare (JD' = 1120-1210, Fig. \ref{plot_flx_total_LC}), 
we explore the correlation between X-ray photon index and flux. We do not see 
any systematic change in the X-ray photon index 
($\Gamma_{X-ray}$) w.r.t. a change in the flux.  The X-ray photon index vs flux plot over the flaring 
period between ``5-8" (see Fig. \ref{plot_flx_total_LC} for labeling) is shown in Fig. 
\ref{orphan_xray} (top) and the estimated correlation coefficient $r_P$ is 0.25 with a 
confidence level of 69~\%. Thus, as per correlation statistics, the X-ray photon index and flux 
are not significantly correlated with each other.  
We also notice that the flaring amplitude is similar at soft and hard X-rays as 
shown in Fig. \ref{orphan_xray} (bottom). The percentage fractional variability is 22.5 and 
25 in the soft and hard X-ray bands, respectively. The comparable fractional variability 
implies that the X-ray flare is equally attributed to emission from the soft and the 
hard X-ray bands.

Although the X-ray light curve of the source is the least sampled one among 
all the multi-frequency light curves, we notice a flare peaking between 
``5" and ``6" [JD' = 1000 to 1200] (see Fig. \ref{plot_flx_total_LC}). However, 
due to the gap in the observations it is 
hard to determine the exact peak time of the flare. If we consider that the maximum in 
the X-ray light curve (say X6) is close to the peak of the flare, then this  
epoch coincides with a minimum in the optical/GeV flux and it is 
observed $\sim 50$~days after the O5/G5 flares (see Fig. \ref{plot_flx_total_LC}).    

\begin{figure}
   \centering
\includegraphics[scale=0.4]{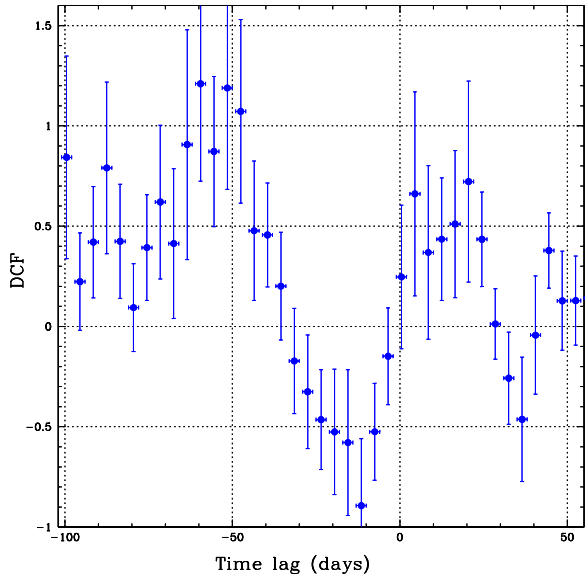}
   \caption{The DCF curve of X-ray vs optical V passband flux using a bin
size of 3 days. }
\label{dcf_Xray_opt}
    \end{figure}

The DCFs of the X-ray light curve with $\gamma$-ray and radio frequency light curves do not 
follow any particular trend as there are very few observations available in the X-ray band. 
A formal X-ray vs optical DCF curve (Fig. \ref{dcf_Xray_opt}) 
shows a peak at a time lag = $-(60 \pm 3)$~days and another peak at $(15 \pm 3)$~days. The large  
DCF error bars are due to sparse data sampling of the X-ray light curve.
In the former case, a negative time lag means that optical variations lead the X-ray ones,
while in the other case the opposite occurs. 
An overall inspection of the light curves in Fig. \ref{plot_flx_total_LC} reveals that 
the optical flare (O5) is observed $\sim 55$~days earlier than the X-ray flare X6, and O6  
appears $\sim 12$~days later. This indicates that the X-ray variability is 
governed by some other effect than the major optical/GeV flares (O5/G5), which appear strongly 
correlated.

\subsection{Radio Spectral analysis}
In the following sections, we will study the spectral variability of S5~0716+714 during 
the different flaring episodes with a focus on the good spectral coverage in the radio bands.

\subsubsection{Modeling the Radio Spectra}
\label{SSA}
\begin{figure*}
   \centering
\includegraphics[scale=0.28, angle=-90]{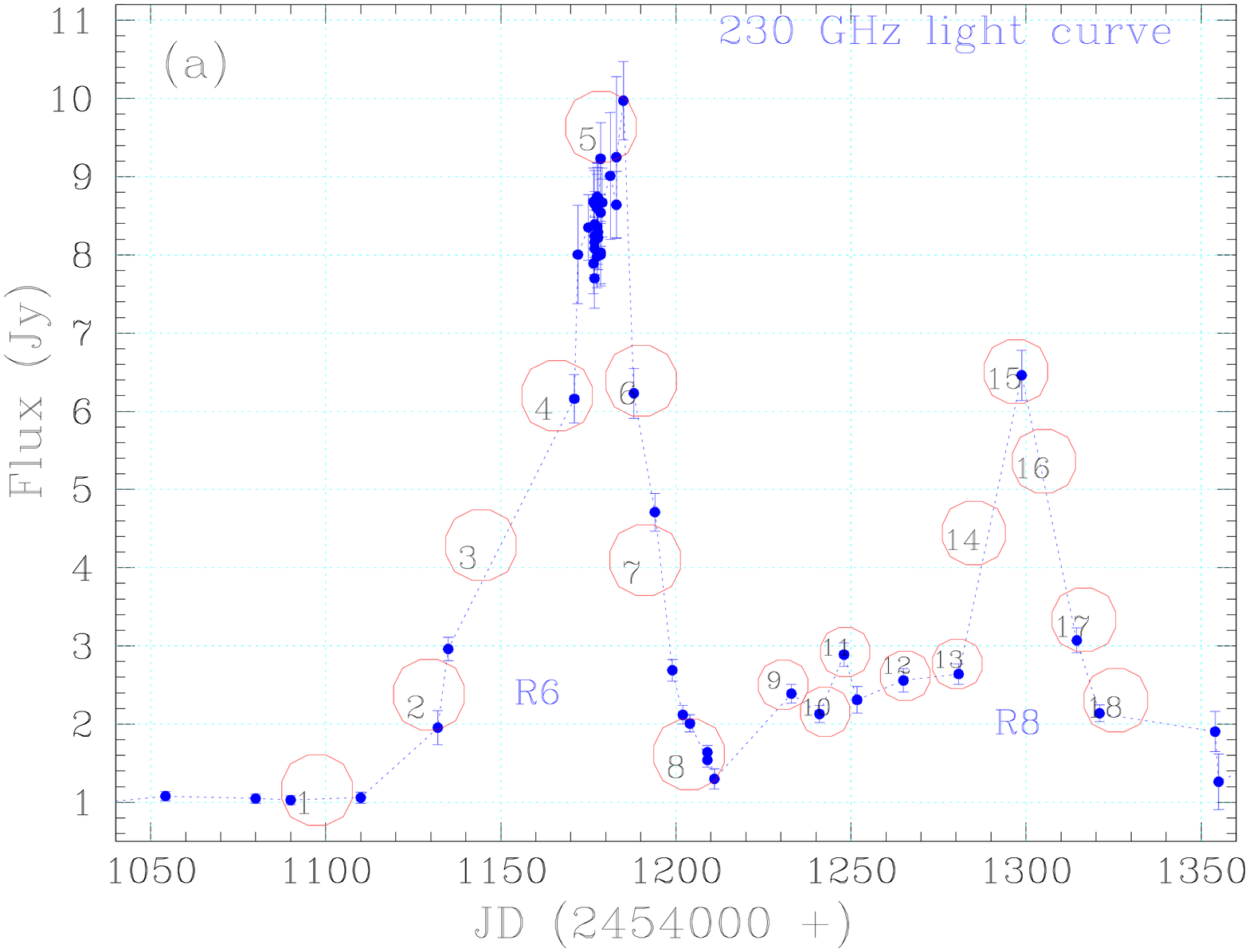}
\includegraphics[scale=0.28, angle=-90]{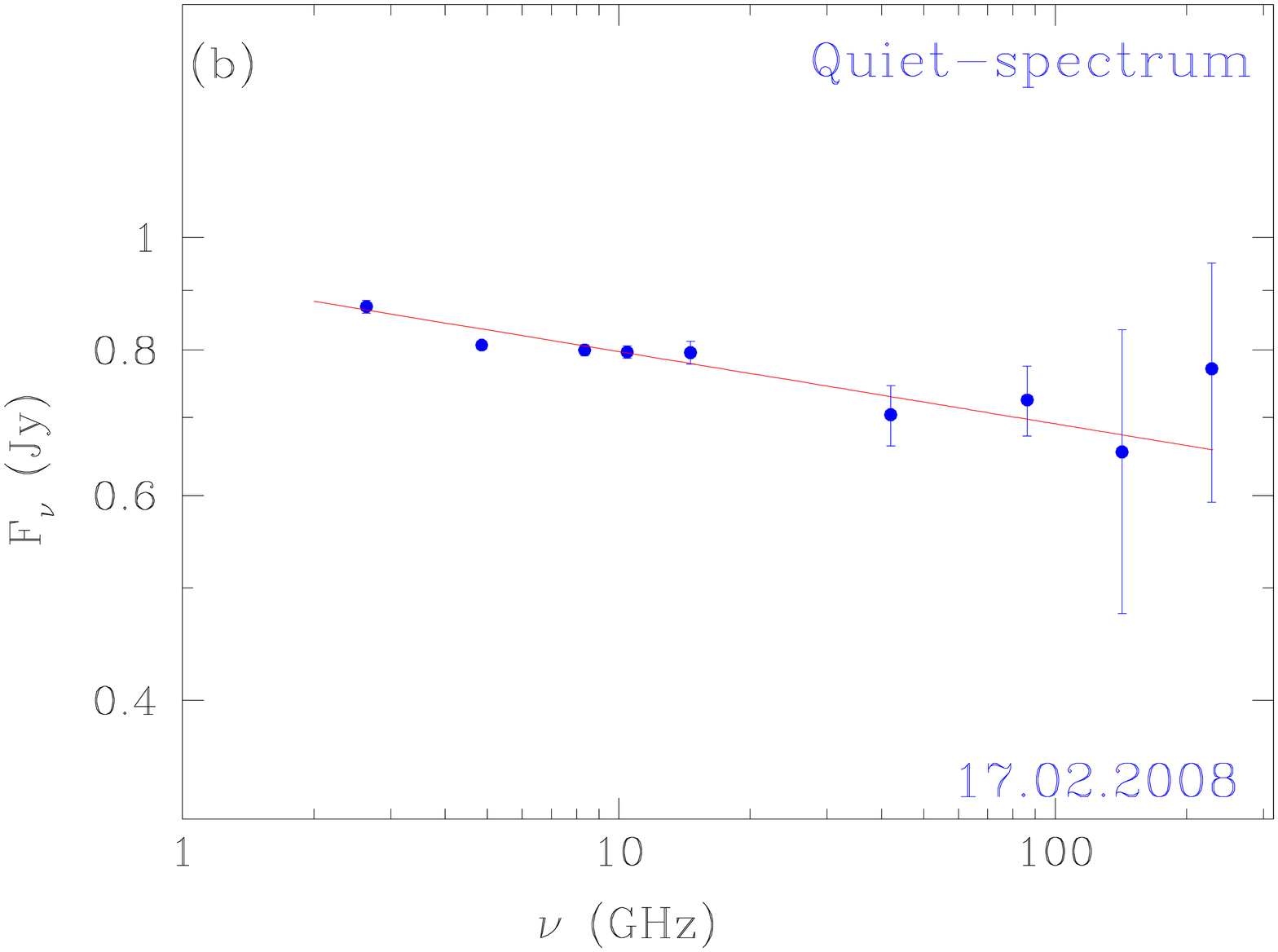}
\includegraphics[scale=0.28, angle=-90]{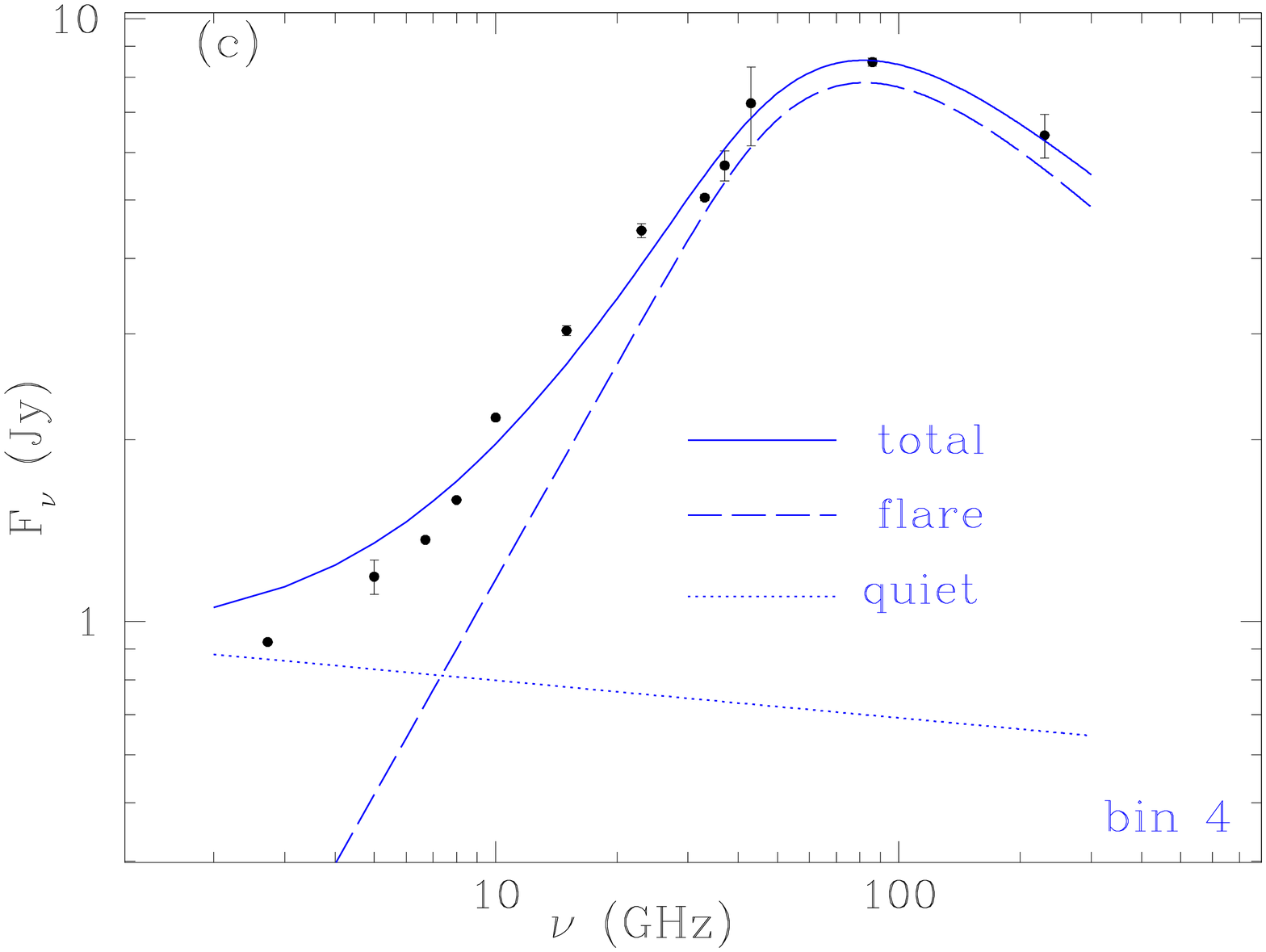}
\includegraphics[scale=0.28, angle=-90]{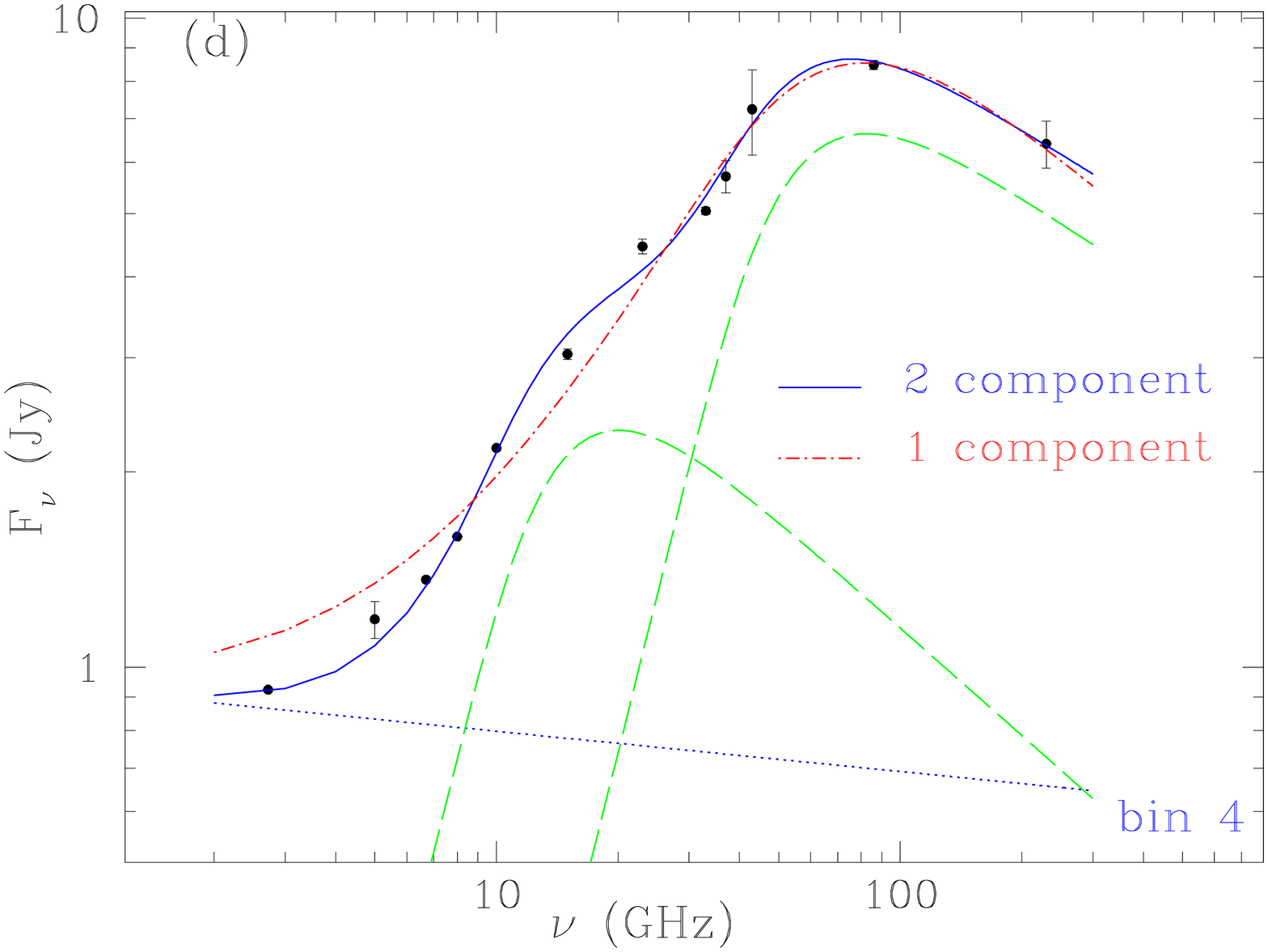}
\includegraphics[scale=0.28, angle=-90]{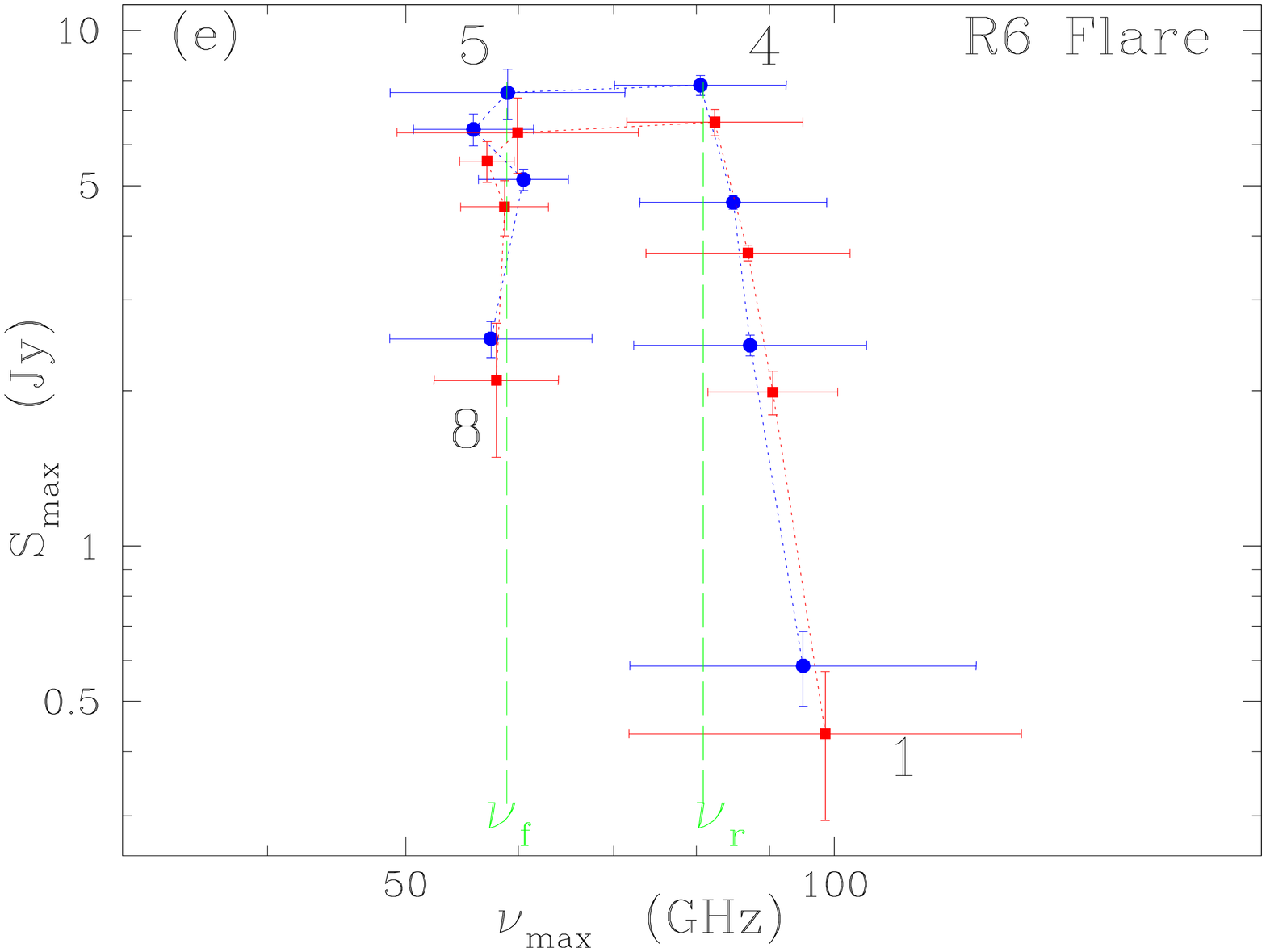}
\includegraphics[scale=0.28, angle=-90]{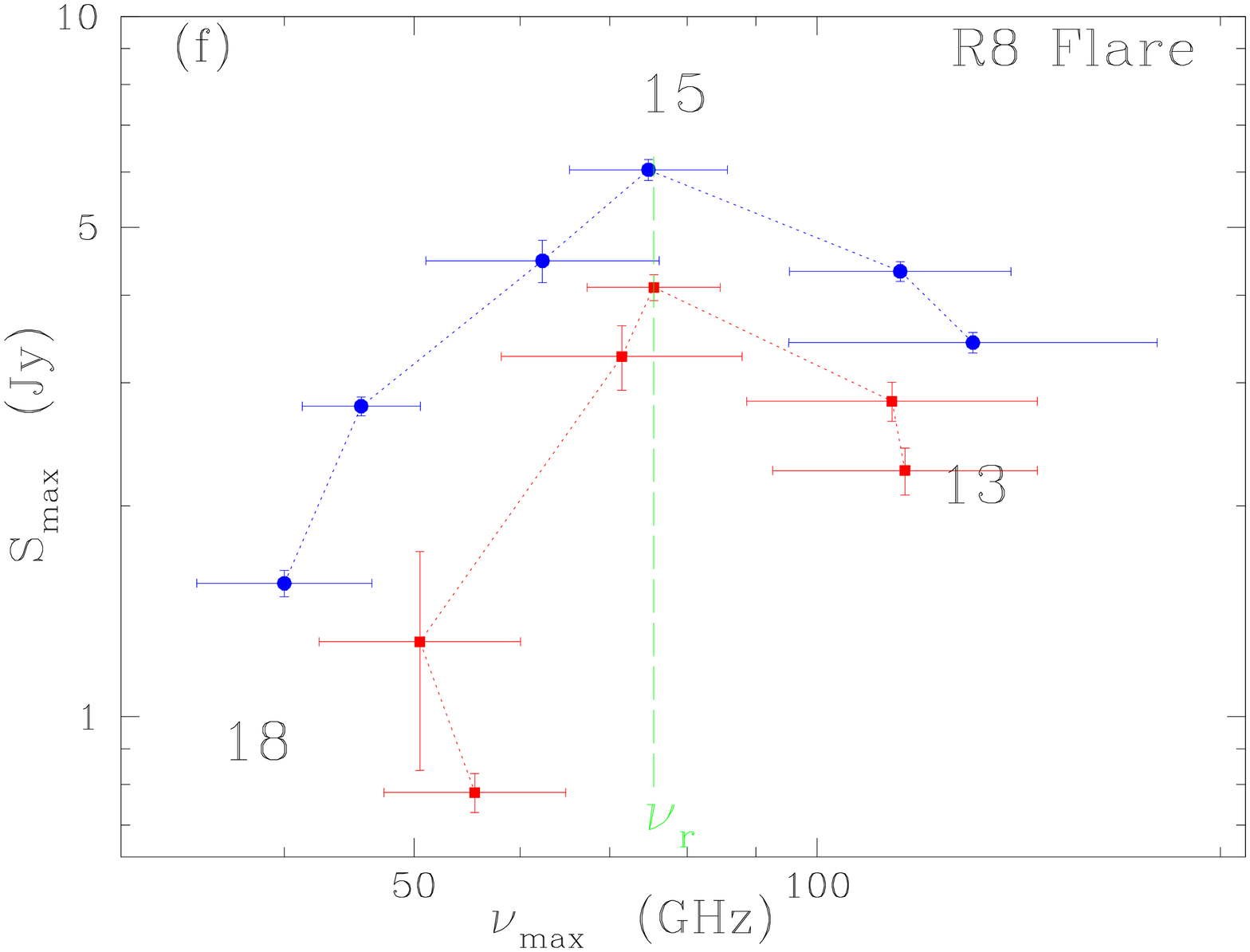}
   \caption{The Evolution of the radio spectra: (a) 230~GHz light curve showing different periods 
over which the spectra are constructed. (b) Quiescent radio spectrum; (c) Results of a single component 
spectral fitting at time bin ``4", the dotted line corresponds to the quiescent spectrum, the dashed one to 
the flaring spectrum and the solid line to the total spectrum. (d) The same spectrum fitted by a two-component 
synchrotron self-absorbed model, with the green dashed line showing the individual components 
and the blue solid line showing a combination of the two. A single component model curve is displayed 
with a dotted-dashed red curve for comparison. (e) $\&$ (f) The time evolution of $S_{max}$ vs $\nu_{max}$ 
for the R6 and R8 radio flares. The spectral evolution extracted using a single-component model is 
shown by blue symbols and the red symbols denote a two-component model (see text for details).}
\label{radio_spectra} 
    \end{figure*}

The multi-frequency radio data allow a detailed study of the spectral evolution of   
the two major radio flares, R6 and R8. We construct quasi-simultaneous radio 
spectra using 2.7 to 230~GHz data. To perform a spectral analysis of the light 
curves simultaneous data points are needed. This is achieved by performing a 
linear interpolation between the flux density values from observations.
A time sampling $\Delta t = 5$~days is selected for the interpolation.
We interpolate the data between the two adjacent observations to predict the flux if the 
data gap is not longer than 5 days; however for longer gaps we drop such data points.   
In Fig. \ref{radio_spectra}, we report the spectral evolution of the R6 and 
R8 radio flares. In this figure, the flux densities are averaged values over the 5 days 
binning period and the uncertainties in the fluxes represent their variation over this 
period.

The observed radio spectrum is thought to result from the superposition of emission from 
the steady state (unperturbed region) and the perturbed (shocked) regions of the jet. We constructed 
the quiescent spectrum using the lowest flux level during the course of our observations. 
Emission from a steady jet is better characterized by a relatively flat spectrum, so we choose the steepest one 
observed on February 17, 2008. The quiescent spectrum is shown in Fig. \ref{radio_spectra} (b). 
The flux densities were fitted by a power law $F(\nu) = C_q (\nu/{\rm GHz})^{\alpha_{q}}$ with  
$C_q = (0.92 \pm 0.02)$~Jy and $\alpha_{q} = -(0.062 \pm 0.007)$.

We fitted the radio spectra using a synchrotron self-absorbed spectrum. A 
synchrotron self-absorbed (SSA) spectrum can be described as \citep[see:][for details]{turler2000, 
fromm2011} : 

\begin{equation}
S_\nu=S_m\left(\frac{\nu}{\nu_m}\right)^{\alpha_t}\frac{1-\exp{\left(-\tau_m\left(\nu/\nu_m\right)^{\alpha_0-\alpha_t}\right)}}{1-\exp{(-\tau_m)}},
\label{snu}
\end{equation}
where $\tau_m\approx3/2\left(\sqrt{1-\frac{8\alpha_0}{3\alpha_t}}-1\right)$ is the optical depth at the turnover 
frequency, $S_m$ is the turnover flux density, $\nu_m$ is the turnover frequency and $\alpha_t$ and 
$\alpha_0$ are the spectral indices for the optically thick and optically thin parts of the spectrum, 
respectively ($S \sim \nu^{\alpha}$).

\begin{table}
\scriptsize 
\caption{Best-fit spectral parameters for the evolution of radio flares using a one-component SSA model }
\begin{tabular}{c c c c c c} \hline
bin   & Time           & $S_m$         &$\nu_m$         &$\alpha_t$         &$\alpha_0$     \\
    & JD' [JD-2454000]  & [Jy]          & [GHz]          &                   &                    \\\hline
 1  &1096-1101     & 0.58$\pm$0.09 &   95.05$\pm$21.78  &   0.70$\pm$0.26   & -1.15$\pm$0.61   \\     
 2  &1130-1135     & 2.45$\pm$0.11 &   87.26$\pm$7.92   &   1.26$\pm$0.18   & -0.37$\pm$0.13    \\    
 3  &1150-1155     & 4.64$\pm$0.14 &   84.92$\pm$5.27   &   1.15$\pm$0.11   & -0.40$\pm$0.09    \\
 4  &1173-1178     & 7.83$\pm$0.34 &   80.52$\pm$4.59   &   1.12$\pm$0.11   & -0.62$\pm$0.12    \\  
 5  &1189-1194     & 7.57$\pm$0.84 &   58.96$\pm$8.05   &   1.37$\pm$0.41   & -0.61$\pm$0.30    \\  
 6  &1197-1201     & 6.42$\pm$0.45 &   55.78$\pm$2.90   &   1.06$\pm$0.12   & -1.48$\pm$0.26    \\  
 7  &1204-1209     & 5.13$\pm$0.24 &   60.49$\pm$2.22   &   0.97$\pm$0.08   & -1.56$\pm$0.18    \\  
 8  &1216-1221     & 2.52$\pm$0.20 &   57.39$\pm$6.03   &   0.70$\pm$0.10   & -1.24$\pm$0.33     \\
 9  &1226-1230     &3.23$\pm$0.17  &   97.03$\pm$12.60  &   0.39$\pm$0.08   & -1.14$\pm$0.71     \\
 10 &1238-1242     &3.39$\pm$0.20  &   106.80$\pm$17.90 &   0.32$\pm$0.05   & -1.30$\pm$0.69     \\
 11 &1267-1272     &3.57$\pm$0.13  &   92.24$\pm$7.94   &   0.43$\pm$0.07   & -0.94$\pm$0.43     \\
 12 &1273-1278     &3.21$\pm$0.15  &   87.92$\pm$14.80  &   0.58$\pm$0.44   & -0.31$\pm$0.12     \\
 13 &1283-1288     & 3.42$\pm$0.11 &   130.70$\pm$32.50 &   0.67$\pm$0.09   & -0.35$\pm$0.17   \\   
 14 &1290-1295     & 4.32$\pm$0.13 &   115.30$\pm$8.14  &   0.69$\pm$0.05   & -0.71$\pm$0.18   \\   
 15 &1298-1303     & 6.04$\pm$0.20 &   74.81$\pm$4.44   &   1.07$\pm$0.10   & -0.47$\pm$0.09    \\   
 16 &1309-1313     & 4.48$\pm$0.31 &   62.35$\pm$9.09   &   1.07$\pm$0.26   & -0.33$\pm$0.17    \\   
 17 &1318-1323     & 2.77$\pm$0.08 &   45.66$\pm$3.06   &   1.56$\pm$0.29   & -0.29$\pm$0.06    \\  
 18 &1340-1345     & 1.55$\pm$0.06 &   40.00$\pm$5.22   &   1.07$\pm$0.27   & -0.18$\pm$0.09   \\\hline
\end{tabular} \\
\label{para_1comp}
\end{table}

For the spectral analysis, we  first subtract the contribution of the quiescent spectrum from 
the data and then used eq. \ref{snu} for fitting.
The uncertainties of the remaining flaring spectrum are calculated taking into account the errors of 
the interpolated data points and the uncertainties of the quiescent spectrum.
We tried two independent approaches to model the radio spectra: (i) a one-component SSA model, 
(ii) a two-component SSA model. \\

\noindent 
{\bf One-component SSA model :} 
 During the fitting process we allowed all four parameters ($S_m, ~\nu_m, ~\alpha_t, ~\alpha_0$) 
(see eq. \ref{snu}) to vary.
In Fig. \ref{radio_spectra} (a) we show the 230 GHz light curve with labels (``numbers"), marking the 
time of best spectral coverage, for which spectra can be calculated. A typical spectrum (for time bin ``4") 
is shown in Fig. \ref{radio_spectra} (c). In Table \ref{para_1comp} we list the spectral parameters of the 
one-component SSA fit for all spectral epochs (bin 1 to 18). 
 In a homogeneous emission region, the spectrum is described by  
characteristic shapes $I_{\nu} \propto \nu^{5/2}$ and $I_{\nu} \propto \nu^{-(s-1)/2}$  for 
the optically thick and thin domain ($s$ is the power law index of the relativistic electrons), respectively.  
Thus, the theoretically expected value of the optically thick spectral index, $\alpha_t$ is 2.5. 
While fitting the spectra 
with a single-component SSA model, we find that $\alpha_t$ varies between 0.32 to 1.56. This  
deviation of $\alpha_t$ from 2.5 indicates that the emission region is not homogeneous, 
and it may be composed of more than one homogeneous components. We also notice that the radio spectra over 
the period between the two radio flares R6 and R8 (from bin9 to bin12) can not be described by
such a spectral model at all.  
Apparently, these spectra seem to be composed of two different components, one peaking 
near 30~GHz (low-frequency component) and the other one at $\sim 100$~GHz (high-frequency component). 
Consequently, we consider a two-component model. \\

\begin{table*}
\centering
\caption{The best fitted spectral parameters over the evolution of radio flares using a two component SSA model }
\begin{tabular}{c c c c c c c c} \hline
bin  &Time       & $\nu_ml$$^{*}$& $S_ml$       &$\alpha_0l$     & $\nu_mh$$^{*}$      & $S_mh$       &$\alpha_0h$        \\
  & JD' [JD-2454000]&   GHz    & Jy            &    & GHz               & Jy              &                \\\hline  
 1  &1096-1101 &20$\pm$0  &0.26$\pm$0.10   &-0.47$\pm$0.59 &   98.52$\pm$32.74  &0.43$\pm$0.13   &-0.51$\pm$0.59   \\
 2  &1130-1135 &20$\pm$0  &0.41$\pm$0.16   &-0.12$\pm$0.13 &   90.00$\pm$15.25  &1.98$\pm$0.11   &-0.12$\pm$0.03   \\
 3  &1150-1155 &20$\pm$0  &1.37$\pm$0.14   &-0.32$\pm$0.08 &   86.95$\pm$6.14   &3.70$\pm$0.13   &-0.23$\pm$0.05   \\
 4  &1173-1178 &20$\pm$0  &2.31$\pm$0.29   &-0.55$\pm$0.20 &   82.41$\pm$4.79   &6.63$\pm$0.39   &-0.41$\pm$0.08   \\
 5  &1189-1194 &20$\pm$0  &2.17$\pm$1.00   &-0.40$\pm$0.47 &   59.90$\pm$8.55   &6.33$\pm$1.05   &-0.62$\pm$0.31   \\
 6  &1197-1201 &20$\pm$0  &2.17$\pm$0.24   &-0.39$\pm$0.12 &   57.01$\pm$1.62   &5.57$\pm$0.50   &-0.72$\pm$0.31   \\
 7  &1204-1209 &20$\pm$0  &1.46$\pm$0.26   &-0.21$\pm$0.11 &   58.66$\pm$2.18   &4.55$\pm$0.55   &-0.68$\pm$0.38   \\
 8  &1216-1221 &20$\pm$0  &1.13$\pm$0.13   &-0.13$\pm$0.05 &   57.88$\pm$3.02   &2.09$\pm$0.60   &-0.91$\pm$1.50   \\
 9  &1226-1230 &18$\pm$1  &0.95$\pm$0.29   &-1.06$\pm$0.69 &   128.20$\pm$7.33  &3.34$\pm$0.10   &-0.46$\pm$0.11   \\
 10 &1238-1242 &18$\pm$1  &1.20$\pm$0.41   &-0.76$\pm$0.38 &   126.10$\pm$8.48  &3.40$\pm$0.13   &-0.55$\pm$0.17   \\
 11 &1267-1272 &22$\pm$1  &0.76$\pm$0.13   &-1.52$\pm$0.67 &   124.40$\pm$4.18  &3.60$\pm$0.07   &-0.38$\pm$0.03   \\
 12 &1273-1278 &23$\pm$1  &0.94$\pm$0.19   &-1.83$\pm$0.98 &   129.90$\pm$8.47  &3.29$\pm$0.12   &-0.33$\pm$0.04   \\
 13 &1283-1288 &20$\pm$0  &1.55$\pm$0.18   &-0.16$\pm$0.06 &   116.30$\pm$12.20 &2.24$\pm$0.17   &-0.20$\pm$0.02   \\
 14 &1290-1295 &20$\pm$0  &1.65$\pm$0.17   &-0.11$\pm$0.03 &   113.70$\pm$15.64 &2.82$\pm$0.18   &-0.28$\pm$0.13   \\
 15 &1298-1303 &20$\pm$0  &2.42$\pm$0.16   &-0.19$\pm$0.03 &   75.51$\pm$3.52   &4.10$\pm$0.17   &-0.41$\pm$0.06   \\
 16 &1309-1313 &20$\pm$0  &2.61$\pm$0.41   &-0.47$\pm$0.75 &   80.93$\pm$9.89   &3.27$\pm$1.88   &-0.40$\pm$0.28   \\
 17 &1318-1323 &20$\pm$0  &1.82$\pm$1.07   &-0.33$\pm$0.40 &   50.10$\pm$5.70   &1.27$\pm$1.51   &-0.21$\pm$0.27   \\
 18 &1340-1345 &20$\pm$0  &0.87$\pm$0.05   &-0.05$\pm$0.01 &   55.70$\pm$7.80   &0.77$\pm$0.05   &-0.26$\pm$0.08   \\\hline 
\end{tabular} \\
$*$ : Index $l$ is for low-frequency component and ~h is for high-frequency component. \\
\label{para_2comp}
\end{table*}

\noindent 
{\bf Two-component SSA model :}
Since the flux densities at cm wavelengths are much higher than the extrapolation 
of the mm-flux with a spectral index $\alpha_t = 2.5$ for the optically thick branch of a 
homogeneous synchrotron source, we assume that besides the mm-submm emitting component, 
there is an additional spectral component which is responsible for the cm emission. 
We therefore fit the radio spectra with a two-component model. This allows us 
to fix $\alpha_t$, and set it to 2.5 for both of the components. We also fix the peak frequency 
of the lower frequency component to 20~GHz, as the low-frequency $\nu_m$ varies between 18 -- 25~GHz 
and the fitting improves only marginally if we allow this parameter to vary. Hence, we study the 
spectral evolution of the radio spectra by fixing $\alpha_t(l)$\footnote{l: low-frequency 
component, ~h: high-frequency component} = $\alpha_t(h) = 2.5$ and $\nu_{ml} = 20$~GHz. 
Such a scenario appears reasonable and is motivated by the idea of a synchrotron self-absorbed 
``Blandford-K\"onigl" jet \citep{blandford1979} and a more variable core or shock component.  
The fitted spectrum using this restricted two-component model is shown in 
Fig. \ref{radio_spectra} (d) and the best fit parameters are given in Table \ref{para_2comp}. 
A variable low-frequency component provides a better fit over bin 9 -- 12. Therefore, we consider  
that both the low- and high- frequency components are varying over the time period between the two flares. 
The two-component SSA model describes the radio spectra much better than a single-component model. We 
therefore conclude that the radio spectra are at least composed of two components, one peaking at 
cm wavelengths and the other at mm-submm wavelengths.

\subsubsection{Evolution of radio flares } 
In the following we adopt a model of spectral evolution as described by \citet{marscher1985} 
which considers the evolution of a traveling shock wave in a steady state jet.  
The typical evolution of a flare in turnover frequency -- turnover flux density ($S_{m}$ -- $\nu_{m}$) 
plane can be obtained by inspecting the 
$R$ (radius of jet)-dependence of the turnover frequency, $\nu_{m}$, and the turnover flux density, $S_{m}$ 
\citep[see][for details]{fromm2011}. During the first stage,  Compton losses are dominant and 
$\nu_{m}$, decreases with increasing radius, $R$, while $S_{m}$, increases. In the second stage, where 
synchrotron losses are the dominating energy loss mechanism, the turnover frequency continues 
to decrease while the turnover flux density remains constant. Both the turnover frequency and 
turnover flux density decrease in the final, adiabatic stage. 
We studied the evolution of the radio flares using the results obtained from both one- and two-component 
SSA models and in each case, we obtained similar results. The evolution of the R6 and R8 
flares in $S_{m}$ -- $\nu_{m}$ plane are shown in Fig. \ref{radio_spectra} (e) -- (f). 
 
In a standard shock in jet model, $S_m \propto \nu_m ^{\epsilon_i}$ \citep{fromm2011,  marscher1985} 
where $\epsilon_i$ depends upon the variation of physical quantities i.e. magnetic field (B), 
Doppler factor ($\delta$) and energy of relativistic electrons \citep[see][for details]{fromm2011, marscher1985}. 
The estimated $\epsilon_i$ values for both the one- and two-component SSA models are given in 
Table \ref{fitted_para_SSA}.  

\citet{marscher1985} predicted a value of $\epsilon_{\rm Compton}$ = $-$2.5 and \citet{fromm2011} obtain  
$-$1.21, whereas \citet{Bjornsson2000} obtained $\epsilon_{\rm Compton}$ = $-$0.43  using a modified 
expression for the shock width. The estimated $\epsilon_{\rm Compton}$ value for the R8 flare lies between 
these values, while for the R6 flare it is too high to be explained by the simple assumptions of a 
standard shock-in-jet model (see Table \ref{fitted_para_SSA}). For the adiabatic stage \citet{marscher1985} derived an exponent 
$\epsilon_{\rm adiabatic} = 0.69$ (assuming $s = 3$) and \citet{fromm2011} found $\epsilon_{\rm adiabatic} = 0.77$. 
We obtain $\epsilon_{\rm adiabatic} \sim 2$ for the R8 flare and $\sim 10$ for the R6 flare which is again too steep.  
The spectral evolution of the R8 radio flare can be well interpreted in terms of a standard shock-in-jet 
model based on intrinsic effects. The rapid rise and decay of $S_m$ w.r.t. $\nu_m$ in the case of the R6 
(see Fig. \ref{radio_spectra}) flare rule out these simple assumptions of a standard shock-in-jet 
model considered by \citet{marscher1985} with a constant Doppler factor ($\delta$). 

We argue that the spectral evolution of the radio flare, R6 (in $S_m - \nu_m$ plane) need to be investigated by 
considering both the intrinsic variation and the variation in the Doppler factor ($\delta$) of the emitting region. 
\citet{qian1996} studied the intrinsic evolution of the superluminal components in 3C~345 with its beaming factor 
variation being taken into account with a typical variation of the viewing angle by 2 -- 8$^{\circ}$.  
In the study of the spectral evolution of the IR-mm flare in 3C~273, \citet{qian1999} found that the bulk 
acceleration of the flaring component improves the fit of the spectral evolution at lower frequencies. 
Therefore, it is worthwhile to include a variation of $\delta$ along the jet axis in our model, which we
parametrize as $\delta \propto R^b$. Such an approach 
could easily explain the large variation in the observed turnover flux density, while the observed turnover 
frequency kept a nearly constant value or changed only slightly \citep{fromm2011}. 

We consider the evolution of radio flares in the framework of dependencies of physical parameters 
$a$, $s$ and $d$  following \citet{lobanov1999}.  Here, $a$, $s$ and $d$ parametrize the variations 
of $\delta \propto R^b$, $B \propto R^{-a}$ and $N(\gamma) \propto \gamma^{-s}$ along jet radius. 
Since it is evident that $\epsilon$ values 
do not differ much for different choices of $a$ and $s$ \citep{lobanov1999}, we assume for simplicity that
$s \approx$ constant and for two extreme values of $a = 1$ and 2, we investigate the variations in $b$.  
The calculated values of $b$ for different stages of evolution of radio flares are given in Table \ref{fitted_para_SSA}. 
{\it It is important to note that the Doppler factor varies significantly along jet radius during the evolution 
of the two radio flares. }


Moreover, the turnover frequency between the Compton and synchrotron stages ($\nu_r$) and the synchrotron
and adiabatic stages ($\nu_f$) 
in the $S_m$ -- $\nu_m$ plane characterize the observed behavior of the radio outbursts \citep{valtaoja1992}.   
In a shock induced flare, the shock strength reaches its maximal development at $\nu_r$ and the decay stage starts 
at $\nu_f$. In Fig. \ref{radio_spectra} (e) -- (f), we display by dashed lines the frequencies $\nu_r$ and $\nu_f$. 
The shock reaches its maximal development at 80~GHz for the R6 flare and at 74~GHz for the R8 flare. The 
observed behavior of the outburst depends on $\nu_r$. In a shock induced flare, the observed frequency 
($\nu_{obs}$) is less than $\nu_r$ 
in the case of low-peaking flares, while $\nu_{obs} > \nu_r$ for high-peaking flares \citep{valtaoja1992}. 
We therefore conclude that both the R6 and R8 radio outbursts are low-peaking radio flares and are in 
quantitative agreement with the formation of a shock and its evolution.

\begin{table*}
\scriptsize 
\caption{Different states of spectral evolution and their characteristics }
\begin{tabular}{c c c c c c c } \hline    
Flare  &Time          & bin     & $\epsilon$      &$\epsilon$          & b            &Stage      \\
       &JD' [JD-2454000] &         &(1 component SSA)&(2 component SSA)   & s=2.2, a=1-2    &      \\\hline 
R6     &1096-1178     &1-4      & -7$\pm$3        &-8$\pm$3            & 0.7      & Compton    \\
       &1178-1194     &4-5      & 0               & 0                  & -0.07    & Synchrotron \\
       &1194-1221     &5-8      & 10$\pm$2        & 11$\pm$3           &2.6       & Adiabatic    \\
R8     &1283-1303     &13-15    & -0.9$\pm$0.1    &-1.2$\pm$0.2        &0.4       & Compton      \\
       &1298-1345     &15-18    & 1.8$\pm$0.2     &2.5$\pm$0.5         &-2        & Adiabatic     \\\hline 
\end{tabular} \\
$\delta \propto R^b$, $B \propto R^{-a}$ and $N(\gamma) \propto \gamma^{-s}$ \\
\label{fitted_para_SSA}
\end{table*}

\subsubsection{Synchrotron Spectral Break } 
\label{spectra_break}
\begin{figure}
   \centering
\includegraphics[scale=0.4]{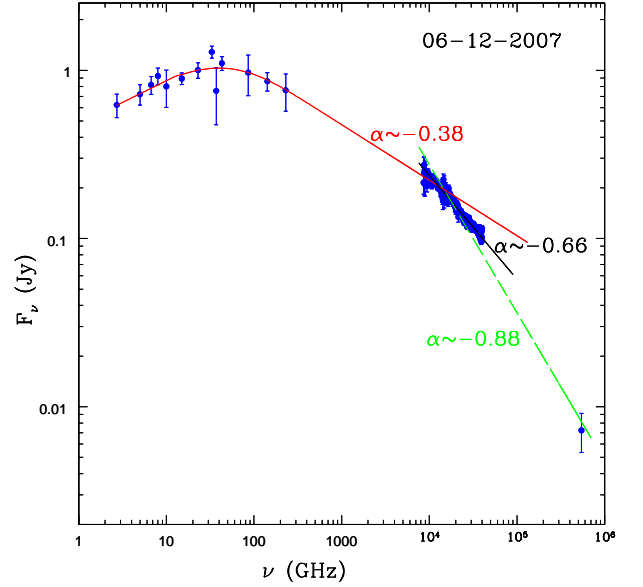}
\caption{Radio-IR spectra using Spitzer observations. The red curve is the best fitted 
synchrotron self-absorbed spectra with a break at (1.3$\pm$0.1)$\times 10^4$ GHz with a 
spectral break $\Delta \alpha = 0.28\pm0.1$. The green line represents the spectral fitting including 
optical data point and this leads to spectral break $\Delta \alpha = 0.51\pm0.09$. }
\label{rad_iras}
\end{figure}

The source was observed at IR frequencies with the Spitzer Space Telescope on December 06, 2007.
We obtained the IRAC+MIPS photometric measurements at 5 -- 40 $\mu m$ from the Spitzer archive 
(http://sha.ipac.caltech.edu/applications/Spitzer/SHA/).  Since the source has been observed at 
radio wavelengths over this period, we combine the cm -- mm and IR observations to construct a more 
complete broadband synchrotron spectrum. The combined radio -- IR spectrum is shown in Fig. \ref{rad_iras}. 
The red curve represents the best fitted synchrotron self-absorbed spectrum with a break at a frequency 
of $\nu_{b} = (1.3 \pm 0.1) \times 10^4$~GHz. The best-fit parameters are: 
$S_m = (1.03 \pm 0.02)$~Jy, $\nu_m = (45.74 \pm 3.12)$~GHz,
$\alpha_t = (0.33 \pm 0.01)$ and the spectral index of the optically thin part ($\alpha_0$) is  
$-(0.38 \pm 0.09)$ and $-(0.66 \pm 0.07)$ above and below the break, respectively.  Hence, modeling of the radio -- IR 
spectrum provides strong evidence for a break in the synchrotron spectrum at $\nu_b \sim 1.3 \times 10^4$~GHz  with a 
spectral break $\Delta \alpha = 0.28 \pm 0.1$. We can also include the optical V passband flux from the AASVO 
(see section \ref{data_opt}) to estimate the spectral break. This leads to a steeper spectral index with  
$\alpha_0 = -0.88 \pm 0.03$ and a break $\Delta \alpha = 0.51 \pm 0.09$.

The spectral break could be attributed to synchrotron loss of the high energy electrons. It is widely accepted that 
synchrotron losses result in a steepening of the particle spectrum by one power and a steepening of the emitted 
synchrotron spectrum by a half-power \citep{reynolds2009, kardashev1962}. Also, synchrotron-loss spectral
breaks differing from 0.5 could be produced naturally in an inhomogeneous source \citep{reynolds2009}.
As $\nu_b$ is mainly determined by synchrotron loss, it depends on the magnetic field strength. One can 
estimate the  minimum-energy magnetic field strength using the following relation given by \citet{heavens1987}: 
\begin{equation}
B_{break} = 2.5 \times 10^{-3} \beta_1^{2/3} L^{-2/3} \nu_b^{-1/3} ~G 
\end{equation}
where $\beta_1 . c$ is the speed of the upstream gas related to the shock, L is the 
length of the emission region in kpc (at $\nu < \nu_b$) and $\nu_b$ is the break frequency in GHz. 
For relativistic shocks $\beta_1$ is close to 1. 
We constrain the length of the emission region $L$ using the variability timescales at 230~GHz as this is the closest 
radio frequency to $\nu_b$. Using $L \leq 0.04 \times 10^{-3}$~kpc (see Section \ref{em_size}), we found $B_{break} \geq 0.14$~G. 
The minimum energy condition implies equipartition of energy, which means $B_{break} \sim B_{eq}$ (equipartition magnetic field).

\subsection{Brightness Temperature, size of emission region and Jet Doppler factors}
\label{brightness_temperature}

\subsubsection{Brightness Temperature $T_{B}^{app}$ } 
The observed rapid variability implies a very compact emission region and hence a high 
brightness temperature if the variations are intrinsic to the source. Assuming 
a spherical brightness distribution for the variable source and that the triggered flux 
variations propagate isotropically through the source, then the light travel time 
argument implies a radius $d \leq c \Delta t$ for the emission region where $\Delta t$
is the time interval of expansion. So, the flux variability observed in radio bands 
allows us to estimate the brightness temperature of the source using the relation 
\citep[see][for details]{ostorero2006, fuhrmann2008}. 

\begin{eqnarray}
T_{B}^{app}=3.47\times10^{5}\cdot \Delta S_{\lambda}\left(\frac{\lambda\,d_{L}}{t_{var,\lambda}
\,(1+z)^{2}}\right)^{2}\,\, K
\label{T_b}
\end{eqnarray}

where $\Delta S_{\lambda}$ is the change in flux density (Jy) over time $t_{var,\lambda}$ (years), 
$d_{L}$ is the luminosity distance in Mpc, $\lambda$ is wavelength in cm and z is the 
redshift of the source. Here 
and in the following calculations we will use $z = 0.31$, which yields a luminosity distance, $d_{L} = 1510$~Mpc 
\citep[see][for details]{fuhrmann2008}.  

Two major outbursts (R6 and R8) are observed in the source at 15~GHz and at higher radio 
frequencies. To calculate the brightness temperature, we have used the rising time of the flares 
(see Table \ref{tab_R6model}) 
separately for the two flares at 15, 37, 86 and 230~GHz, as these are  
the best sampled light curves. The radio flares follow a slow rising and fast decaying 
trend, so we calculate $t_{var}$ for both the rising and the decay phases of the flares. The calculated 
$t_{var}$ for the two radio flares are listed in column 2 of Table \ref{tab_dopp} and the  
apparent brightness temperatures ($T_{B}^{app}$) are in column 3.

\subsubsection{Doppler factor from variability timescales $\delta_{var}$  } 
The calculated $T_{B}^{app}$  is one to two orders of magnitude higher than the 
IC-limit $T_{B, IC}^{limit}$ of $T_B \sim$ 10$^{12}$ K \citep{kellermann1969} at 
all frequencies up to 230~GHz. We notice that $T_{B}^{app}$ exceeds 
$T_{B, IC}^{limit}$ even at short-mm bands.  The excessive brightness 
temperature can be interpreted by relativistic boosting of the radiation, which gives 
to a lower limit of the Doppler factor of the emitting region 

\begin{equation}
\delta_{var} = (1+z)~\Big [\frac{T_{B}^{app}}{10^{12}} \Big ] ^{\frac{1}{3 + \alpha}}
\end{equation}    
Here $\alpha$ is the spectral index of the optically thin part of the radio spectrum. We obtained 
$\alpha_{thin} = -0.23$ to $-0.91$ for the R6 radio flare and $\alpha_{thin} = -0.20$ to $-0.41$ for 
the R8 flare (see Section 4.1 for details). The calculated $\delta_{var}$ values are listed in 
column 4 of Table \ref{tab_dopp}. We obtain $\delta_{var} \geq 14$ for the two radio flares.

\begin{table}
\caption{Variability Brightness temperatures }
\begin{tabular}{l c c c c }  
           &              &  R6 flare              &                     &    \\\hline               
Frq. (GHz)  &t$_{var}^{a}$   &T$_{B}^{app}$            & $\delta_{var}$  & $\theta$   \\
(GHz)      &(days)       &(10$^{12}$ K)            &                     &(mas)        \\\hline
15         &61           &154                      & 10                  &0.091             \\
37         &65           &62                       & 7                   &0.068             \\
86         &60           &13                       & 3                   &0.027             \\
230        &50           &3                        & 2                   &0.015             \\
           &             &                         &                     &                  \\\hline
           &             &  R8 Flare               &                     &                  \\\hline
15         &37           &307                      & 14                  &0.077             \\
37         &18           &109                      & 9                   &0.025             \\
86         &25           &55                       & 7                   &0.021             \\
230        &10           &5                        & 3                   &0.004             \\\hline 
\end{tabular}  \\
a: the lower value corresponds to the rising phase of flare while the higher to the decaying phase.
\label{tab_dopp}
\end{table}

In addition, we can also use the intrinsic brightness temperature limit based on the equipartition between particle energy and field energy \citep{scott1977}: $T_{B,eq}\sim\,5\times10^{10}$\,K 
which is derived on the basis of an argument that this limit better reflects the stationary 
state of a synchrotron source which for many sources yields 
$T_B \lesssim\,10^{11}$\,K \citep[e.g.]{readhead1994}. 
In this case, the calculated Doppler factor values using the equipartition limit, 
$\delta_{var,eq}=(1+z)\sqrt[3+\alpha]{T_{B}^{app}/5\times10^{10}}$, become higher by a factor of 4.47 
i.e. $\delta_{var,eq} = 4.47 \times \delta_{var}$.

\subsubsection{Size of the emission region $\theta$  } 
\label{em_size}
One can obtain the size of the emission region using the calculated Doppler factors ($\delta$) and 
variability time scales ($t_{var}$) : 
\begin{eqnarray}
\theta=0.173\,\frac{t_{var}}{d_{L}}\,\delta(1+z)\,\,  mas  
\label{dim}
\end{eqnarray}

The angular size $\theta$ calculated using $\delta_{var,IC}$ are listed column 5 of Table \ref{tab_dopp}. 
We obtain that the estimated value of the angular dimension $\theta$ lies between 0.004 -- 0.09 mas. Again, 
$\theta$ will be a factor of 4.47 higher if we use $\delta_{var,eq}$.  
In linear dimensions, the size of the emission region ranges between (0.6 -- 12.1)$\times10^{17}$ cm.

\subsubsection{Inverse Compton Doppler factor $\delta_{IC}$ } 
One can constrain the Inverse Compton Doppler factor ($\delta_{IC}$) 
by comparing the expected and observed fluxes at high energies 
\citep[see][for details]{ghisellini1993}. The IC Doppler factor is defined as 
\begin{eqnarray}
\delta_{IC}=\left[f(\alpha)S_{m}(1+z)\right]^{(4-2\alpha)/(10-6\alpha)}\left(\frac{ln(\nu_{c}/\nu_{m})\nu_{\gamma}^{\alpha}}{S_{\gamma}\theta^{(6-4\alpha)}\nu_{m}^{(5-3\alpha)}}\right)^{1/(10-6\alpha)}\,\,.
\label{doppler_IC_2}
\end{eqnarray}
where $\nu_{c}$ is the synchrotron high frequency cut-off in GHz, $S_{m}$ the flux 
density in Jy at the synchrotron turnover frequency $\nu_{m}$, $S_{IC}$ the observed 
$\gamma$-ray flux in Jy (assumed to arise from the IC process) at $\nu_{\gamma}$ in keV, 
$\alpha$ is the spectral index of the optically thin part of the spectrum, $\theta_{\nu}$ the 
source's angular size in mas and $f(\alpha)\simeq 0.14 - 0.08 \alpha$. The apparent variability 
size is calculated using eq. \ref{dim}. For the high energy 
cut-off we follow \citet{fuhrmann2008} and use $\nu_{c} \sim 5.5 \times 10^5$~GHz.  

For these  calculations, we used $t_{var}$ equals 9~days, which is the fastest variability 
timescale for the R6 flare at 86~GHz. $\alpha$ is obtained from the SSA modeling (see Section \ref{SSA}) 
 and the estimated values are given in Table \ref{tab_doppIC}. 
$\delta_{IC}$ is calculated for the same four time bins which we use to model the broadband SEDs of the 
source (see Section \ref{spectra_all}). The estimated values for $\delta_{IC}$ are given in 
Table \ref{tab_doppIC}. We find that during the 
four different activity states of the source $\delta_{IC} \geq 20$.   \\

\begin{table}
\caption{ Brightness temperature}
\begin{tabular}{l c l  }   
Time bin & $\delta_{IC}$              & Used parameters                         \\\hline  
Bin1&  $\delta_{IC, 0.5 KeV}$ $>$ 11  &  $S_{70}$ = 3.71 Jy, $\alpha$ = -0.74,  \\
    &                                 &  $S_{0.5 KeV}$ =2.97$\times 10^{-6}$ Jy \\
    &  $\delta_{IC, 7.5 KeV}$ $>$ 20  &  $S_{80}$ = 3.71 Jy, $\alpha$ = -0.74,  \\
    &                                 &  $S_{70.5 KeV}$ =8.27$\times 10^{-8}$ Jy \\
    &  $\delta_{IC, 100 MeV}$ $>$ 14  &  $S_{80}$ = 3.71 Jy, $\alpha$ = -0.74,  \\
    &                                 &  $S_{100 MeV}$ =1.17$\times 10^{-11}$ Jy \\
Bin2&  $\delta_{IC, 0.5 KeV}$ $>$ 11  &  $S_{40}$ = 1.68 Jy, $\alpha$ = -0.52,  \\
    &                                 &  $S_{0.5 KeV}$ =2.97$\times 10^{-6}$ Jy \\
    &  $\delta_{IC, 7.5 KeV}$ $>$ 14  &  $S_{40}$ = 1.68 Jy, $\alpha$ = -0.52,  \\
    &                                 &  $S_{7.5 KeV}$ =4.57$\times 10^{-8}$ Jy \\
    &  $\delta_{IC, 100 MeV}$ $>$ 14  &  $S_{40}$ = 1.68 Jy, $\alpha$ = -0.52,  \\
    &                                 &  $S_{100 MeV}$ =1.50$\times 10^{-10}$ Jy \\
Bin3&  $\delta_{IC, 0.5 KeV}$ $>$ 14  &  $S_{82}$ = 9.89 Jy, $\alpha$ = -0.76,  \\
    &                                 &  $S_{0.5 KeV}$ =3.85$\times 10^{-6}$ Jy \\
    &  $\delta_{IC, 7.5 KeV}$ $>$ 15  &  $S_{82}$ = 9.89 Jy, $\alpha$ = -0.76,  \\
    &                                 &  $S_{7.5 KeV}$ =2.97$\times 10^{-7}$ Jy \\
    &  $\delta_{IC, 100 MeV}$ $>$ 17  &  $S_{82}$ = 9.89 Jy, $\alpha$ = -0.76,  \\
    &                                 &  $S_{100 MeV}$ =2.94$\times 10^{-11}$ Jy \\
Bin4&  $\delta_{IC, 0.5 KeV}$ $>$ 12  &  $S_{78}$ = 3.85 Jy, $\alpha$ = -0.78,  \\
    &                                 &  $S_{0.5 KeV}$ =2.97$\times 10^{-6}$ Jy \\
    &  $\delta_{IC, 7.5 KeV}$ $>$ 12  &  $S_{78}$ = 3.85 Jy, $\alpha$ = -0.78,  \\
    &                                 &  $S_{7.5 KeV}$ =2.97$\times 10^{-6}$ Jy \\
    &  $\delta_{IC, 100 MeV}$ $>$ 12  &  $S_{78}$ = 3.85 Jy, $\alpha$ = -0.78,  \\
    &                                 &  $S_{100 MeV}$ =2.08$\times 10^{-11}$ Jy \\\hline 
\end{tabular}  \\
\label{tab_doppIC}
\end{table}

\subsubsection{Gamma-ray Doppler factor $\delta_{\gamma}$ } 
It is also possible to obtain a limit on the Doppler factor $\delta$ by considering that the 
high-energy $\gamma$-ray photons can collide with the softer radiation to produce e$^{\pm}$
pairs with the assumption that the bulk of the high-energy emission ($\gamma$-rays and X-rays) is 
produced in the same emission region. The cross-section of this process is maximized 
at $\sim \sigma_{T}/5$ \citep[see][for details]{svensson1987}, where $\sigma_{T}$ is the Thomson 
scattering cross-section. This leads to  a lower limit on $\delta$ with the
requirement that $\tau_{\gamma \gamma} (\nu) < 1$ \citep{dondi1995, finke2008}:
\begin{equation}
\delta_{\gamma} > \big [ \frac{2^{a-1} (1+z)^{2-2a} \sigma_T D_L^2}{m_e c^4 t_{var}} \epsilon f_{\epsilon^{-1}}^{syn} \big ]^\frac{1}{6 - 2a}
\end{equation}  
where $a$ is the power law index of the synchrotron spectrum i.e. $f_{\epsilon}^{syn} \propto \epsilon^{a}$, 
$\sigma_T$ is  the scattering Thomson cross-section, $m_e$ is the electron mass, $\epsilon_1 = E/(m_e c^2)$ 
is the dimensionless energy of a $\gamma$-ray photon with energy E for which the optical depth of the emitting region $\tau_{\gamma\gamma} = 1$. For the highest energy GeV (207~GeV)  
photon observed in the source (Rani et al. 2012), 
we obtain  $\epsilon = 207 \, {\rm GeV}/(5.11 \times 10^{-4} \, {\rm GeV}) = 4 \times 10^4$ and 
$\epsilon^{-1} = 2.4 \times 10^{-6}$.  Using 
$f_{\epsilon^{-1}}^{syn} = 3.88 \times 10^{-11}$~ergs~cm$^{-2}$~s$^{-1}$, we obtain $\delta_{\gamma} \geq 9.1$. 
The detection of the source at above 400~GeV \citep{anderhub2009} constrains $\delta_{\gamma} \geq 9.8$

\subsubsection{Magnetic field from synchrotron self-absorption }
It is also possible to constrain the magnetic field using the standard 
synchrotron self-absorption expressions. Following \citet{marscher1987},  
an expression for the magnetic field B in a homogeneous synchrotron self-absorbed region 
is given by: 
\begin{eqnarray}
B_{SA}\,[G]\,=10^{-5} b(\alpha) \theta^{4}\nu_{m}^{5}S_{m}^{-2}\left(\frac{\delta}{1+z}\right),
\label{B}
\end{eqnarray}
where $b(\alpha)$ depends on the optically thin spectral index $\alpha_{thin}$ 
\citep[see Table 1 in][]{marscher1987}, $S_m$ is the flux density, $\theta$ is 
the source's angular size at the synchrotron turnover frequency $\nu_{m}$ and 
$\delta$ is the Doppler factor.
The size of the emitting region responsible for the observed variations can be constrained 
using mm-VLBI measurements of the core region of S5~0716+714 by \citet{bach2006}:
$\theta < 0.04$~mas. Using $b(\alpha) = 3.13$, $S_m = 3.89$~Jy, $\nu_m = 80$~GHz, 
we calculate a lower limit of the magnetic field $B_{SA}$ in the range of (0.0078--0.0198)$\delta$ mG.
Using $\delta \geq 7$ at $\nu_m \sim 80$~GHz (see Table \ref{tab_dopp}), 
we obtained $B_{SA} \geq$ 0.05 to 0.14~G. 
The size of the emission region constrained using the causality arguments, $\theta \sim 0.0027$~mas 
at $\nu_m \sim 80$~GHz (see Table \ref{tab_dopp}) gives $B_{SA} \geq 0.03$~G.  These calculations 
constrain $B_{SA} \geq 0.03$ -- 0.14~G

\subsubsection{Equipartition Magnetic field and Doppler factor}
The equipartition magnetic field $B_{eq}$, which minimizes the total energy 
$E_{tot}=(1+k)E_{e}+E_{B}$ (with relativistic particle energy $E_{e}\sim B^{-1.5}$
and energy of the magnetic field $E_{B}\sim B^{2}$), is given by the following
expression  \citep[e.g.][]{bach2005, fuhrmann2008}:
\begin{eqnarray}
B_{eq}&=&\left(4.5\cdot(1+k)\,f(\alpha,\nu_{a},\nu_{b})\,L\,R^{3}\right)^{2/7}
\label{B_min}
 \end{eqnarray}
here $k$ is the energy ratio between electrons and heavy particles, $L$ is the synchrotron luminosity 
of the source given by $L=4\pi\,d_{L}^{2}(1+z)\int^{\nu_{a}}_{\nu_{b}}{S\,d\nu}$, $R$ is the size of the 
component in cm, $S_{m}$ is the synchrotron peak flux in Jy, $\nu_{m}$ is the synchrotron peak frequency
in GHz and $f(\alpha,\nu_{a},\nu_{b})$ is a tabulated function 
depending on the upper and lower synchrotron frequency

cutoffs $\nu_{a},\nu_{b}$. Using $\nu_{a}=10^{7}$\,Hz, $\nu_{b}=5.5\cdot10^{14}$\,Hz, and  
$f(-0.5, 10^7, 10^{11}) = 1.6 \times 10^7$, we obtain    
\begin{eqnarray}
B_{eq} = 5.37\times 10^{12}\left(k ~S_{m}\,\nu_{m}\,d_{L}^{2}\,R^{-3}\right)^{2/7} ~ G
\label{B_min}
 \end{eqnarray}
Using $B_{eq} \geq 0.14$ (see Section \ref{spectra_break}), $S_m = 3.89$~Jy, $\nu_m = 80$~GHz, 
$R = 2.90 \times 10^{16}$ -- $1.2 \times 10^{18}$~cm (estimated using 
$t_{var} = 25$~days at $\nu_m = 86$~GHz), the above expression yields $k = 1$. 
A small value of $k$  implies that the jet is mainly composed of electron-positron 
plasma. 

Equation \ref{B} and \ref{B_min} give different 
dependencies of the magnetic field on $\delta$, i.e. $B_{SA} \sim \delta$ and $B_{eq}\sim\delta^{2/7\alpha+1}$.
This yields $B_{eq}/B_{SA}=\delta_{eq}^{2/7\alpha}$. Adopting the above numbers,
we obtain Doppler factors $\delta_{eq,B}$ in the range of 14 -- 20 (for $\alpha$ = -(0.35 to 0.7)).  \\

\subsubsection{Comparison of the estimated parameters}
The apparent brightness temperature $T_B$ obtained from the day-to-day variations exceeds the theoretical 
limits by several orders of magnitudes. Although $T_B$ decreases towards the mm-bands, it is still 
higher than the IC-limit ($10^{12}$ K). $T_B$ exceeds $10^{14}$~K at 15~GHz and $10^{12}$~K at 230~GHz. 
We have obtained lower limits to the Doppler factor of the source using different methods as discussed 
in the earlier sections. These methods reveal a range of consistent 
lower limits to the Doppler factor with $\delta_{var} \geq 14$, 
$\delta_{IC} \geq 20$,  $\delta_{\gamma} \geq 10$ and $\delta_{eq,B} \geq 20$. 
Comparing the Doppler factor estimates obtained with different methods seems 
to suggest that $\delta \geq 20$. An independent 
approach to estimate $\delta$ is spectral modeling of the broadband SEDs, and this gives $\delta = 25$ 
(see Section \ref{spectra_all}), which 
is in agreement with the former values.  These limits are in good agreement with the  
estimates based on the recent kinematical VLBI studies of the source \citep{bach2006} and  the IC Doppler 
factor limits obtained by \citet{fuhrmann2008}. As  $\delta_{eq,B}$ agrees fairly well 
with the $\delta$ values derived from the other methods, we conclude that the emission 
region is in a state of equilibrium.   

The estimated magnetic field value from the broadband spectral 
modeling lies between 0.05 and 1~G.  A break in the optically thin power-law slope at 
a wavelength of $\sim 23 \, \mu$m constrains the equipartition magnetic field to 
$B_{eq} \geq 0.36$~G.  We obtained $B_{SA} \geq 0.14$~G from the 
synchrotron self-absorption calculation.     
The size of the emission region ($\theta$) derived on the basis of causality 
arguments lies between 0.004 -- 0.091 mas which agrees fairly well with the 
size of emission region constrained using mm-VLBI measurements \citep{bach2006}.

\subsection{The Complete spectral energy distribution}
\label{spectra_all}
\begin{figure*}
   \centering
\includegraphics[scale=0.30, angle=-90]{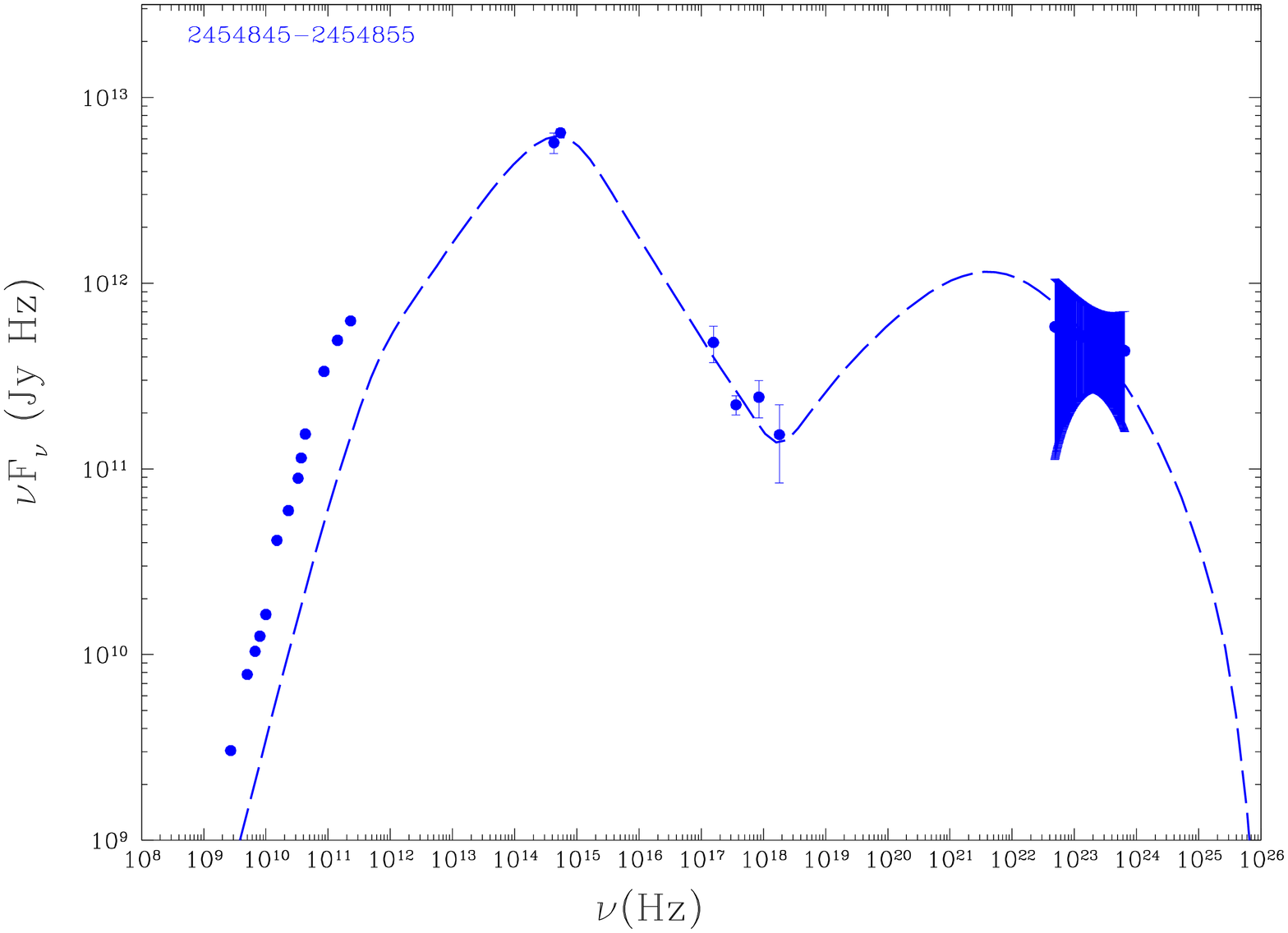}
\includegraphics[scale=0.30, angle=-90]{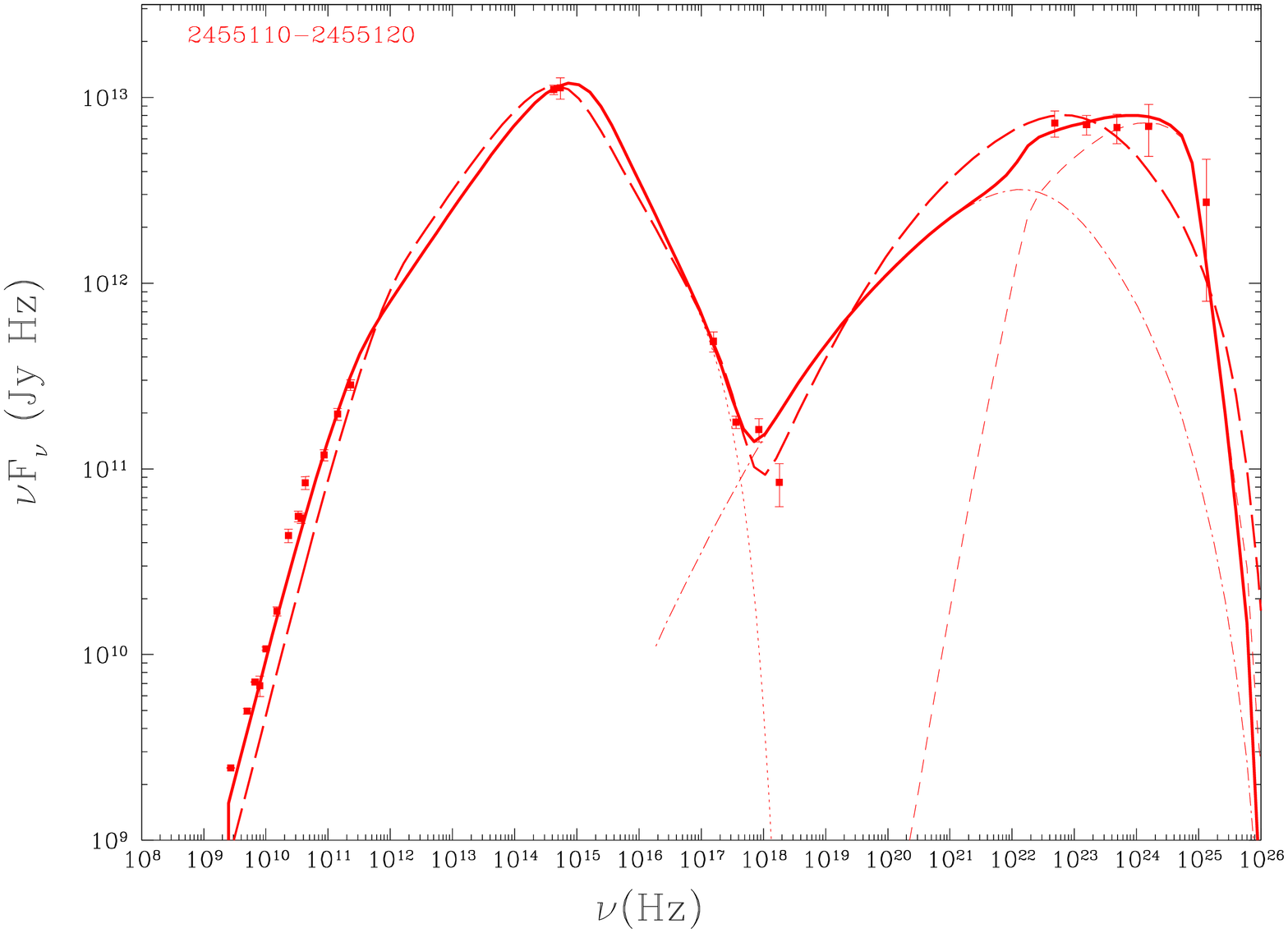}
\includegraphics[scale=0.30, angle=-90]{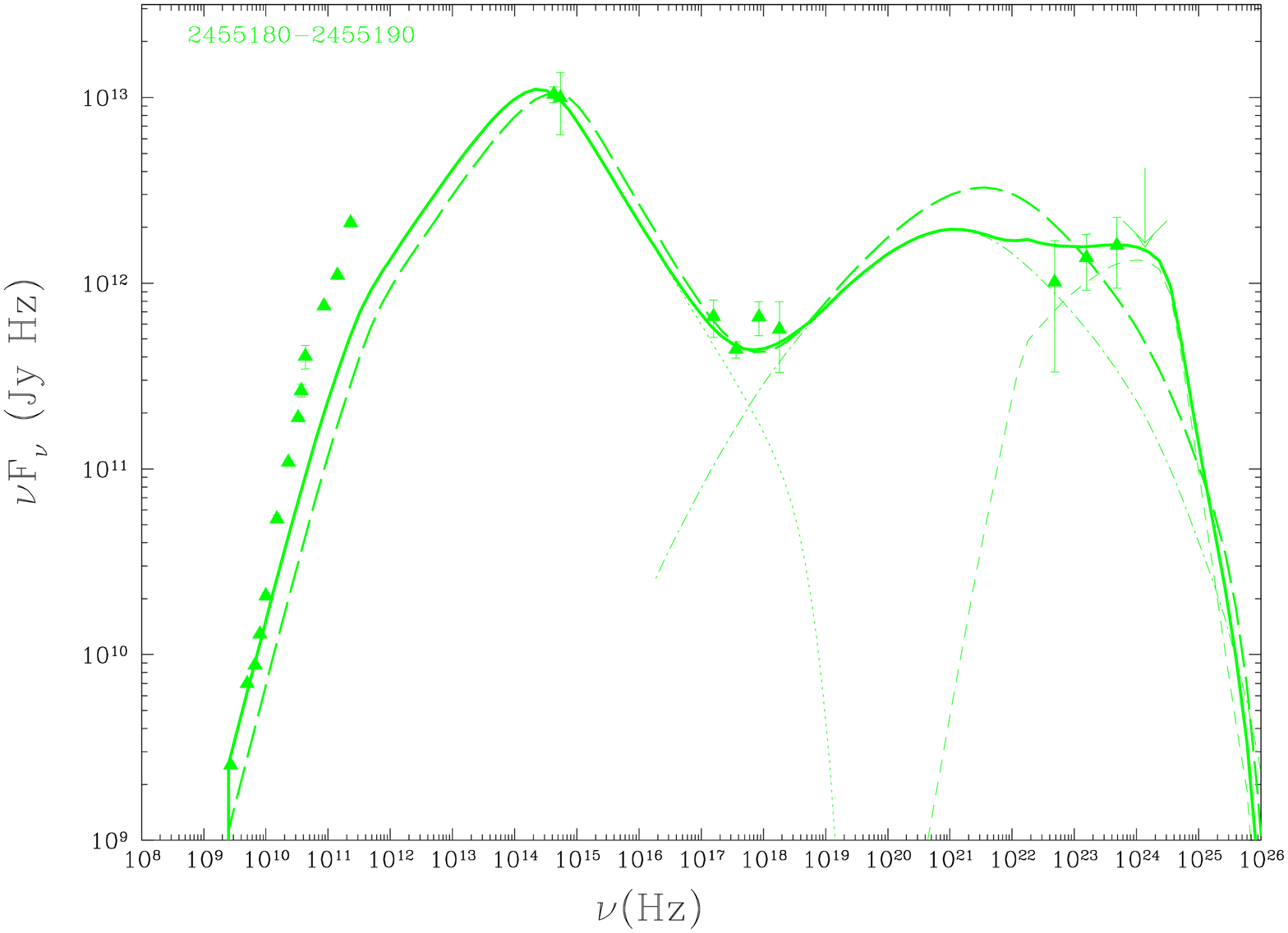}
\includegraphics[scale=0.30, angle=-90]{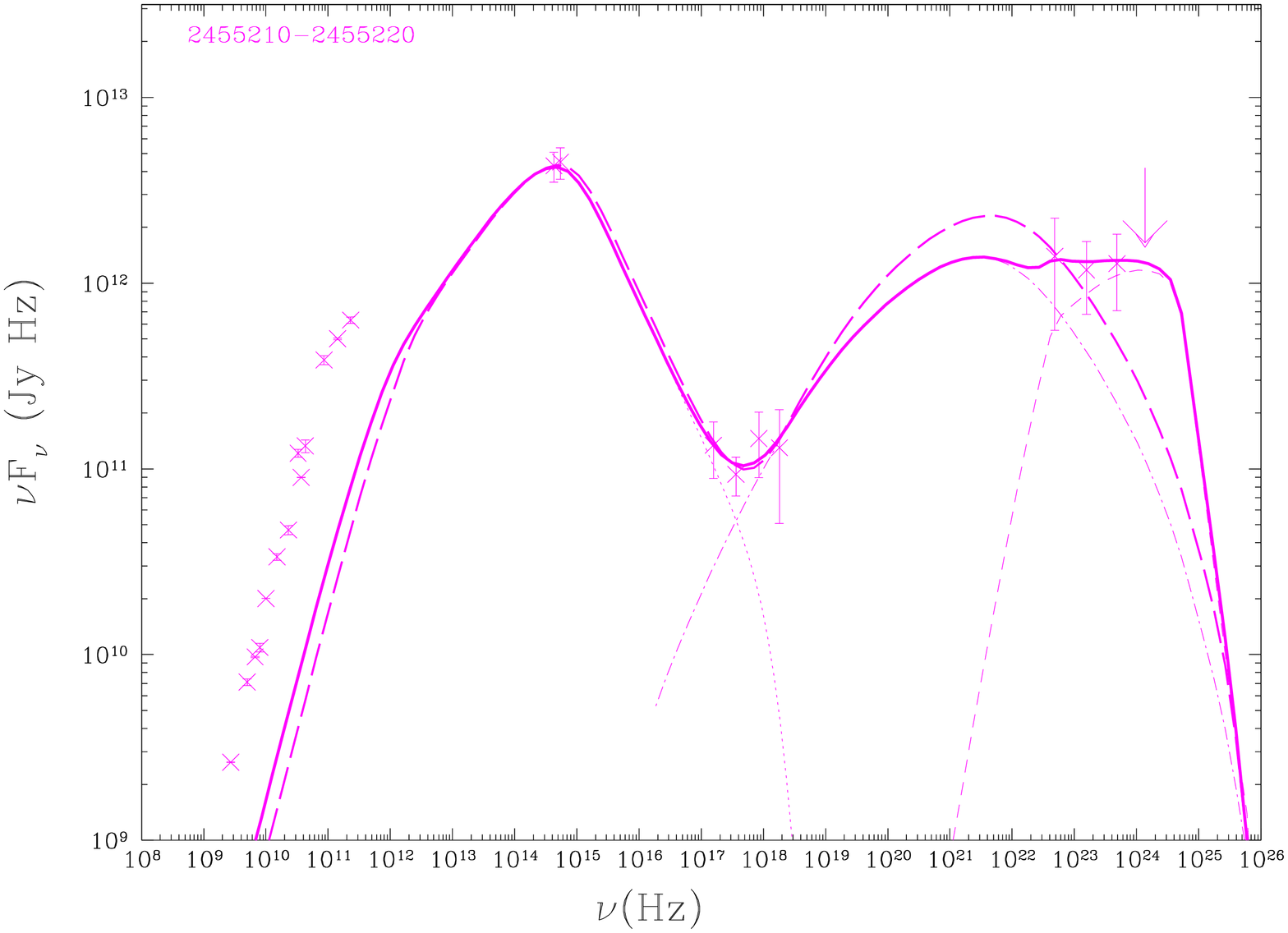}
   \caption{Broad band SEDs of S5~0716+714. Each SED is constructed
using 10-day averaged multi-frequency data. The error bars represent the variation of flux
over 10 days in each bin. Pure SSC models are shown as thick dashed curves. For EC fits, 
the total model SEDs are the thick solid curves; the synchrotron components are dotted, 
the SSC components are dot-dashed, and the EC components are thin dashed curves. }

\label{sed_full}
    \end{figure*}

The broadband monitoring of the source over several decades of frequencies allows us 
to construct multiple quasi-simultaneous SEDs. The SEDs of the source constructed 
over 4 different periods of time are shown in Fig. \ref{sed_full}. These time bins
($t_{bin}$) reflect different brightness states of the source and each time bin has a 
width of 10 days. The variation in flux over the bin width is shown as error bars 
in the SED plots. We construct the broadband SEDs for the following activity periods : \\
Bin1\footnote{we use `bin' for radio spectra and `Bin' for radio to GeV spectra}: 
: Radio-mm(steady), optical(high), X-ray(steady), GeV(low) \\ 
Bin2 : Radio-mm(low), optical(flaring), X-ray(low), GeV(flaring) \\
Bin3 : Radio-mm(flaring), optical(flaring), X-ray(flaring), GeV(low) \\
Bin4 : Radio-mm(steady), optical(low), X-ray(steady), GeV(steady) \\

\begin{table*}
\caption{Parameters of SSC and EC fits to SED of S5 0716+714 }
\begin{tabular}{l c c c c  c c c} \hline
Parameters   & Bin1 [JD'=845-855] &\multicolumn{2}{c}{Bin2} [JD'=1110-1120] &\multicolumn{2}{c}{Bin3} [JD'=1180-1190] &\multicolumn{2}{c}{Bin4} [JD'=1210-1220]  \\ 
             &  SSC             & SSC      &  EC     &   SSC   &  EC     &  SSC    &  EC     \\\hline
$\gamma_{1}$ & 2.5$\times 10^3$ & 1.1$\times 10^3$    & 4.0$\times 10^3$    & 2.5$\times 10^3$   & 1.8$\times 10^3$   &  3.0$\times 10^3$   &  2.5$\times 10^3$   \\
$\gamma_{2}$ & 1.0$\times 10^5$ & 2.6$\times 10^5$    & 6.5$\times 10^5$   & 2.0$\times 10^3$    & 2.0$\times 10^5$    &  1.0$\times 10^5$   &  1.0$\times 10^5$    \\
$q$          & 3.10             & 3.20      & 3.40     & 3.15    & 3.10     &  3.45   &  3.45    \\
$\eta$       & 25               & 100      & 25      & 25      & 25      &  25     &  25       \\
B (G)        & 1                & 0.05     & 0.7     & 0.9     & 0.95    &  0.8    &  1        \\
$\Gamma$     & 25               & 25       & 25      & 25      & 25      &  25     &  25        \\
$R_{b}$ (cm) & 1.25$\times 10^{16}$  &1.7$\times 10^{17}$ &2.0$\times 10^{16}$ &1.4$\times 10^{16}$ & 2.0$\times 10^{16}$ & 7.5$\times 10^{15}$ & 7.5$\times 10^{15}$  \\
$\theta$ (degree)  & 2.29           & 2.29     & 2.29    & 2.29    & 2.29    &  2.29   &  2.29       \\
L$_e$[$10^{44}$] (erg s$^{-1})$     &1.33   & 26.99    & 4.15    & 4.298   & 4.31    &  3.09   &  2.48       \\
$e_{B}$      & 1.61                 & 0.063    & 1.11    & 0.87    & 1.59    &  0.27 &  0.53       \\
T$_{ext}$ K  & --                   & --       & Ly-$\alpha$  & --    & Ly-$\alpha$  &  --     &  Ly-$\alpha$          \\
E$_{ext}$ (erg cm$^{-3})$ & --      & --       & 1.7$\times 10^{-5}$  & --      & 3.0$\times 10^{-6}$  &  --     &  1.0$\times 10^{-5}$       \\\hline
\end{tabular} \\
$\gamma_{1}$, $\gamma_{2}$ : Low- and High-energy cutoff \\
$q$ : Injection electron spectral index \\ 
$\eta$ : Electron escape timescale parameter \\
B (G) : Magnetic field at z=0  \\ 
$\Gamma$ : Bulk Lorentz factor \\
$R_{b}$ (cm) : Blob radius  \\
$\theta$ (degree) : Observing angle  \\
L$_e$[$10^{44}$] : Electron power \\
$e_{B}$ : Magnetic field equi-partition parameter \\
T$_{ext}$ : External radiation peak photon energy \\ 
E$_{ext}$ : External radiation field energy density  
\label{ssc_ec_par}
\end{table*}

The double-humped structures of the broadband SEDs can usually be modeled by 
both leptonic and hadronic models \citep[e.g.][]{br2012}. 
Here we have used a quasi-equilibrium version of a leptonic one-zone jet model as described by 
\citet{br2012}.    
In this model, the observed radiation is assumed to be originating from the ultra-relativistic 
electrons (and positrons) in a spherical emission region of co-moving radius R$_{B}$ propagating
with relativistic speed $\beta_{\Gamma} c$ ($\Gamma$ is bulk Lorentz factor) along the jet, 
which is offset by an angle $\theta$ w.r.t the line-of-sight.  
We fix $\theta$ to be such that  the bulk Lorentz factor, $\Gamma$ 
equals the Doppler factor, $\delta$, which, for highly relativistic motion ($\Gamma \gg 1$) 
implies $\theta = 1/\Gamma$.
The emitting electrons are assumed to be instantaneously accelerated into a power-law distribution 
of electron energy, $E_e = \gamma m_e c^2$, of the form $Q(\gamma) = Q_{0} \gamma^{-q}$ with $q$ being the 
injection electron spectral index and  $\gamma_1$ and $\gamma_2$ are the low- and the high-energy cutoffs.

An equilibrium in the emission region between particle injection, radiative cooling, and escape of particles
from the emission region yields a temporary quasi-equilibrium state described by a broken 
power law. The particle escape is parametrized through an escape timescale parameter 
$\eta  > 1$ so that $t_{esc} = \eta R/c$. The balance between the particle escape and radiative cooling
will lead to a break in the equilibrium particle distribution at a break Lorentz factor 
$\gamma_{b}$, where $t_{esc} = t_{cool}(\gamma)$. The cooling timescale is calculated  
self-consistently taking into account  synchrotron, SSC and EC cooling.
Depending on whether $\gamma_{b}$ is greater than or less than $\gamma_{1}$, the
system will be in the slow cooling or fast cooling regime, respectively, 
leading to different spectral indices of the equilibrium electron 
distribution \citet{bottcher2002a}. 

In the fitted model the number density of injected particles is normalized to the 
resulting power in ultra-relativistic electrons propagating along the jet given by,
\begin{equation}
L_e = \pi R_{e}^{2} \Gamma^{2} \beta_{\Gamma} c m_e c^2 \int_1^	\infty \! \gamma 
\, n(\gamma) \, \mathrm{d} \gamma
\end{equation} 

The magnetic field is considered as a free parameter in the emission region. The Poynting 
flux along jet is $L_{B} = \pi R_{e}^{2} \Gamma^{2} \beta_{\Gamma} c u_{B}$, where 
$u_{B} = B^{2}/(8 \pi)$ is the magnetic field energy density. 
The equipartition parameter $e_B = L_B/L_e$ is calculated for each fitted model. 

After evaluation of the quasi-equilibrium particle distribution in the emission
region, our code calculates the radiative output from the synchrotron, SSC, 
and EC emissions self-consistently with the radiative cooling rates. 
The external radiation field, which serves as seed photons for EC scattering,
is assumed to be isotropic in the stationary AGN rest frame. Its spectrum can
be chosen to be a thermal blackbody with temperature $T_{ext}$ and radiation energy 
density $u_{ext}$, or a line-dominated spectrum (or a combination of the two). 
The direct emission from this external radiation field is added to the emission 
from the jet to yield the total modeled SED, which we fit to the observations. 

We first tried to fit the SEDs with a pure SSC model, as this has fewer free parameters than
the EC version of the model. However, except for the SED of bin 1 (see below), pure SSC models
typically fail to reproduce the {\it Fermi}/LAT spectra of the SEDs. 
Also a detailed study of the $\gamma$-ray spectrum of the source (Rani et al. 2013),
shows a  significant correlation between detection of the high energy GeV photons and change in
spectral slope below and above the break energy, which suggests that BLR opacity
has a significant impact on the observed spectral breaks. 
Therefore, we included an external radiation component, as outlined above, to produce SSC+EC fits.

The fitted models are shown in Fig. \ref{sed_full} and the best-fit parameters are 
given in Table \ref{ssc_ec_par}. The pure SSC model does a moderately good job in 
describing the SEDs of the low states, although the $\gamma$-ray spectra appear systematically
too steep. The SED of Bin1 is well fitted with the SSC model, while for the other time bins 
an EC component is required to fit the GeV spectra. The high-state is very problematic for 
the SSC model as it would require a much lower magnetic field (far below equipartition) and --
in the case of Bin 2 -- a very large emission region, in conflict with the often observed
intraday optical variability. All the low-state fits are possible with parameters close to 
equipartition between relativistic electrons and the magnetic field.
A model including external Compton generally does a better job in reproducing the entire 
SEDs (including the $\gamma$-ray spectrum), if one uses an external radiation field dominated 
by Ly-$\alpha$ emission from a putative broad line region (BLR). 
For S5 0716+714, we found that the radiation field energy density of this external field
varies between 10$^{-6}$ to 10$^{-5}$ ergs cm$^{-3}$, which is a factor of $\sim$1000 lower
than what we expect for a typical quasar. However, this is a reasonable value for a BL Lac
like S5 0716+714 which is known to have a featureless spectrum. Furthermore, this low BLR energy 
density value naturally explains the origin of $\gamma$-ray spectral breaks observed in the source. 
Moreover, the low BLR energy density is consistent with the non-detection of emission lines. 
Parameters close to 
equipartition can be used for all time bins, including the high states.

At first glance the fits look good, but in more detail the fit to the radio data for some bins is relatively
poor. In the EC model, the model fits the cm-radio data quite well, but is much below the mm data for Bin3. 
The model for Bin4 does not fit the radio data at all (see Fig. \ref{sed_full}). So, in general the model 
under-predicts the radio flux at mm and cm bands. This indicates the possibility of a missing spectral 
component at cm-mm wavelengths. We have seen in Section 
3.3.1 that a two-component model better describes the radio spectra. Therefore, we conclude that an additional 
synchrotron component is required to fit the broadband SED at mm to cm wavelengths.

\section{Discussion }
The densely sampled multi-frequency observations of the BL Lac object S5~0716+714 over 
the past three years allow us to study its broadband flaring behavior and  
probe into the physical process, location and size of the emission regions. 
We found a direct connection between GeV and optical flares, and major flares propagate 
down to radio wavelengths. The radio outbursts seem to be smeared out at 10~GHz and lower 
frequencies. An orphan X-ray flare lags the major optical-GeV flare (O5-G5) by 
$\sim 55$~days and the X-ray emission is produced by both synchrotron and 
inverse-Compton mechanisms. It seems that 
the interaction of shocks with the underlying jet structure might be responsible for optical 
and high energy emission, and opacity plays a key role in the time-delayed emission 
at radio wavelengths.

\subsection{Broadband correlation alignment}
\label{broad_corr}
\begin{figure}
   \centering
\includegraphics[scale=0.4]{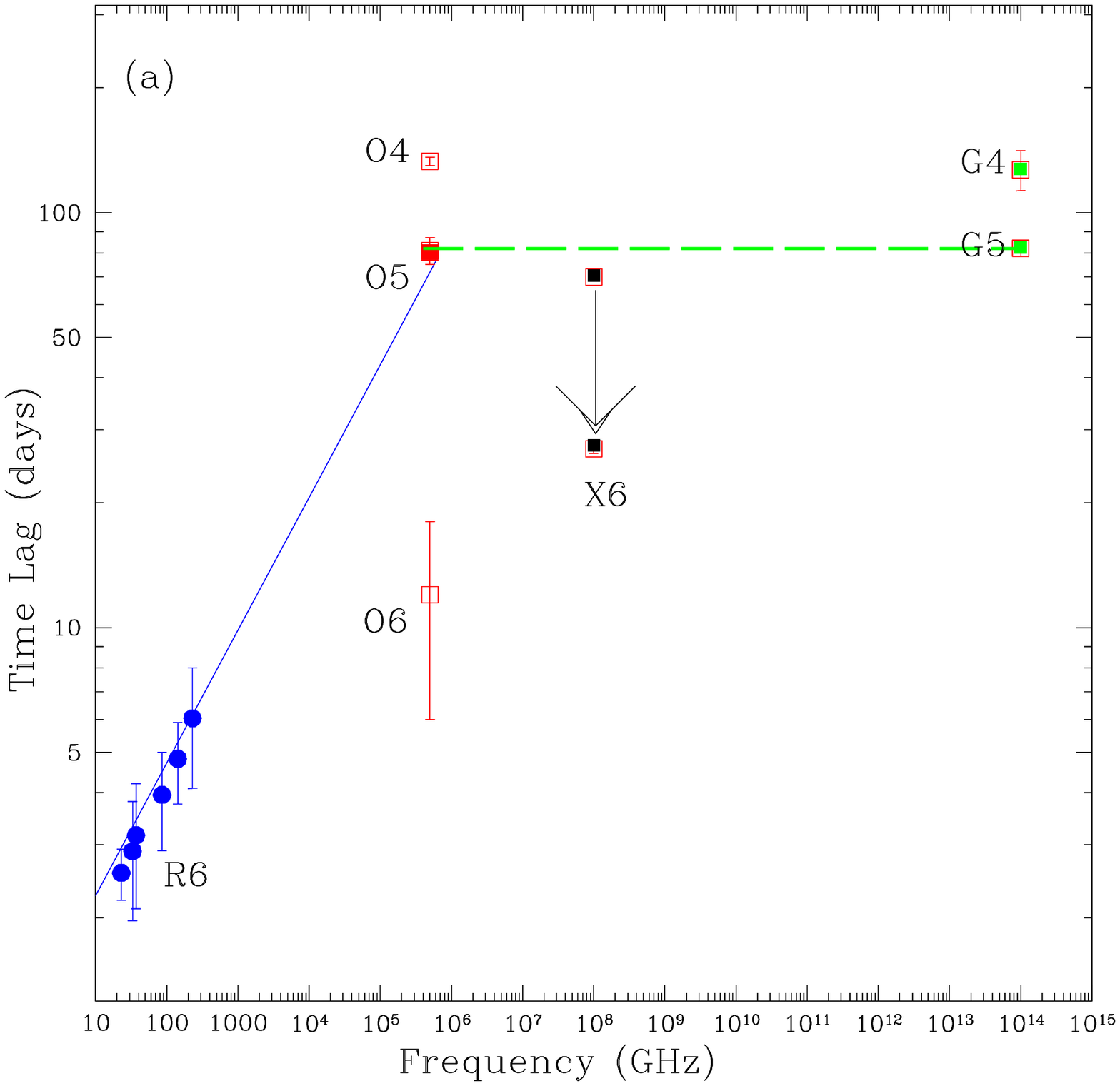}
\includegraphics[scale=0.4]{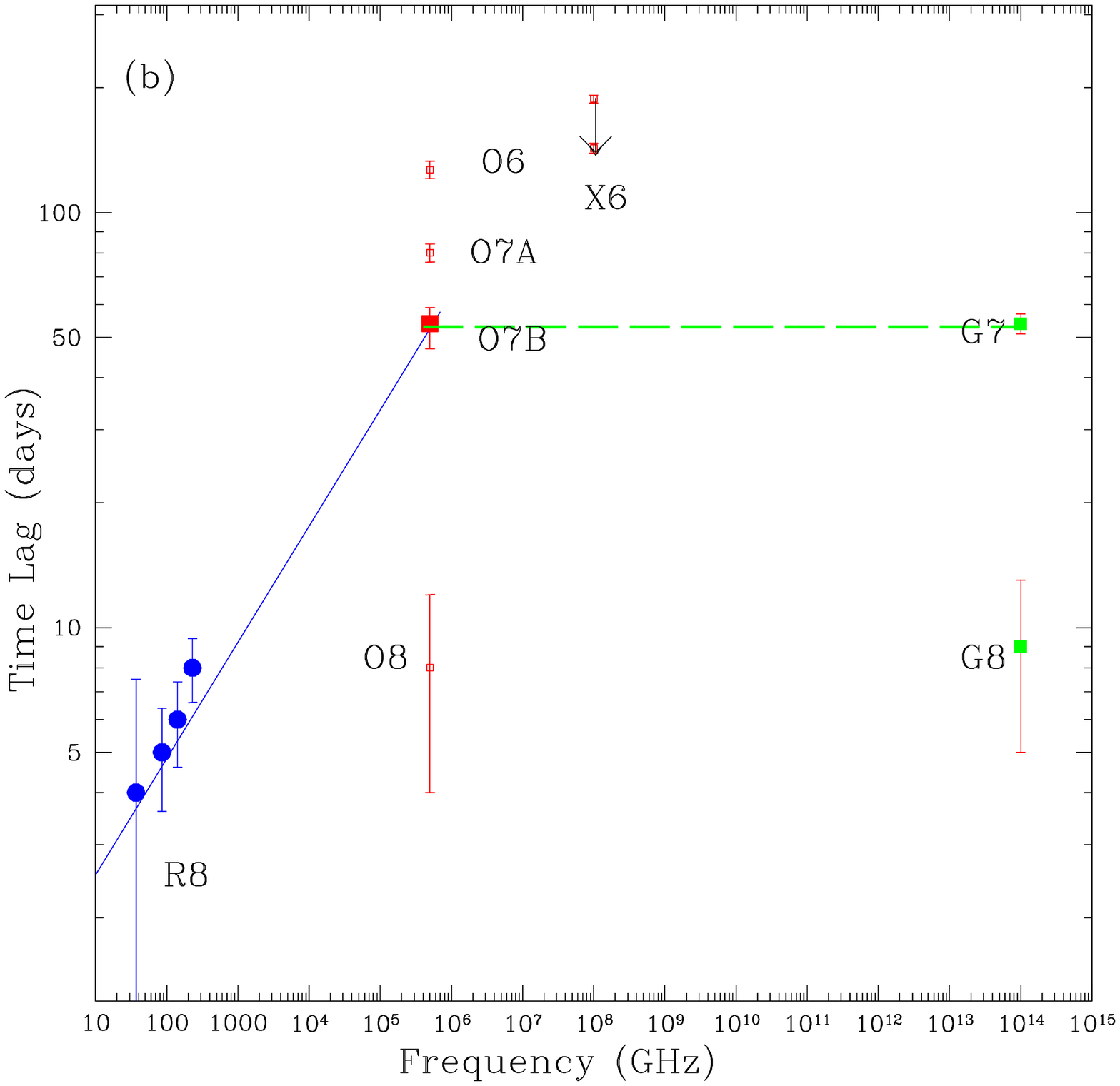}
   \caption{(a) A plot of time lag measurements vs frequency using 15 GHz as the
reference frequency for the radio flare R6. The best fitted power law at
radio and mm frequencies is extended up to the optical wavelengths. The near by optical, 
X-ray and GeV flares are shown with their possible time lags w.r.t. R6.
(b) A similar plot for the R8 flare (see Fig. \ref{plot_flx_total_LC} for flare labeling) 
In both plots, the dashed lines indicate the SSC process with simultaneous optical--$\gamma$-ray 
events.
}
\label{lag_frq_FlareA}
    \end{figure}

Following the broadband cross-correlation analysis (\ref{sec_corr}), we plot the estimated time lag as a 
function of frequency in Fig. \ref{lag_frq_FlareA}. Fig. \ref{lag_frq_FlareA} (a) shows the plot of the 
time lag measurements at different frequencies for the R6 flare using 15~GHz as the reference frequency 
(see Fig. \ref{plot_flx_total_LC}). It has become evident in Section \ref{sec_radio_dcf} that the 
time lags (w.r.t. 15~GHz) increase with frequency and 
follow a power-law as a function of frequency with a slope $\sim 0.3$. 
If we extend the fitted power law  to optical frequencies, then the R6 flare meets the O5 flare, 
which is observed $\sim 60$~days earlier than the R6 flare. 
A formal cross-correlation between optical and radio frequency light curves indicates a 
significant correlation with a delay of $\sim 60$~days at radio wavelengths. 
The dashed line in Fig. \ref{lag_frq_FlareA} (a) connects the simultaneous optical -- GeV flares. 
The  optical-GeV correlation shows no time lag among the flares at the two frequencies, 
i.e. O5 correlates with G5 and O4 with G4; but there is no respective GeV flare for O6. 
The nearby optical, X-ray and GeV flares are shown with their possible time lags w.r.t. R6. 
The allowed time range of the peak of the X-ray flare is marked 
with an arrow.  In Figure \ref{lag_frq_FlareA} (b), we show a similar plot for the R8 flare.
Both of these figures provide a one-to-one connection of the broadband flares based on our  
analysis.

\subsection{Origin of Optical variability}
During our observations, the source was highly active at optical frequencies showing multiple 
flares roughly separated from each other by $\sim 60$ --- 70~days, superimposed on a long-term 
variability trend at a $\sim 350$~day timescale. 
The periodogram analysis reveals two significant peaks at $\sim 63$ and 359~day timescales. A more 
robust analysis using the power spectrum density method implies that the significance of a detection 
of a quasi-periodic signal at the frequency corresponding to these timescales is only 2 $\sigma$. Thus, 
the significance of detection remains marginal. However, it is important to note that periodic variations 
at a year timescale has also been observed earlier in the source \citep{raiteri2003}. 

During the two years of our observations, we found that the long-term variability amplitude of the source 
remains almost constant at about 1.3 magnitudes. A constant variability amplitude 
can be interpreted in terms of variations of the Doppler boosting factor \citep{raiteri2003}. 
The change in $\delta$ can be due to either a viewing angle ($\theta$) variation or a change of the 
bulk Lorentz factor ($\Gamma$) or maybe a combination of both. We notice that a change in $\delta$ by a 
factor of 1.2 can be easily interpreted as a few degree variation in $\theta$, while it requires a noticeable 
change of the bulk Lorentz factor. 
We therefore propose that the geometry significantly affects the long-term 
flux base-level modulations. 
Such variations are very likely originating as a relativistic shock 
traces a spiral path through the jet \citep{marscher1996}.

\subsection{Origin of $\gamma$-rays}
The source displays substantial activity at $\gamma$-rays during the high optical activity period. 
This is to be expected in leptonic models, as the same electrons radiating the optical synchrotron 
photons would emit $\gamma$-rays through the inverse Compton scattering process. Here, we observe 
a similar flaring behavior at the two frequencies. We also found that the 
flux variations at optical and GeV frequencies are significantly correlated with each other (on weekly 
timescales) and corresponding to each optical flare ``O1" to ``O9" (except O6) there is a local 
maximum ``G1" to ``G9" at GeV frequencies.  In addition to that the variability timescales 
(both short and long) are also comparable 
for optical and GeV light curves. We note that the ratio between the high and low $\gamma$-ray flux levels 
is about 15, while in the optical band the
same ratio is of the order of 3.7. Thus, the $\gamma$-ray flux density appears to vary as the square of change in 
the optical flux density. This reflects a quadratic dependence of the GeV flux variations compared to optical variability. 
This favors a SSC interpretation.  However, we would also like point out that a weak EC contribution is also 
required in order to model the GeV spectrum of the source. 


\subsection{Opacity and delay at radio wavelengths}
The source reaches a maximum in simultaneous optical -- GeV flaring activity at ``5" (see Fig. \ref{plot_flx_total_LC}, flares: O5-G5). This maximum 
coincides with the beginning of a major radio outburst ``R6" at 230~GHz. The R6 radio flare 
is observed $\sim 65$~days later than the optical flare O5 at 230~GHz. The R6 flare is 
followed by another radio flare, R8, with a moderate level of flux activity between the two.   
The two major 230~GHz radio outbursts (R6 and R8) are smoothed and delayed at lower radio frequencies till 
15~GHz. The flaring activity seems to be completely washed out at $\le 10$~GHz. The estimated time lag 
(using 15~GHz as reference frequency) at each frequency as a function of frequency follows a power law 
with a slope $\sim 0.3$. Delayed emission at lower frequencies is a clear indication of opacity 
effects due to synchrotron self-absorption \citep{kudryavtseva2011}. 

As per the cross-correlation analysis, the optical -- radio variability is found to be 
significantly correlated, with the flux variations at optical frequencies leading those 
at radio bands by $\sim 60$~days (see Section \ref{dcf_opt_rad}). Most earlier studies 
on the radio -- optical correlation have shown that the radio events lag behind the optical 
ones by several weeks or months \citep[e.g.][and references therein]{tornikoski1994, clements1995, 
villata2007, raiteri2003, jorstad2010, agudo2011}. 
Similar variability timescales ($\sim 90$ and 180~days) at optical and radio frequencies 
again hint at a co-spatial origin of variability. It is worth pointing out that the long 
term variability timescales are common at optical (and also at $\gamma$-rays) 
and radio frequencies. The fast repetitive optical/$\gamma$-ray flares are not observed at 
radio wavelengths. Therefore, it is not unreasonable to suggest that the long term 
variability features observed at optical/GeV frequencies 
propagate down to radio frequencies with a time lag of $\sim 60$~days.    
As we notice in Section \ref{SSA}, the two radio outbursts are low-peaking flares. 
Thus, a 60~day time delay between the optical and radio activity emphasizes the optical 
flares being the precursor of the radio flares \citep{valtaoja1992}.

\subsection{Origin of the orphan X-ray flare}
When the source is flaring at optical -- GeV frequencies, it is quiet at X-rays. 
Although it is hard to locate the exact peak time of the X-ray flare, it is 
obvious that the maximum of the X-ray flux peaks $\sim 50$~days later than the 
major optical -- GeV flares (O5-G5) (see Fig. \ref{plot_flx_total_LC}). 
At this epoch, the source was in a relatively steady state at optical/GeV frequencies 
while there is another bright optical flare lagging the X-ray maximum by  
$\sim 10$~days. We also notice that the fractional variability of the source is comparable 
at soft (22.5~\%) and hard (25~\%) X-rays. 
Interestingly, we do not find any significant correlation among the X-ray 
spectral index and flux. This may be due to the poor data sampling or may be intrinsic to the source. 
The concave shape of the X-ray spectrum (see Section \ref{spectra_all}),  
suggests that the X-ray emission shows a combination of synchrotron and inverse-Compton mechanisms 
which could prevent the source from exhibiting any steepening or hardening trend during the flare.

A similar orphan X-ray flare was also observed in the blazar 3C 279 by \citet{abdo2010c} with X-ray 
flaring activity lagging optical -- GeV flares by 60~days. The authors argued that X-ray photons are 
produced further down to the jet 
compared to optical -- GeV photons. \citet{hayashida2012} argued   
in the context of a two component model; the X-ray flare is produced by the low-frequency component which is 
less variable compared to the high-frequency component.
Although we do not completely understand the origin of the orphan X-ray flare in S5~0716+714,  
it appears possible that the X-ray emission is not co-spatial with the optical/$\gamma$-ray emission in this 
event.  We notice some low level flux activity 
(mini flare, say R7) in between the two major radio flares (``R6" and ``R8", see Fig. \ref{plot_flx_total_LC}). 
While modeling the radio spectra of the source, we also noticed that a two-component model better describes 
the synchrotron spectra of the source over this period. 
This indicates that either multiple shocks are hitting the emission region which at first produces the major 
flare ``O5/G5-R6", then ``X6/O6-R7" and later ``O7/G7-R8", or  the radiation is contributed by two synchrotron 
components with the low-frequency component producing the X-ray flare.

\section{Conclusions}
In this paper we presented the results of the radio to $\gamma$-ray monitoring of S5~0716+714 from 
April 2007 to January 2011. The source was very active at optical and higher frequencies.  
Two major radio outbursts were observed during this high activity period. 
From the rapid rise and decay, we derive variability brightness temperatures exceeding the IC limit, 
which at least for mm flares is a very unique behavior.  

A long-term variability trend ($\sim 350$~days timescale) is visible in the optical light curves which is 
superimposed with repetitive variations on shorter time scales ($\sim 60$~day). 
A comparison of the various flaring episodes of S5~0716+714 strongly indicates a one-to-one correlation 
between the strength of the $\gamma$-ray emission and the strength of the optical emission.
A quadratic dependence of the amplitude of the $\gamma$-ray variability with respect to that 
of the optical favors an SSC explanation. 

The high-energy (optical -- GeV) flares propagate down to radio frequencies with a time delay of 
$\sim 65$~days following a power-law dependence on frequency with a slope $\sim 0.3$. 
This indicates that opacity plays a key role in producing time delays among light curves at 
optically thin and thick wavelengths. Since the radio outbursts are low-peaking flares, such 
a long time lag is only possible in the case of optical flares being the precursors of radio 
ones. The evolution of the radio flares are in agreement with the 
generalized shock model proposed by \citet{valtaoja1992}.
The evolution of the flare in the turnover frequency -- turnover flux density ($\nu_m - S_m$) plane shows 
a very steep rise and decay over the Compton and adiabatic stages with a slope too high to be expected from 
intrinsic variations, requiring an additional Doppler factor variation along the jet. 
For the two flares, we notice that $\delta$ changes as  $R^{0.7}$ during the rise and as $R^{2.6}$ during 
the decay of R6 flare. The evolution of the R8 flare is governed by $\delta \propto R^{0.4}$ during the
rising phase and $\delta \propto R^{-2.0}$ during the decay phase of the flare.

An orphan X-ray flare is observed $\sim 50$~days after the major optical -- GeV flares. 
The detection of an isolated X-ray flare challenges the simple one-zone emission models, rendering 
them too simple. The lack of substantial observations over the flaring epoch makes it even more 
complicated to understand.  We found that this flare has equal contributions from both the synchrotron 
and the high-energy (inverse-Compton in a leptonic model interpretation) emission mechanisms.

We model the broadband SEDs of the source using two different versions of leptonic models: 
a pure SSC and SSC+EC. 
We found that the low activity states of the source are well described by a pure SSC model while an EC contribution is required 
to reproduce the SEDs for high states.  The SSC+EC model returns magnetic field parameter value closer to equipartition, 
providing a satisfactory description of the broadband SEDs. We found that satisfactory model
fits can be achieved if the external radiation field is dominated by Ly-$\alpha$ emission from the 
broad-line region. This model nicely describes the broadband SEDs of the source at optical 
and higher frequencies, but  under-predicts the cm -- mm spectra at least for few time periods. A separate synchrotron component
seems required to fit the cm -- mm radio fluxes. This may also provide a hint towards the origin 
of the orphan X-ray flare.

\begin{acknowledgements}
The {\it Fermi}/LAT Collaboration acknowledges the generous support of a number of agencies
and institutes that have supported the {\it Fermi}/LAT Collaboration. These include the National
Aeronautics and Space Administration and the Department of Energy in the United States, the
Commissariat \`a l'Energie Atomique and the Centre National de la Recherche Scientifique / Institut
National de Physique Nucl\'eaire et de Physique des Particules in France, the Agenzia Spaziale
Italiana and the Istituto Nazionale di Fisica Nucleare in Italy, the Ministry of Education, 
Culture, Sports, Science and Technology (MEXT), High Energy Accelerator Research Organization 
(KEK) and Japan Aerospace Exploration Agency (JAXA) in Japan, and the K.\ A.\ Wallenberg 
Foundation, the Swedish Research Council and the Swedish National Space Board in Sweden.

This research is partly based on observations with the 100-m telescope of the 
MPIfR (Max-Planck-Institut f\"ur Radioastronomie) at Effelsberg. This work has 
made use of observations with the IRAM 30-m telescope. IRAM is supported by 
INSU/CNRS (France), MPG (Germany) and IGN (Spain).
The Submillimeter Array is a joint project between the Smithsonian
Astrophysical Observatory and the Academia Sinica Institute of Astronomy
and Astrophysics and is funded by the Smithsonian Institution and the
Academia Sinica.  The observations made use of the Noto telescope operated by 
INAF - Istituto di Radioastronomia.

BR gratefully acknowledges the travel support the COSPAR Capacity-Building Workshop fellowship program.
IN was supported for this research through a stipend from the International Max
Planck Research School (IMPRS) for Astronomy and Astrophysics at the Universities of Bonn and Cologne.
N.M. is funded by an ASI fellowship under contract number I/005/11/0.
We also acknowledge the Swift Team and the Swift/XRT monitoring program efforts, as well as analysis 
supported by NASA Fermi GI grants NNX10AU14G. 
LX is supported by the National Natural Science Foundation of China (No.11273050). 
Work at UMRAO has been supported by a series of grants from the NSF and from NASA. Support 
for operation of the observatory was provided by the University of Michigan.
IA acknowledges funding by the ``Consejer{\'i}a de Econom{\'i}a, Innovaci{\'o}n y Ciencia" of the Regional 
Government of Andaluc{\'i}a through grant P09-FQM-4784, an by the ``Ministerio de Econom{\'i}a y Competitividad" 
of Spain through grant AYA2010-14844.
The Mets\"ahovi team acknowledges the support from the Academy of Finland
to our observing projects (numbers 212656, 210338, 121148, and others). 
We thank Ivan Agudo for his contribution at the 30m telescope and for discussion. 
We would like to thank Marcello Giroletti, the internal referee from {\it Fermi}/LAT team for 
his useful suggestions and comments. We thank the referee for several helpful suggestions. 
We would also like to thank Jeff Hodgson for the help in finalizing the text. 
\end{acknowledgements}


\end{document}